\documentclass[longauth]{aa}

\newcommand{\uv}{\ensuremath{( u, v)}\xspace}

\newcommand{\fso}{\ensuremath{f_0^*}\xspace}
\newcommand{\denv}{\ensuremath{d_\mathrm{env}}\xspace}

\newcommand{\modif}[1]{#1}
\newcommand{\modiff}[1]{#1}
\newcommand{\modifff}[1]{#1}
\usepackage{siunitx}
\usepackage{graphicx}
\usepackage{color}
\usepackage{ amssymb }
\usepackage{dcolumn}
\newcolumntype{d}[1]{D{.}{\cdot}{#1} }
\usepackage{txfonts}
\begin{document} 

\title{A family portrait of disk inner rims around Herbig\,Ae/Be stars:\thanks{Based on observations made with ESO Telescopes at the La Silla Paranal Observatory under program ID 190.C-0963.} }
\subtitle{Hunting for warps, rings, self shadowing, and misalignments in the inner astronomical units.}


\author{J. Kluska
          \inst{1,2}
          \and
          J.-P. Berger\inst{3} \and
          F. Malbet\inst{3} \and
          B. Lazareff\inst{3} \and
          M. Benisty\inst{4,3} \and
          J.-B. Le Bouquin\inst{3} \and 
O.~Absil \inst{5}\fnmsep\thanks{F.R.S.-FNRS Research Associate} \and F. Baron\inst{6} \and A. Delboulbé\inst{3} \and G. Duvert\inst{3} \and A. Isella\inst{7} \and L. Jocou\inst{3} \and A. Juhasz\inst{8}, S. Kraus\inst{2} \and R. Lachaume\inst{9,10} \and
F. Ménard\inst{3} \and R. Millan-Gabet\inst{11} \and J. D. Monnier\inst{12} \and T. Moulin\inst{3} \and K. Perraut\inst{3} \and C. Pinte\inst{13, 3} \and S. Rochat\inst{3}  \and F. Soulez\inst{14} \and M. Tallon\inst{14} \and W.-F. Thi\inst{15} \and E. Thiébaut\inst{14} \and W. Traub\inst{16} \and G. Zins\inst{17}
     }

  \institute{Instituut voor Sterrenkunde (IvS), KU Leuven, Celestijnenlaan 200D, 3001 Leuven, Belgium\\
 \email{jacques.kluska@kuleuven.be}
         \and
         University of Exeter, School of Physics and Astronomy, Stocker Road, Exeter, EX4 4QL, UK
         \and
         Univ. Grenoble Alpes, CNRS, IPAG, F-38000 Grenoble, France
         \and
         Unidad Mixta Internacional Franco-Chilena de Astronomía, CNRS/INSU UMI 3386 and Departamento de Astronomía, Universidad de Chile, Casilla 36-D, Santiago, Chile
         \and
Space sciences, Technologies, and Astrophysics Research (STAR) Institute, University of Li\`ege,Li\`ege, Belgium
         \and 
         Department of Physics and Astronomy, Georgia State University, Atlanta, GA, USA
         \and
         Department of Physics and Astronomy, Rice University, 6100 Main Street, Houston, TX 77005, USA  
         \and 
         Institute of Astronomy, University of Cambridge, Madingley Rd, Cambridge CB3 0HA, UK
         \and 
         Instituto de Astrofísica, Facultad de Física, Pontificia Universidad Católica de Chile, Av. Vicuña Mackenna 4860, 7820436 Macul, Santiago, Chile
         \and 
         Max-Planck-Institut für Astronomie, Königstuhl 17, D-69117 Heidelberg, Germany
         \and 
         NASA Exoplanet Science Institute, MS 100-22, California Institute of Technology, Pasadena, CA 91125, USA
         \and 
         Department of Astronomy, University of Michigan, 1085 S. University Avenue, Ann Arbor, MI 48109, USA
         \and 
         Monash Centre for Astrophysics (MoCA) and School of Physics and Astronomy, Monash University, Clayton, Vic 3800, Australia
         \and 
         Univ Lyon, Univ Lyon1, Ens de Lyon, CNRS, Centre de Recherche Astrophysique de Lyon UMR5574, F-69230, Saint-Genis-Laval, France
         \and 
         Max-Planck-Institut für extraterrestrische Physik, Giessenbachstrasse 1, 85748, Garching, Germany
         \and 
         Jet Propulsion Laboratory, M/S 321-100, 4800 Oak Grove Drive, Pasadena, CA, 91109, USA
         \and 
         European Southern Observatory, Casilla 19001, Santiago 19, Chile}

   \date{Received ; accepted }

 
  \abstract
   {The innermost astronomical unit (au) in protoplanetary disks is a key region for stellar and planet formation, as exoplanet searches have shown a large occurrence of close-in planets that are located within the first au around their host star.
   }
   {We aim to reveal the morphology of the disk inner rim using near-infrared interferometric observations with milli-arcsecond resolution provided by near-infrared multitelescope interferometry.} 
   {We provide model-independent reconstructed images of 15 objects selected from the Herbig AeBe survey carried out with \modifff{PIONIER at the Very Large Telescope Interferometer (VLTI)}, using the semi-parametric approach for image reconstruction of chromatic objects (SPARCO). \modif{We propose a set of methods to reconstruct and analyze the images in a consistent way.}}
   {\modif{We find that 40}\% of the systems (\modif{6/15}) are centrosymmetric \modif{at the angular resolution of the observations}. \modif{For the rest of the objects,} we find evidence for asymmetric emission due to moderate-to-strong inclination of a disk-like structure for $\sim$30\% of the objects (5/1\modif{5}) and noncentrosymmetric morphology due to a nonaxisymmetric and possibly variable environment (4/1\modif{5}, \modif{$\sim$}2\modif{7}\%). Among the systems with a disk-like structure, 20\% (3/1\modif{5}) show a resolved dust-free cavity. \modif{Finally, we do not detect extended emission beyond the inner rim. }}
   {The image reconstruction process is a powerful tool to reveal complex disk inner rim morphologies, which is complementary to the fit of geometrical models. At the angular resolution reached by near-infrared interferometric observations, most of the images are compatible with a \modif{centrally peaked} emission \modif{(no cavity)}. 
   \modif{For the most resolved targets, image reconstruction reveals morphologies that cannot be reproduced by generic parametric models (e.g., perturbed inner rims or complex brightness distributions). Moreover, the nonaxisymmetric disks show that t}he spatial resolution probed by optical interferometers makes the observations \modif{of the near-infrared emission (inside a few au)} sensitive to temporal evolution with a time-scale down to a few weeks. \modif{ The evidence of nonaxisymmetric emission that cannot be explained by simple inclination and radiative transfer effects requires alternative explanations, such as a warping of the inner disks.}
   \modif{Interferometric observations can therefore be used to follow the evolution of the asymmetry of those disks at a\modiff{n} au or sub-au scale.} }


   \maketitle
%

\section{Introduction}

The study of protoplanetary disks around young stars is fundamental to understanding planet formation:
The initial conditions for planet formation are indeed determined by the disk properties, its dust and gas densities, its composition, structure, and dynamics. 
Once formed, the planets in turn influence the disk structure, potentially causing gaps, warps, and a variety of features \modif{that were revealed by millimeter-interferometry \citep[e.g.,][]{Perez2014,Andrews2018} and scattered-light imaging \citep[e.g.,][]{Benisty2015,Benisty2018,Avenhaus2018}}. 
\modif{However, these observing techniques cannot access the morphologies of the inner disk regions (located within  $\mathbf{5}\,\mbox{au}$ around the central star) as they cannot reach a sufficiently high angular resolving power ($\sim$ milli-arcsecond) even though ALMA starts to reach sub\,10\,au resolution for some targets}.
These inner regions are nonetheless of fundamental importance because they \modif{are the place} where most of the planets are believed to form in or migrate to \citep{Mordasini2009a,Mordasini2009b,Alibert2011}.
\modif{Recent numerical studies have shown that the inner astronomical units of disks might harbor a strong viscosity transition between a so-called dead-zone, which is nearly impermeable to a magnetic field, and the hotter ionized disk front, showing active accretion powered by magneto-rotational instability \cite[see, e.g.,][]{Kretke2009,Flock2016}. This inner edge of the \modiff{dead-zone} might be the preferential location for dust pile-up and planetesimal formation through different instability mechanisms \citep{Kretke2009, Flock2017,Ueda2019} and an effective planet filtering frontier \citep{Faure2016}.}


\modif{Further evidence for the presence of perturbed inner disks comes from} their known photometric and spectral variability \modif{associated with the presence of a circumstellar disk}, in particular, in the infrared.
Many authors have shown that 60\% to 100\% of stars with disks are variable \citep[e.g.,][]{Espaillat2011,McGinnis2015,Poppenhaeger2015,Flaherty2016} and that the variability patterns are diverse (scale height variation, variable accretion, or ejection).
Observations of the outer disk can also hint at variability in the close-in regions, as shown recently in the case of a rotating shadow in the disk of \object{TW Hya} \citep{Debes2017} or \object{HD 135344B} \citep[][]{Stolker2017}.
In many cases, the period of the variability is compatible with Keplerian periods for radii up to the first inner astronomical unit around the star. 
\modif{Directly constraining the structure of the inner disks is an important key to better understand how planets form. This requires spatially resolving the inner astronomical units to possibly detect a companion or a disk perturbation \citep{Flock2017}}%


The morphology of disks in the innermost regions can be constrained with multitelescope interferometry in the infrared-wavelength range. 
With a baseline of 140\,m, such as the longest one currently available at the Very Large Telescope Interferometer (VLTI), and by observing in the near-infrared (1.65\,$\mu$m), a resolution of $\sim0.3$\,au can be reached for young stars belonging to the closest star forming regions (like Taurus at 140\,pc). 
The first interferometric near-infrared size measurements showed that the emission sizes are proportional to the square root of the luminosity of the central star \citep[e.g.,][]{Monnier2002}, suggesting that the near-infrared \modif{observations} probe the dust sublimation front of the disk.
Interestingly, some objects, which were studied more extensively, display a radially extended emission inside the theoretical dust sublimation radius that cannot be explained by an inner rim alone. 
The presence of a gaseous disk \citep{Kraus2008} or refractory dust \citep{Tannirkulam2008,Benisty2011} inside the conventional silicate sublimation radius was suggested to account for both an extended emission and a large flux excess at near-infrared wavelengths.
Recent interferometric studies of the Br$_\gamma$ hydrogen line have pointed towards the presence of a disk wind in several objects \citep[e.g.,][]{Malbet2007,Garcia2015,Kurosawa2016}.
Such a wind can lift dust off  from the disk and create an additional excess emission in the near-infrared \citep{Bans2012}.
In order to study the general properties of these regions, a comprehensive survey was needed.

Recently, \citet[hereafter L17]{Lazareff2017} observed a large sample of intermediate-mass stars in the  $H$\,band continuum in a systematic and homogeneous fashion, with the Precision Integrated-Optics Near-infrared Imaging ExpeRiment (PIONIER) instrument, \modifff{which was the first one to combine four telescopes at the VLTI}. 
The observations of 51 Herbig\,AeBe stars were analyzed using parametric models to constrain the main characteristics of the dust emission at the au-scale. 
The main findings from this study are as follows: (1) The sublimation temperature of the dust emitting in the H-band is higher ($\sim1800\,\mathrm{K}$) than the classical silicate sublimation temperature ($\sim1500\,\mathrm{K}$); (2) the inner rims are smooth and radially extended, and (3) their brightness asymmetry is suggestive of inclination effects that make the rim at the far side brighter that the side closer to the observer.
\modif{Those findings are confirmed by a survey of Herbig stars in the $K$-band performed with the GRAVITY instrument at the VLTI \citep{Perraut2019}}

\modif{There is ample observational and numerical evidence that inner disks might show a much more complex morphology than the one used in L17 to analyze interferometric observations. The putative presence of instabilities, vortices \citep{Fung2014,Faure2015, Flock2017}, or ring-like structures related to \modiff{dead-zone} dust pile-up self shadowing \citep{Ueda2019} justifies the need for a different approach than parametric modeling}.
In this paper, we analyze a subsample of 1\modif{5} objects observed by L17, with good enough \uv-coverage to allow for image reconstruction in order to obtain model-independent information on the disk morphology.
This approach aims at investigating any structure that cannot be accounted for by simple geometrical model fitting \modif{and is a novel way to analyze the dataset without being dependent on a geometrical model. 
We therefore compared the results of the image reconstructions \modif{with} those from parametric model fitting.
We also searched for additional structures at the inner disk rim or outside of it that could be linked with dynamical instabilities, perturbations by a companion, detection of an inner cavity, or disk self-shadowing. 
For that purpose, we employed several image analysis tools that could be used in future optical and infrared interferometric imaging surveys.}
To perform image reconstruction, we used the Semi Parametric Approach for image Reconstruction of Chromatic Objects \citep[SPARCO;][]{Kluska2014} that enables the use of the spectral channels altogether to separate the central star from the environment.

The paper is organized as follows. We describe the \modif{observations} and data selection \modiff{in} Sect.\ref{sec:dataimgrec} and the image reconstruction process that is applied to each individual object in Sec.\,\ref{sec:imgrec}. We also present the reconstructed images of the extended structure in Sec.\,\ref{sec:results} with associated analysis graphs. In Sec.\,\ref{sec:discussion}, we classify the objects and we discuss the origins of the detected asymmetries.


\section{Observations and target selection}

\label{sec:dataimgrec}

 \begin{figure*}[th]
   \centering
   \includegraphics[width=19cm]{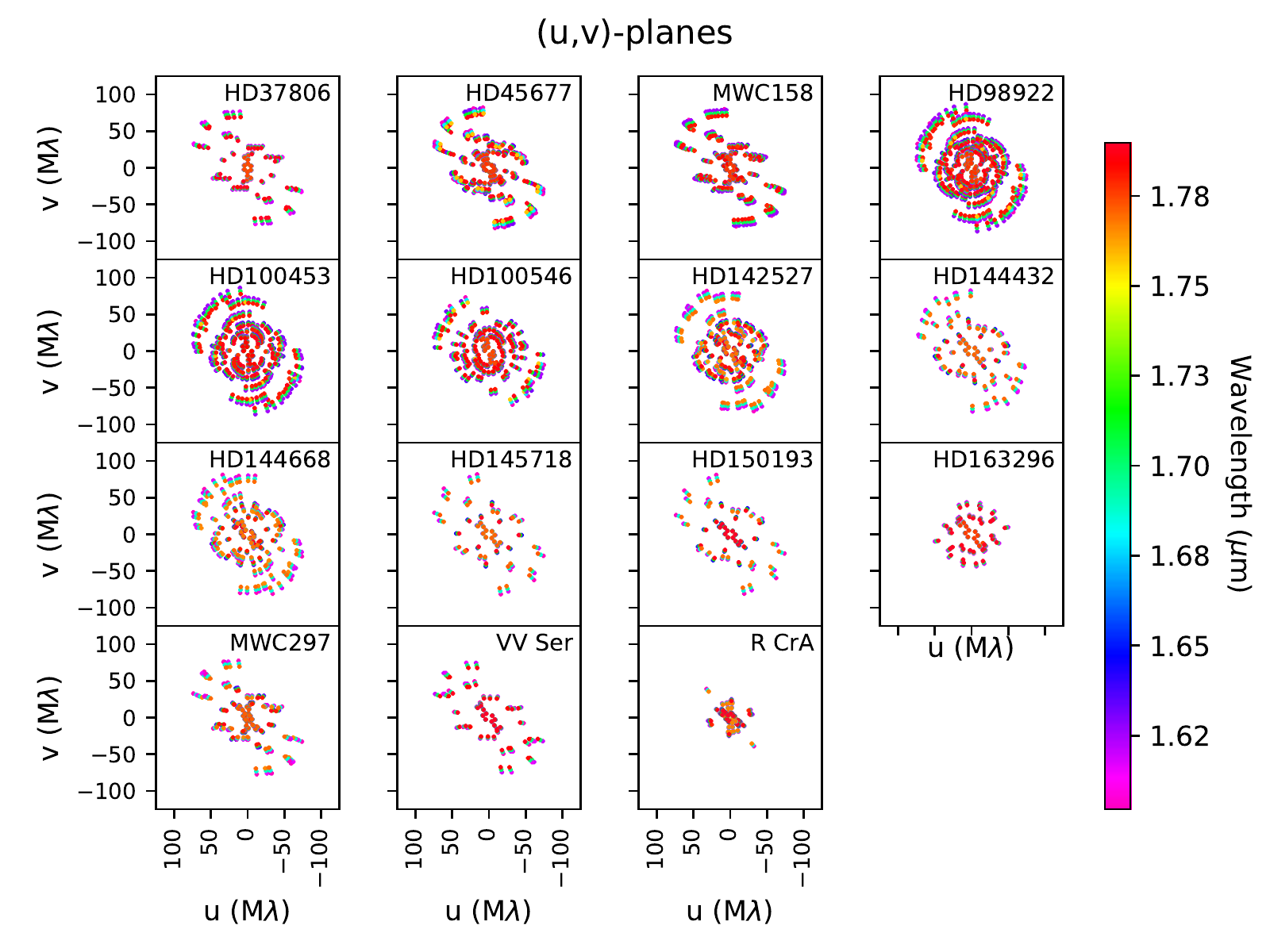}
      \caption{\uv coverage of the targets. The colored \modif{circles} represent the \uv-coverage; the colors correspond to the wavelength (blue for 1.55$\mu$m, orange for 1.65$\mu$m, and red for 1.75$\mu$m).}
         \label{fig:uvplans}
   \end{figure*}

\begin{table*}[!t]
\caption{\modiff{General information on our targets} }
\begin{center}
\begin{tabular}{l c c c c c r r c }
 \hline
  \hline
 Object & RA & DEC & Age & distance & $\log$ L$_*$ &  T$_\mathrm{eff}$ & M$_*$ & Reference  \\
& & & [Myrs] & [pc]  & [L$_\odot$]  & [K] & [M$_\odot$] &  \\
 \hline
\object{HD37806} & 05 41 02.29 & -02 43 00.7 &  1.6 &428$^{+11}_{-11}$ & 2.2 & 10475 & 3.1 &  1 \\[0.1cm]
\object{HD45677} & 06 28 17.42 & -13 03 11.1 &  0.6  &621$^{+13}_{-12}$ & 2.8 &16500 &  4.7    &  1 \\ [0.1cm]
\object{MWC158} & 06 51 33.40 & -06 57 59.4 &  0.6 &380$^{+10}_{-9}$&  2.5 & 9450 & 4.2 &  1 \\ [0.1cm]
\object{HD98922} & 11 22 31.67 & -53 22 11.5 &  0.2 &689$^{+16}_{-16}$ & 3.0 & 10500 & 6.2  &  1 \\ [0.1cm]
\object{HD100453}& 11 33 05.58 & -54 19 28.5 &  6.5 &104$^{+1}_{-1}$  &  0.8 & 7250 & 1.3&  1 \\ [0.1cm]
\object{HD100546}& 11 33 25.44& -70 11 41.2 &  5.5 &110$^{+1}_{-1}$  & 1.4 & 9750 & 2.1&  1 \\ [0.1cm]
\object{HD142527}& 15 56 41.89 & -42 19 23.2 &  6.6 & 157$^{+1}_{-1}$ & 1.0 & 6500 & 1.6 &  1 \\ [0.1cm]
\object{HD144432}& 16 06 57.95& -27 43 09.8 &  5.0 & 155$^{+2}_{-1}$ & 1.0 & 7500 & 1.4 &  1 \\ [0.1cm]
\object{HD144668} & 16 08 34.29 &-39 06 18.3 &  2.7 &161$^{+2}_{-2}$&  1.7 & 8500 & 2.4  &  1 \\ [0.1cm]
\object{HD145718}& 16 13 11.59 & -22 29 06.6 & 9.8 & 152$^{+2}_{-2}$   & 0.9 & 8000 & 1.6 &  1 \\ [0.1cm]
\object{HD150193} & 16 40 17.92 & -23 53 45.2 &  5.5 &151$^{+2}_{-2}$ &  1.4 & 9000 & 1.9 &  1 \\[0.1cm]
\object{HD163296} & 17 56 21.29 & -21 57 21.9 &  7.6 &102$^{+1}_{-1}$&  1.2 & 9250 & 1.8&  1 \\ [0.1cm]
\object{MWC297} & 18 27 39.53 & -03 49 52.1 &  0.03 &376$^{+12}_{-12}$ & 4.6 & 24500 & 16.9  &  1 \\ [0.1cm]
\object{VV Ser}& 18 28 47.86 & +00 08 39.9 &  2.8 & 420$^{+8}_{-13}$ & 2.0 & 13800 & 2.9 & 1\\ [0.1cm]
\object{R CrA} &19 01 53.69 & -36 57 08.2 &  1.5 &95$^{+6}_{-6}$ &  2.1 & 9550 & 3.5 & 2\\ 
\end{tabular}
\end{center}
\tablebib{(1)~\citet{Vioque2018}; (2)~\citet{Sissa2019}.
}
\label{tab:targets}
\end{table*}%

The dataset was extracted from the VLTI Large Program on Herbig Ae/Be stars (190.C-0963, PI: Berger). 
\modif{This program made use of PIONIER, which is a four-beam interferometric combiner working in the $H$-band at 1.65$\mu$m \citep[][]{Lebouquin2011}.
The presented dataset was observed between \modif{2012 Dec 19} and \modif{2013 June 06} and comprises 15 targets \modiff{(see Table\,\ref{tab:targets})}.
The associated datasets are shown in the Appendix in Fig.\,\ref{fig:data}, \ref{fig:data2}, \ref{fig:data3}, and \ref{fig:data4}.}
We refer to L17 for details regarding data reduction. 
This large program has been divided in two subprograms: survey observations on a large sample to assess the statistics (L17) and a detailed study on a small object sample to perform image reconstruction on the most resolved targets (this work).
\modif{An interferometer measures partial information on the visibility, a complex number containing the fringe contrast, and phase that relates to the image via a Fourier transform.}
The dataset considered in this paper shows a clear decrease of visibility amplitudes \modif{with baseline, meaning that the targets are spatially resolved}.
The three spectral channels across the $H$-band display a similar trend where the blue channels have higher squared visibility values because there is a higher stellar contribution \citep[e.g.,][]{Kluska2014}.

The quality of an image reconstruction depends on the \modif{ratio between the smallest angular resolution element and the size of the emission} and on the completeness of the \uv-plane.
\modif{During the observations, targets that appeared to be well resolved were more extensively observed to optimize the \uv-coverage and perform image reconstructions.}
The \uv-coverages of the selected 1\modif{5} targets are displayed in Fig.\,\ref{fig:uvplans}.




\section{Image reconstruction approach}
\label{sec:imgrec}

\modif{The image reconstruction is performed in a Bayesian framework where the best image minimizes a mathematical distance ($\mathcal{J}$), which is a combination of two terms:
\begin{equation}
\mathcal{J}= \mathcal{J}_\mathrm{data} + \mu \mathcal{J}_\mathrm{rgl},
\end{equation}
where $\mathcal{J}_\mathrm{data}$ is the distance to the data (here the $\chi^2$), $\mathcal{J}_\mathrm{rgl}$ is the regularization function (here the quadratic smoothness function defined as $\mathcal{J}_\mathrm{rgl}=\Sigma(x-Sx)^2$ where $S$ is a smoothing operator implemented via finite differences\footnote{\modiff{They are also known as the discrete Laplace operator.} It is computed by using the values of a pixel and its direct neighbors, such as $Sx_\mathrm{i,j} = \frac{1}{8} (x_\mathrm{i-1,j} + x_\mathrm{i+1,j} + x_\mathrm{i,j-1} + x_\mathrm{i,j+1}) + \frac{1}{2} x_\mathrm{i,j}$. At the image borders and corners, the nonexisting pixels are replaced by the central one in the equation.}), and $\mu$ is the regularization weight that determines the balance between the two distances.
The quadratic smoothness regularization is one of the most reliable regularizations \citep{Renard2011}. Also, the fact that this regularization is quadratic limits the effects of local minima.
The value of the regularization weight ($\mu$) depends on the (u, v)-coverage and on the morphology of the target \citep[][]{Renard2011}.
}

The image reconstruction process is the one used in \citet{Kluska2016} and \citet{Hillen2016}, coupling the SPARCO method \citep{Kluska2014} with the multiaperture image reconstruction algorithm \citep[MiRA][]{Thiebaut2008}. 
\modif{
In this paper, we used a simple model for the star in the form of a point source with a visibility of unity across all baselines.
The final \modif{complex} visibility ($V_\mathrm{tot}$) is therefore computed by this relation:
\begin{equation}
V_\mathrm{tot} = \frac{ \fso\big(\lambda/\lambda_0\big)^{-4} + \big(1-\fso\big)\big(\lambda/\lambda_0\big)^{\denv} V_\mathrm{img} }{ \fso\big(\lambda/\lambda_0\big)^{-4} + \big(1-\fso\big)\big(\lambda/\lambda_0\big)^{\denv} },
\end{equation}
where $\lambda_0=1.65\mu$m is the reference wavelength, $\lambda$ is the wavelength of observation (depending on the observed spectral channel), $\fso$ is the stellar-to-total flux ratio (between 0 and 1), $\denv$ is the spectral index\footnote{The spectral indices are defined as $d = \frac{d\log F_\lambda}{d\log \lambda}$.} of the circumstellar environment (\denv=-4 is the Rayleigh-Jeans regime; \denv=1 corresponds to the spectral index of a black-body with a temperature of 1400\,K), and $V_\mathrm{img}$ is the visibility of the image computed via a Fourier Transform.
}

\modif{It is important to note that \fso and \denv are called the chromatic parameters and\modifff{, together with the regularization weight $\mu$,} they have to be determined for each target separately.
In order to do so, we} briefly recall the main steps involved. \modifff{First, t}\modif{o determine the stellar-to-total flux ratio $\fso$ and the spectral index of the circumstellar environment $\denv$, we performed a grid of reconstructions on those parameters. We then selected the parameters with the lowest $\mathcal{J}$ (Appendix\,\ref{app:fsde}).}
\modifff{Second, t}o determine the best \modif{regularization} weight $\mu$ of the quadratic smoothness regularization, we used the L-curve method \citep[Appendix\,\ref{app:lcurve};][]{Lcurve}. 
This method allows one to find the regularization weight that is optimal between a regime dominated by the data, but which is very noisy, and a regime dominated by the regularization, but for which the image does not fit the data.
\modif{The two regimes form two asymptotes and the value in the knee of the curve represents the optimal value of the regularization.}
\modif{We investigated the effect of different regularization weights for two targets in our dataset with different angular sizes (HD45677 and HD100453, see Appendix\,\ref{app:lcurve} and Fig.\,\ref{fig:Lcurve2}). The effect of the regularization weight does not change the overall image features.}
\modifff{Last, a} bootstrap process was then applied (Appendix\,\ref{app:boot}) to assess the significance of the image features.

\begin{table}[!t]
\caption{Parameters of the \modif{image} reconstructions (see text for details).
}
\begin{center}
\begin{tabular}{l c c c c c}
 \hline
  \hline
 Object &  pixel & \fso & \denv & $\log\mu$ & $\chi^2_\mathrm{r}$    \\
 & size  & & & & \\
 & [mas] & & & & \\
 \hline
\object{HD37806}  & 0.1 & 28.9\%  & 0.71 & 8.7 & 1.16  \\ 
\object{HD45677}  & 0.25 & 43.5\% & 1.73 & 10.5 & 1.78  \\ 
\object{MWC158}   & 0.1 & 21.8\% & 0.06 & 9.7 & 1.92 \\ 
\object{HD98922}  & 0.1 & 30.3\% & 1.25 & 8.7 & 1.01 \\ 
\object{HD100453} & 0.1 & 60.3\% & 1.19 & 10.0 & 0.81\\ 
\object{HD100546} & 0.1 & 47.2\% & -0.54 & 8.8 & 1.04\\ 
\object{HD142527} & 0.1 & 50.3\% & -0.36 & 9.8 & 1.51  \\ 
\object{HD144432} & 0.1 & 52.7\% & -0.63 & 8.8 & 0.42 \\ 
\object{HD144668} & 0.1 & 41.8\% & -0.04 & 9.8 & 1.16   \\
\object{HD145718} & 0.1 & 69.6\% & -1.13 & 10 & 0.96 \\
\object{HD150193} & 0.1 & 40.1\% & -1.44 & 9 & 0.58  \\
\object{HD163296} & 0.2 & 36.9\% & -0.27 & 8.7 & 0.81  \\ 
\object{MWC297}   & 0.1 & 15.2\% & 2.43 & 10.7 & 1.66 \\ 
\object{VV Ser}   & 0.1 & 32.2\% & -2.28 &9 & 1.17 \\
\object{R CrA}    & 0.\modif{1} & \modif{14.9}\% & \modif{-2.05} & 9& \modif{2.24}  \\

\end{tabular}
\end{center}
\label{tab:param}
\end{table}%

For each object, Table\,\ref{tab:param} lists the pixel size, the chromatic parameters, the regularization weight $\log\mu$, and the corresponding reduced $\chi^2_\mathrm{r}$ of the image\footnote{This is only for illustrative purposes as the process does not minimize the $\chi^2_\mathrm{r}$ (see Eq.\,1).}.

Once the images were computed, we used the following \modifff{three} tools to analyze them.
\modifff{First, a}symmetry maps: Those maps contain \modif{the part of the emission} that is not point-symmetric with respect to the central star.
\modifff{Second, r}adial profiles: These profiles are \modifff{made by} azimuthally averag\modifff{ing the} \modif{brightness of the deprojected image}.
\modifff{Third, a}symmetry factors ($f_\mathrm{asym}$ and $\phi_\mathrm{asym}$): They describe the strength and the orientation of the asymmetry with respect to the central star.

\modif{
To justify the pertinence of these tools for our analysis, we tested them on a dataset generated from a radiative transfer model of a star with a disk (see Appendix\,\ref{sec:tools}).}

\section{Results}
\label{sec:results}

\subsection{Reconstructed images}

\begin{figure*}[t]
\centering
\includegraphics[width=15cm]{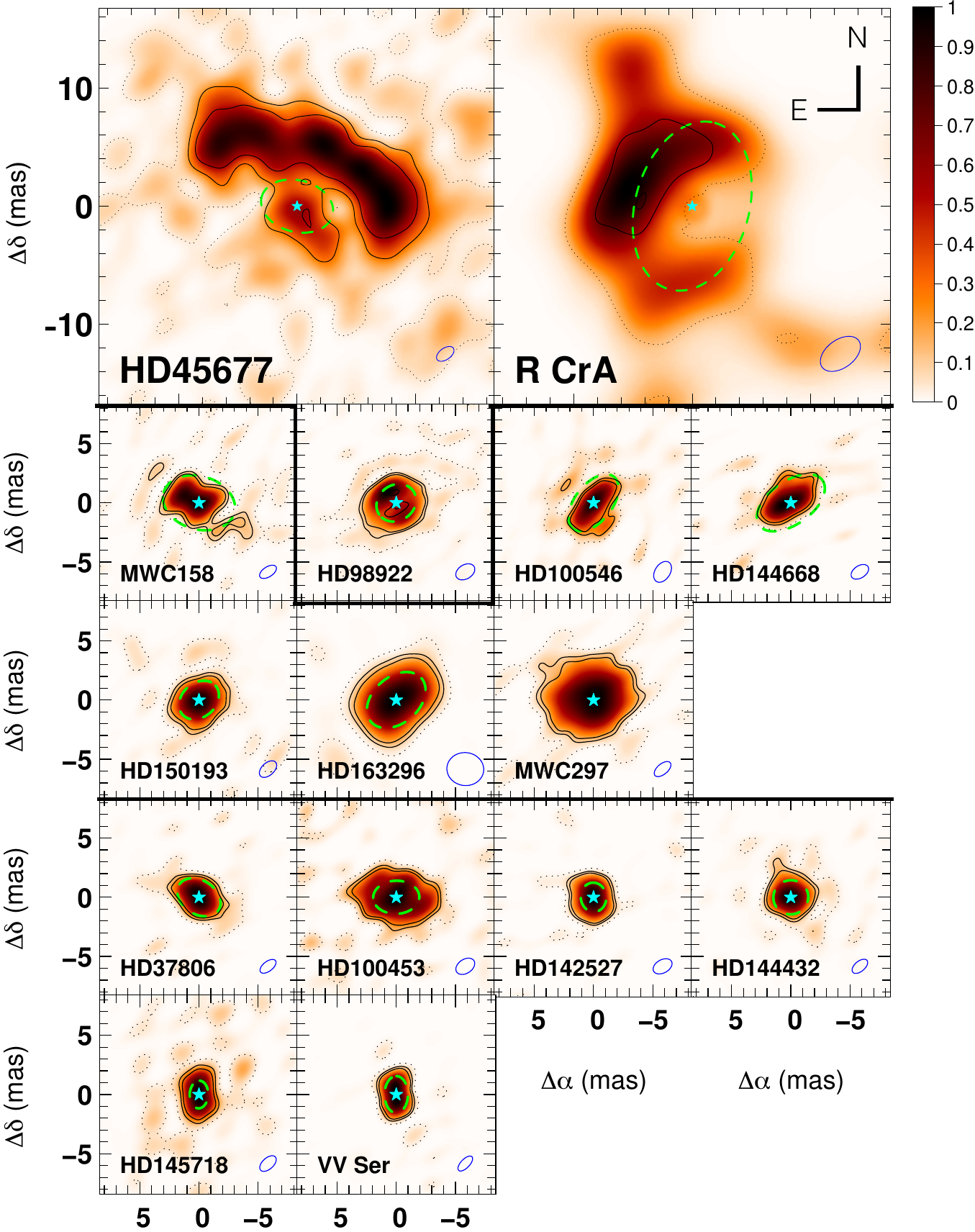}
\caption{Reconstructed images of the Herbig\,AeBe targets \modiff{separated in three groups. Top: the centrally depressed objects; middle: the centrally peaked and asymmetric objects; bottom: the centrally peaked and asymmetric objects.} The cyan star marks the position of the central star, which has been subtracted, the black contours indicate the significance at 1 (dashed line), \modif{3,} and \modif{5}-$\sigma$ (solid lines), and the blue ellipse shows the size of the beam. 
The green dashed ellipses represent the theoretical dust sublimation radius (see Sect.\,\ref{sec:Rsub}).
}
\label{fig:images}
\end{figure*}

Fig.\,\ref{fig:images} displays the reconstructed images for all the Herbig\,AeBe stars of our selection. 
\modiff{Most of the targets have a $\chi^2_\mathrm{r}$ that is around unity. Four of them, however, have relatively larger $\chi^2_\mathrm{r}$ (HD45677, MWC158, MWC297, and R\,CrA). This could be due to the break of a hypothesis used in the image reconstruction, for example that the flux in the image has the same spectral index or that the object is static during the observations.
The break of the latter hypothesis is discussed in Sect.\,\ref{sec:RTcompare}.}
Three targets, \object{HD\,45677}, \object{HD\,98922,} and \object{R\,CrA}, show the \modif{presence of a cavity (}ring-like morphology\modif{)} with a \modif{brightness} deficit close to the central star.
The other objects show an elongated morphology.

Five images exhibit flux at large distances from the star (\object{HD\,45677}, \object{MWC\,158}, \object{HD\,98922}, \object{HD\,100453,} and \object{HD\,145718}), but their significance per pixel is less than 3-$\sigma$. 
\modif{This flux} can either result from image reconstruction artifacts due to the lack of observations at small spatial frequencies (\object{HD\,45677}) or from a drop in the squared visibility curve at very short baselines (\object{MWC\,158}, \object{HD\,98922}, \object{HD\,100453}, \object{HD\,145718}) caused by an \modif{over-resolved} component \citep[][L17]{Monnier2006,Pinte2008,Anthonioz2015,Kluska2016,Klarmann2016}.
The morphology of the \modif{over-resolved} component cannot be determined by the current dataset because of the lack of very short baselines.
Without any additional spatial information (from, e.g., single telescope aperture masking observations), the reconstruction of the \modif{over-resolved} flux distribution is highly influenced by the \uv-plane and is not treated further in this work.


\subsection{Radial profiles}

\label{sec:radipro}

\begin{figure*}
\centering
\includegraphics[width=16cm]{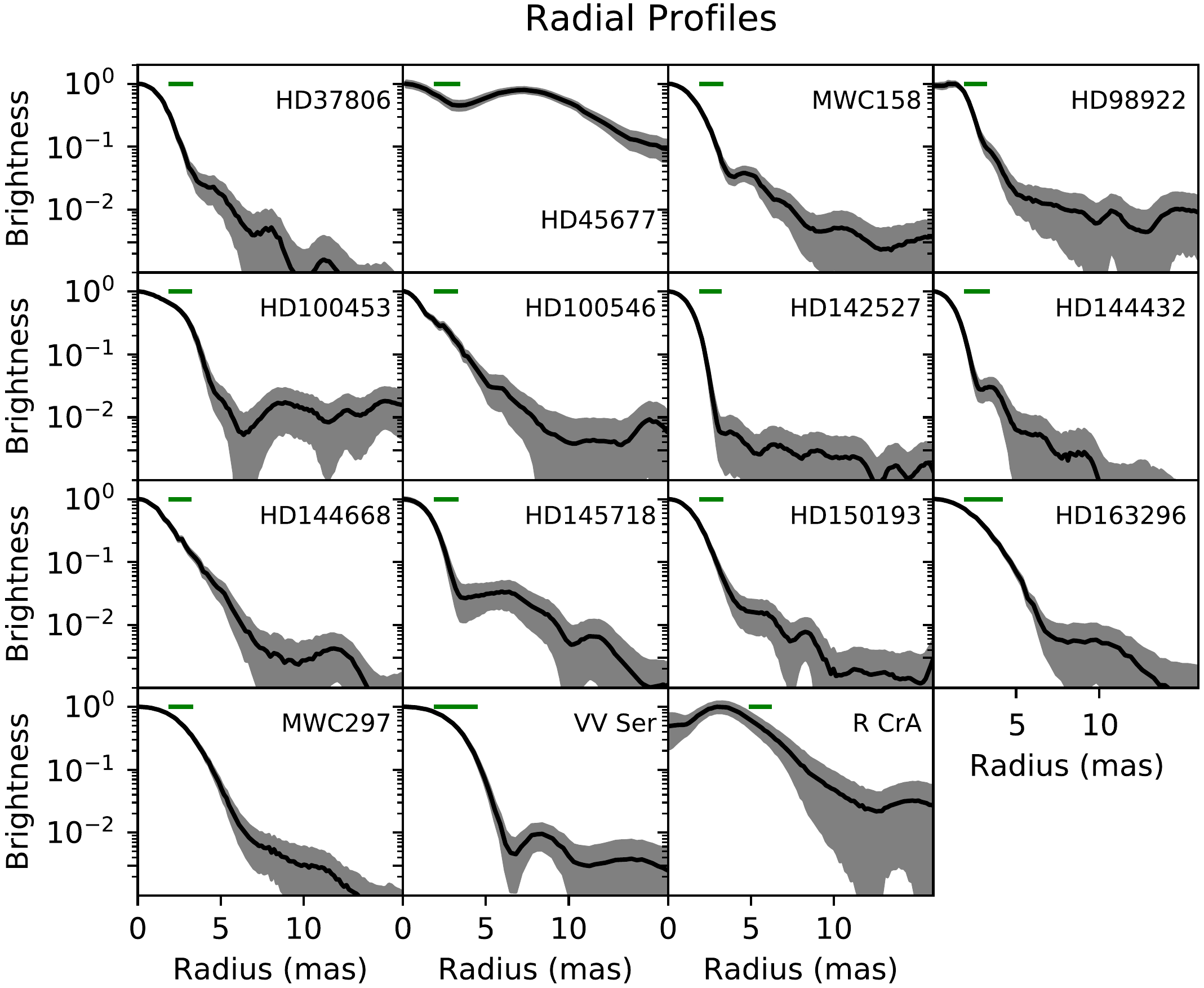}
\caption{Radial profiles of the reconstructed images of Fig.~\ref{fig:images}. Black solid lines: radial profiles from images with 5-$\sigma$ error bars in \modif{color shades. The green line shows the azimuthally averaged size of the interferometric beam.}}
\label{fig:radi}
\end{figure*}

To probe the radial structure of our targets, we computed radial profiles for each image. 
To correct for the object orientations, we used the inclinations and position angles we obtained by fitting the images in the Fourier space (see Appendix\,\ref{app:incPA})
We \modif{segregated} the 1\modif{5} targets into two categories based on the following profiles.

\subsubsection{Centrally depressed} \modif{This category comprises the targets with an inner cavity seen in their radial profile (see Fig.\,\ref{fig:radi}).} 
\object{HD\,45677}, \object{HD\,98922,} and \object{R\,CrA} show centrally depressed profiles, that is, increasing with radius towards a maximum and then decreasing. 
We note that the profile of \object{HD\,45677} displays a decrease before an increase of the \modif{brightness} close to the star, indicating the presence of an additional component inside the central depression that is marginally resolved. 

\subsubsection{Centrally peaked} The emission is resolved but no inner cavity is detected at the angular resolution of our observations (the radial profile are monotonically decreasing). We note that 8\modif{0}\% of the objects, that is, the 1\modif{2} targets (\object{HD\,37806}, \object{MWC\,158}, \object{HD\,100453}, \object{HD\,100546}, \object{HD\,142527}, \object{HD\,144432}, \object{HD\,144668}, \object{HD\,145718}, \object{HD\,150193}, \object{HD\,163296}, and \object{MWC\,297}) display a monotonically decreasing profile.


\subsection{Asymmetry analysis}
\label{sec:asym}


 \begin{figure*}[t]
   \centering
   \includegraphics[width=15cm]{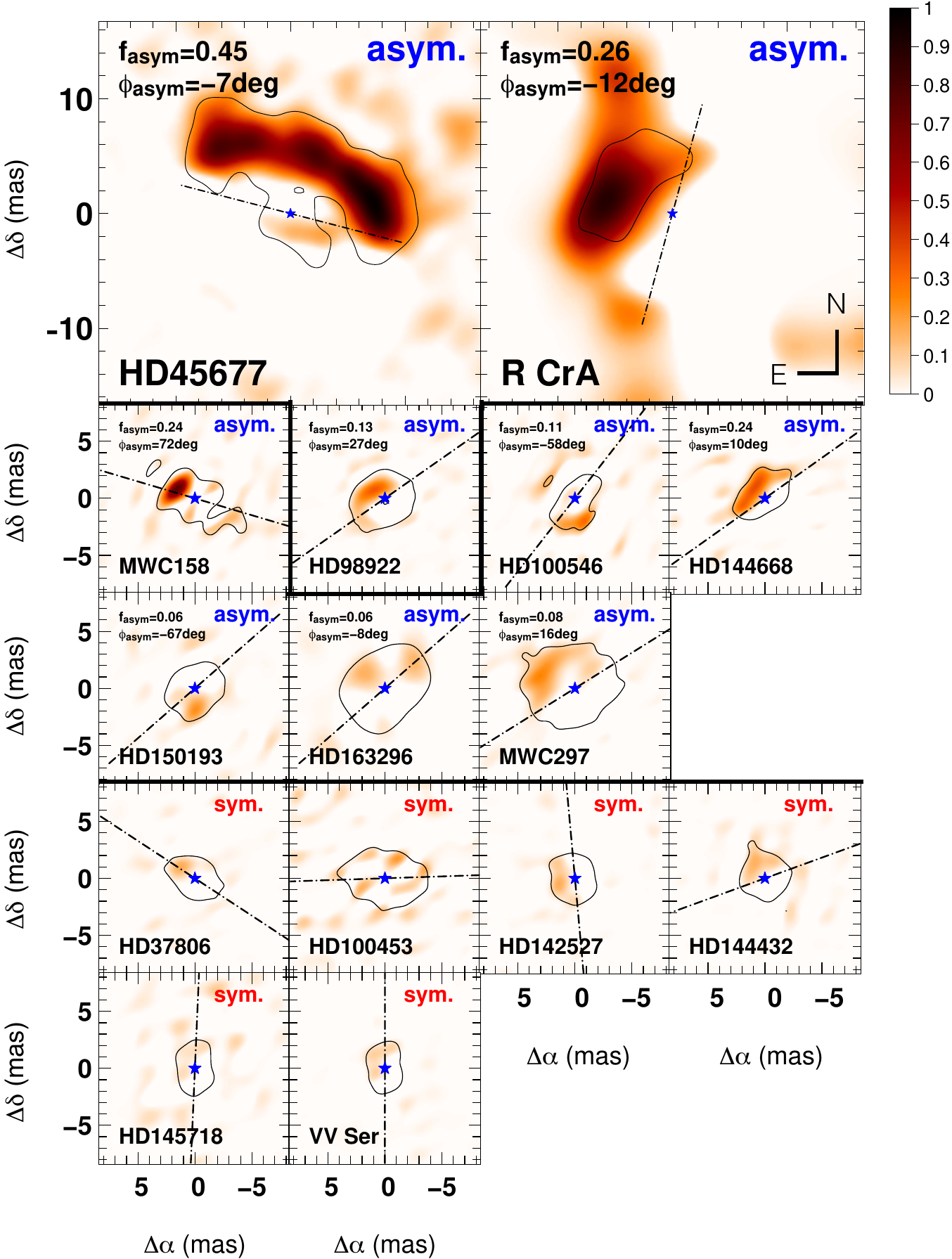}
   \caption{Asymmetry maps of the targets \modiff{classified the same way as in Fig.\,\ref{fig:images}}. The blue star represents the central star, the solid black contour represents the \modif{3}-$\sigma$ significance profile, and the dot-dashed line represents the orientation of the major-axis from parametric fitting. The targets with three closure phase points or more with a significance of 5-$\sigma$ are indicated as ``asym."}
   \label{fig:asym}
 \end{figure*}

\begin{table}[!t]
\caption{\modif{Asymmetry factors \modiff{and orientations} for the objects, which have a significant closure phase signal.} \modiff{The east of north position angle of the asymmetry is given by PA+$\phi_\mathrm{asym}$-90$^\circ$. } }
\begin{center}
\begin{tabular}{l c c c c}
 \hline
  \hline
 Object& $f_\mathrm{asym}$ & $\phi_\mathrm{asym}$ & \modiff{PA} & \modiff{inc.}\\
  & & [$^\circ$] & \modiff{[$^\circ$]} & \modiff{[$^\circ$]} \\
 \hline
 \object{HD45677}  & 0.45$\pm$0.01 & \modiff{-}7$\pm$1  &   \modiff{76$\pm$11}  & \modiff{45$\pm$2} \\ 
  \object{MWC158} &  0.24$\pm$0.01 & 72$\pm$2 & \modiff{74$\pm$4}   & \modiff{44$\pm$3} \\
 \object{HD98922}  & 0.13$\pm$0.01  & 27$\pm$6  & \modiff{125$\pm$9} & \modiff{31$\pm$5} \\ 
 \object{HD100546} & 0.11$\pm$0.01   & \modiff{122}$\pm$16 & \modiff{142$\pm$2} & \modiff{56$\pm$2}  \\
 \object{HD144668} &  0.24$\pm$0.01 & 10$\pm$2 & \modiff{125$\pm$1} & \modiff{56$\pm$1} \\
 \object{HD150193} &  0.06$\pm$0.01 & \modiff{113}$\pm$20& \modiff{131$\pm$4} & \modiff{40$\pm$2}  \\
 \object{HD163296} &  0.06$\pm$0.01 & \modiff{-}8$\pm$23 & \modiff{131$\pm$1} & \modiff{46$\pm$1} \\
 \object{MWC297} &  0.08$\pm$0.01 & 16$\pm$6 & \modiff{122$\pm$8} & \modiff{38$\pm$5} \\
 \object{R CrA} &  0.26$\pm$0.02 & \modiff{-}12$\pm$5 & \modiff{165$\pm$8} &  \modiff{49$\pm$6} \\
\end{tabular}
\end{center}
\label{tab:skewness}
\end{table}%

To investigate any departures from axisymmetry in our targets, we computed the asymmetry maps for each image (Fig.\,\ref{fig:asym}). \modif{We also computed asymmetry factors $f_\mathrm{asym}$ and $\phi_\mathrm{asym}$ (Table \ref{tab:skewness}), which quantify the level of asymmetry and its orientation. 
These factors are described in Appendix\,\ref{sec:tools}. We note that 
$f_\mathrm{asym}$ describes the amplitude of the asymmetry. 
A $\phi_\mathrm{asym}$ \modiff{close to} 0$^\circ$ or \modiff{$\pm$180$^\circ$} means a preferential orientation along the minor axis, \modiff{whereas} a $\phi_\mathrm{asym}$ close to $\pm$90$^\circ$ \modiff{indicates an asymmetry} along the major axis}.

All of the images \modif{deviate, to some extent, from point symmetry.}
In order to study the significant asymmetries only, we focus on the objects that have a significant nonzero closure phase signal.
\modif{It is important to note that 60\%} of our sample (9/15; \object{HD45677}, \object{MWC158}, \object{HD98922}, \object{HD100546}, \object{HD144668}, \object{HD150193}, \object{HD163296}, \object{MWC297,} and \object{R CrA}) present at least three nonzero closure phase measurements beyond 5-$\sigma,$ which indicate a departure from point-symmetry. 
In the rest of the asymmetry analysis, we focus on those objects only.

The objects that \modif{are centrally depressed} present a clear asymmetry: They are brighter on one side of the major-axis than on the other.
The radial profile of \object{HD45677} (see Fig.\,\ref{fig:radi}) has two radial components: \modif{a half-}ring and an emission very close to the star.
In the asymmetry map, the inner \modif{centrally peaked} emission asymmetry is reversed with respect to the \modif{centrally depressed} one. 
The \modif{half-}ring \modif{brightness} maximum \modiff{is} towards the northwest and the inner part maximum is in the opposite direction (i.e., towards the southeast; see Fig.\ref{fig:asym} top-left). 

Two \modif{centrally peaked} objects (\object{HD144668} and \object{MWC297}) clearly show the same kind of asymmetry map as the \modif{centrally depressed} objects even though their inner rims are not resolved. 
These asymmetry maps are similar to the model one (shown in Fig.\,\ref{fig:asymmap}).
Their morphology is therefore likely to be rim-like as well.  
\object{MWC158}, \object{HD100546}, \object{HD150193,} and \object{HD163296} have an irregular asymmetry map, which is probably due to a more complex inner rim morphology \citep[as demonstrated for \object{MWC158},][]{Kluska2016}. 
\modif{The asymmetry factors for these objects can be found in Table\,\ref{tab:skewness} and leads us to distinguish the following two trends. }


\label{sec:azimpro}


\subsubsection{Asymmetry direction perpendicular to the emission position angle}
The clearest example of such an asymmetry is the ring of \object{HD\,45677}. 
This ring is very asymmetric when describing an arc in the image and in the asymmetry map.
\modif{This is reflected} in the asymmetry factor\modif{s} of \modif{$f_\mathrm{asym}=0.45\pm0.01$ and $\phi_\mathrm{asym}=-7^\circ\pm1^\circ$}, which \modif{have} the \modif{highest amplitude} among our targets.
The other asymmetric objects that have an \modif{asymmetry angle ($\phi_\mathrm{asym}$) \modiff{less} than 30$^\circ$ from 0$^\circ$} are \object{HD98922}, \object{HD144668}, \object{HD163296}, \object{MWC297,} and \object{R CrA}.
\object{R CrA} has the second largest asymmetry \modif{amplitude factor ($f_\mathrm{asym}=0.26\pm0.02$)} and has a similar asymmetry map to \object{HD45677}.
The asymmetry map of \object{HD144668} displays flux on one side of the major-axis and the asymmetry \modif{amplitude} is \modif{relatively high} ($f_\mathrm{asym}$ = 0.24$\pm$0.01).
\object{HD98922} has the highest asymmetry angle of this subsample ($\phi_\mathrm{asym}$=27$^\circ\pm$6$^\circ$). This is probably due to the presence of flux on the southeast part as displayed by the asymmetry map.
\object{MWC~297} shows an asymmetry map and asymmetry factor\modif{s that are} compatible with \modif{a brightness enhancement on one side} of the major axis.
\modif{Finally, we note that \object{HD163296} has a low asymmetry angle, but it displays a relatively complex asymmetry map with several maxima and minima.}
\modiff{It is interesting to mention that HD142527 and HD144432, despite having only two and one closure phase measurements at 5-$\sigma$ over zero, have a $f_\mathrm{asym}$ of 0.10$\pm$0.01 and 0.11$\pm$0.01 and a $\phi_\mathrm{asym}$ of 171$^\circ\pm$9$^\circ$ and 1$^\circ\pm$11$^\circ$, respectively. This, together with their asymmetry maps, would make them compatible with an asymmetry direction that is perpendicular to the emission position angle.}

\subsubsection{Asymmetry direction not oriented perpendicular to the position angle}

\modif{We find that three out of the fifteen objects showing nonaxisymmetric emission display a skewness orientation that cannot be related to the disk position angle in a clear way. 
In this group, \object{MWC\,158} is the source displaying the most extreme asymmetry factors ($f_\mathrm{asym}=0.24\pm0.01$ and $\phi_\mathrm{asym}=72^\circ\pm2^\circ$), but it was already identified as a temporally variable source in previous interferometric observations \citep{Kluska2016}. 
The emission of \object{HD100546} and \object{HD150193} show skewness that cannot be related to any of the two major or minor axes.}

\section{Discussion}
\label{sec:discussion}

 To discuss and interpret the images, we \modif{classify the images (Sect.\,\ref{sec:classify}) before discussing individual characteristics of the circumstellar emission, such as the detection of an inner cavity (Sect.\,\ref{sec:hole}) or extended structures (Sect.\,\ref{sec:extended}). We also compare the most resolved targets \modif{with} the parametric models from L17 (Sect.\,\ref{sec:compm1}). 
Finally, we discuss some features of the reconstructed images with respect to the dust sublimation radius (Sect.\,\ref{sec:Rsub}), compare the asymmetry factors with those of axi-symmetric models (Sect.\,\ref{sec:fasym}), discuss the origin of the noncentrosymmetry (Sect.\,\ref{sec:RTcompare}) as well as inner and outer disk misalignment (Sect.\,\ref{sec:misal}) .} 


\subsection{Classification}
\label{sec:classify}

\begin{table}[!t]
\caption{Classification of the objects related to their morphological properties in the radial (horizontal) and azimuthal direction (vertical). }
\begin{center}
\begin{tabular}{l|l|l}
 & \modif{Centrally} & \modif{Centrally } \\
 &\modif{peaked} &  \modif{depressed} \\
 \hline
 Sym\modif{metric} &   \object{HD37806}& \\
 &\object{HD100453}&\\
  & \object{HD142527}&\\
   & \object{HD144432} & \\
 & \object{HD145718}&\\
  & \object{VV Ser} & \\
 \hline
 \modif{Asymmetry }  &\object{HD144668} & \object{HD45677}\\
 \modif{perpendicular} & \object{HD163296}& \object{HD98922}\\
 \modif{to the position angle}&\object{MWC297} & \object{R CrA}\\
 \hline
 \modif{Asymmetry not }& \object{MWC158} &\\
 \modif{perpendicular}& \object{HD100546}&\\
 \modif{to the position angle}& \object{HD150193} &
\end{tabular}
\tablefoot{Asym. incl.: Asymmetric objects with inclination-like effect; Asym. irreg.: Asymmetric objects with an irregular azimuthal pattern.}
\end{center}
\label{tab:summ}
\end{table}

\modif{To synthetize all the morphological features that we retrieved, we classify the targets based on the observational properties (see Table\,\ref{tab:summ}), that is, on the radial profiles and azimuthal analysis (Sect.\,\ref{sec:radipro} and \ref{sec:azimpro}).}
The \modif{centrally peaked} objects that are marginally resolved appear to be centrosymmetric.  
The lack of significant nonzero closure phase signals can be interpreted as a strong flux contribution from the central star, which would smooth out the asymmetry from the disk.
\modif{Another interpretation would be that a} low disk inclination reduces the contrast between the far and the near sides of the rim or that we lack precision to detect the asymmetry. 
In support of the high stellar-to-total flux ratio hypothesis, we find that \modif{four} out of the \modif{six} symmetric objects have a strong contribution of the stellar flux (\fso \textgreater 50\%), while all the noncentrosymmetric objects have a weaker contribution from the star (\fso \textless 50\%). 
The objects showing a ring-like emission, which we associate to the dust sublimation rim, show azimuthal features that are compatible with inclination effects.
\modif{Three} \modif{centrally peaked} objects \modif{(HD144668, HD163296, and MWC297)} also show an inclination-like signature.
They are likely to have an inner disk cavity as well but it is not resolved by our observations \modif{because of their angular size, radial extensions, and/or disk inclination}. 
\modif{The (non)detections of inner cavities are discussed in more detail in Appendix\,\ref{sec:innerhole}}.

\modif{T}he objects showing irregular asymmetry do not show an inner cavity.
This \modif{can be due to a lack of angular resolution (see Appendix\,\ref{sec:innerhole}), but it might also} indicate that a ring-like morphology is not possible with such irregular brightness distributions.
Having a clear inner rim would lead to inclination effects in the case of a strong or moderate inclination angle.
Two of these objects (\object{HD100546} and \object{HD150193}) still need to be confirmed as irregular \modif{since their $\phi_\mathrm{asym}$ has relatively large error bars (see Sect.\,\ref{sec:RTcompare}).}
\modif{However, \object{HD\,100546} is surrounded by a perturbed circumstellar environment with gaps \citep{Benisty2010,Tatulli2011} and possible bars and a warp in the first tens of astronomical units \citep{Mendigutia2017,Walsh2017}, which possibly explain the irregular structure we detect.}
\modif{MWC158} is \modif{already} known to have a strongly variable environment \citep[\object{MWC158};][]{Kluska2016}.

\subsection{\modif{Detection of inner cavities}}
\label{sec:hole}
\begin{figure}
\centering
\includegraphics[width=9cm,angle=0]{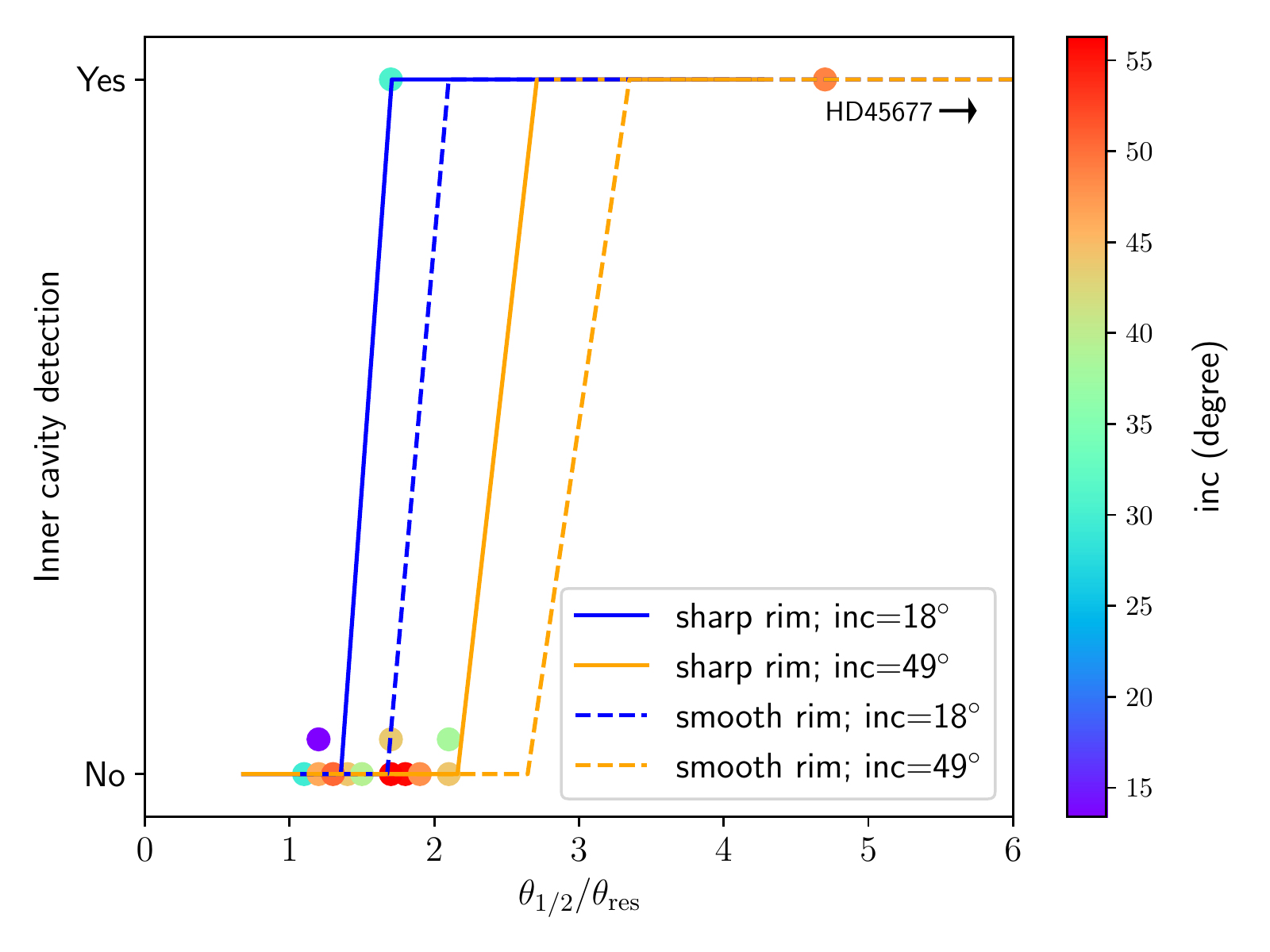}
\caption{\modif{Inner hole detection in image reconstructions from (a) (continuous lines) synthetic observations of radiative transfer models and (b) (dots) our targets. Models with two radial distributions and two inclinations are included, see legend and text. Some dots are offset to avoid overlap. }}
\label{fig:InnerHole}
\end{figure}

\modif{The detection of a central cavity in the brightness distribution can be impacted by various factors, one of them being the ratio between the object's size and the reached angular resolution. As a measure of the angular resolution, we used one-half of the smallest fringe angular wavelength achieved in the observations: $\theta_\mathrm{res} = \lambda_\mathrm{min}/2B_\mathrm{max}$, and we examined the detectability of the cavity as a function of the ratio $\theta_\mathrm{1/2} / \theta_\mathrm{res}$  where $\theta_\mathrm{1/2}$ is the isophotal half-flux radius.
The isophotal half-flux radius ($\theta_\mathrm{1/2}$) is defined as the size of an ellipse, which has the same orientation ($inc$ and PA) as the imaged target and that contains half of the image flux. }


\modif{
We do not detect an inner cavity in twelve out of the fifteen targets despite them having a relatively extended circumstellar emission  compared with the smaller angular scale resolved by the (u, v)-coverage ($\theta_\mathrm{res}$).
In order to understand this, we made image reconstructions on radiative transfer models \citep[using MCFOST;][]{Pinte2006} of an A0 star surrounded by a disk having a sharp \citep{Isella2005} and a radially extended \citep[or smooth;][]{Tannirkulam2007} inner rim (more details in Appendix\,\ref{sec:lessonsRT}) for two inclinations: one almost pole-on ($inc=18^\circ$) and a relatively high inclination ($inc=49^\circ$) corresponding to the higher inclinations in our sample (see Table\,\ref{tab:incPA}).
We show the detection of the inner cavity as a function of the $\theta_\mathrm{1//2}/\theta_\mathrm{res}$ ratio on Fig.\,\ref{fig:InnerHole}. 
We did not plot the detection against the inner cavity size because we want to compare it with our actual observations.
In the actual dataset, in case of an undetected inner cavity, its size is unknown. }

\modif{We can see that for pole-on orientations, the ratios of $\theta_\mathrm{1/2}/\theta_\mathrm{res}$ are comparable though higher for smooth rim models as expected.
For higher inclinations, the inner cavity of the models were resolved for ratios of $\theta_\mathrm{1/2}/\theta_\mathrm{res}$ between two and three with sharp rim models, which were resolved at lower values.
The inclination of a disk, therefore, has a strong influence on the detection of the inner disk cavity.}

\modif{
We also plotted our targets on this diagram so we can discuss the inner cavity (non)detections.
The inner cavity of two (HD45677 and R CrA) of the most resolved targets were detected because of their size.
HD98922 is at the theoretical limit of detection, but its pole-on orientation and likely sharp inner rim contribute to the inner cavity detection. 
Several of the nondetections are probably due to a small size compared with the achieved angular resolution (HD37806, HD142527, HD144432, HD163296, and VV Ser, with $\theta_\mathrm{1/2}/\theta_\mathrm{res}\leq1.5$).
Several targets are more resolved ($\theta_\mathrm{1/2}/\theta_\mathrm{res}$ > 1.5), but most of them have high inclination ($i$>45$^\circ$).
There are two targets (HD100453 and  MWC297) that are relatively well resolved ($\theta_\mathrm{1/2}/\theta_\mathrm{res}$ = 2.1) and with inclinations of 44.2$^\circ$ and 38.3$^\circ,$ respectively, for which no inner cavity is detected.
For these targets, it is more likely that the radially extended emission from the inner rim causes the nondetection, which is compatible with conclusions from L17.
For the other targets, the reasons for the nondetection are unclear (e.g., lack of angular resolution, absence of inner cavity, or high inclination) and longer baselines are needed to investigate the presence of inner cavities.}

\subsection{\modif{Detection of extended structures}}
\label{sec:extended}

\modif{
In the reconstructed images (Fig.\/\ref{fig:images}) in our sample, it appears that there are no significant feature\modiff{s} (above 3-$\sigma$) outside the inner emission.
From the azimuthally averaged radial profiles, there is also no significant emission outside the core emission.
Here, we discuss the reason for the lack of extended structures in the images.
To place an upper limit on the detection of such an emission, we added a ring to the restored images of our targets, generated a new dataset with the same (u,v)-plane, reconstructed an image, and plotted the radial profile.
We did that with two different ring radii (5 and 10\,mas) and different fluxes (from 1\% to 15\% of the total flux at 1.65$\mu$m, the ring has the same spectral index as the image). }

\begin{figure}
\centering
\includegraphics[width=9cm,angle=0]{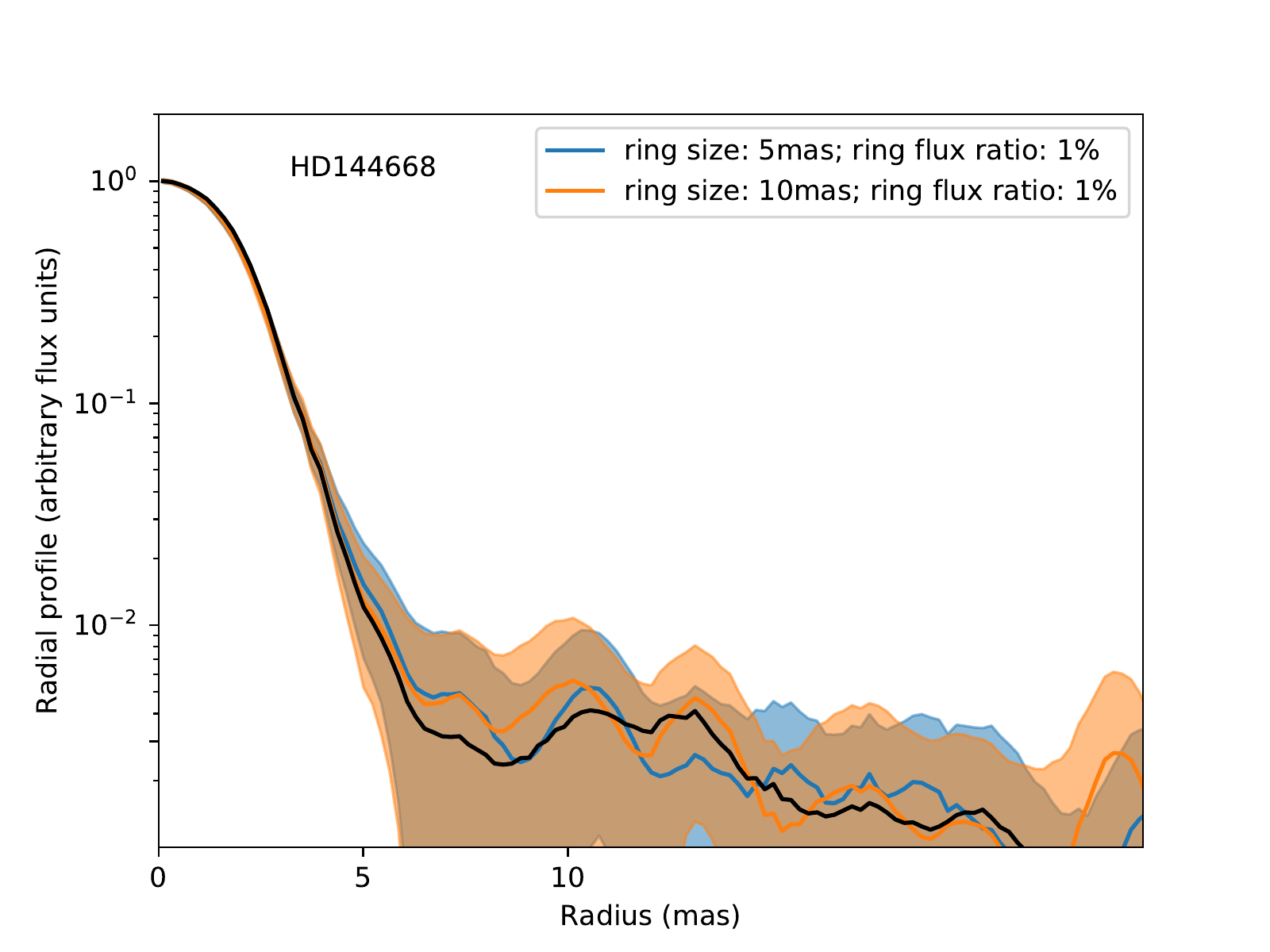}
\includegraphics[width=9cm,angle=0]{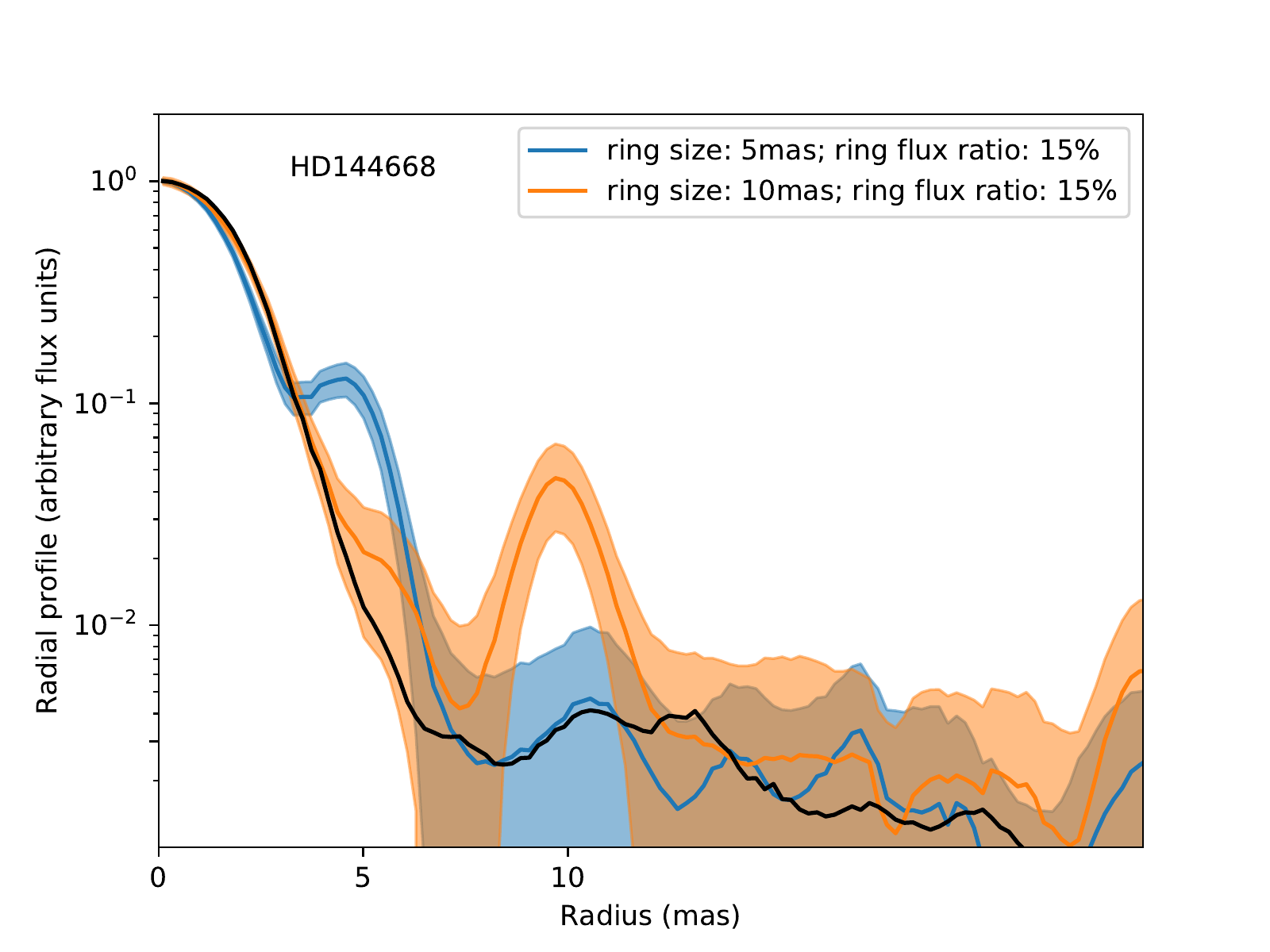}
\caption{\modif{Radial profiles of reconstructed images on synthetic data where a ring has been added. The black curves show the radial profiles in the images
    restored from the original data.  The blue and orange curves show
    the radial profiles in the restored images with an additional ring
    of radius 5\,mas and 10\,mas, respectively. The shaded areas show the limits at $\pm$1 standard deviation of the profiles (of the same color). The standard deviation of the profile from the original data is not shown. 
Top: the ring has a flux of 1\% of the total flux at 1.65$\mu$m. Bottom: the ring has a flux of 15\% of the total flux at 1.65$\mu$m.}}
\label{fig:extendedflux}
\end{figure}

\modif{
The radial profiles on reconstructed images using the dataset of one representative target, HD144668, are displayed in Fig.\,\ref{fig:extendedflux} for ring-to-total flux ratios of 1\% and 15\%.
The profiles with an added ring at 1\% are within $\pm$1 standard deviation of the original profile.
This is true for most of the reconstructions up to 10\% flux.
For the 15\% flux ring, the detection in the radial profiles are at $\sim$5\,sigma and $\sim$2\,sigma for the 5 and 10\,mas radius rings, respectively.
No bump detected is detected in the radial profiles on our targets \modiff{and the average detection limit between 5 and 15\,mas is around 10\% (excluding the most resolved targets HD45677 and R\,CrA).}
We can therefore rule out any structure that is larger than the core emission that contributes to a $\sim$1\modiff{0}\% flux level or less. We note that \citet{Ueda2019} predict that, under certain conditions, dust pile-up at the dead-zone inner boundary might cast a shadow on the inner astronomical units, which would lead to a second apparent ring-like structure of a few astronomical units in diameter. However, the predicted flux levels are on the order of the percent at best. We clearly lack the dynamical range to detect such a faint structure.
The near-infrared emission in our images is therefore dominated by three sources: the star, the core emission close to the star, and, for some targets, an over-resolved emission for which we can only evaluate the integrated flux through model fitting (L17).
Several possible origins can explain the latter, such as scattered light from the external parts of the disk (see, e.g., \cite{Benisty2010}) or emission from quantum heated particles as pointed out by \citet[]{Klarmann2016}. 
}

\subsection{\modif{Comparison with model fitting from L17}}
\label{sec:compm1}


\begin{figure*}[th]
\centering
\includegraphics[width=7cm]{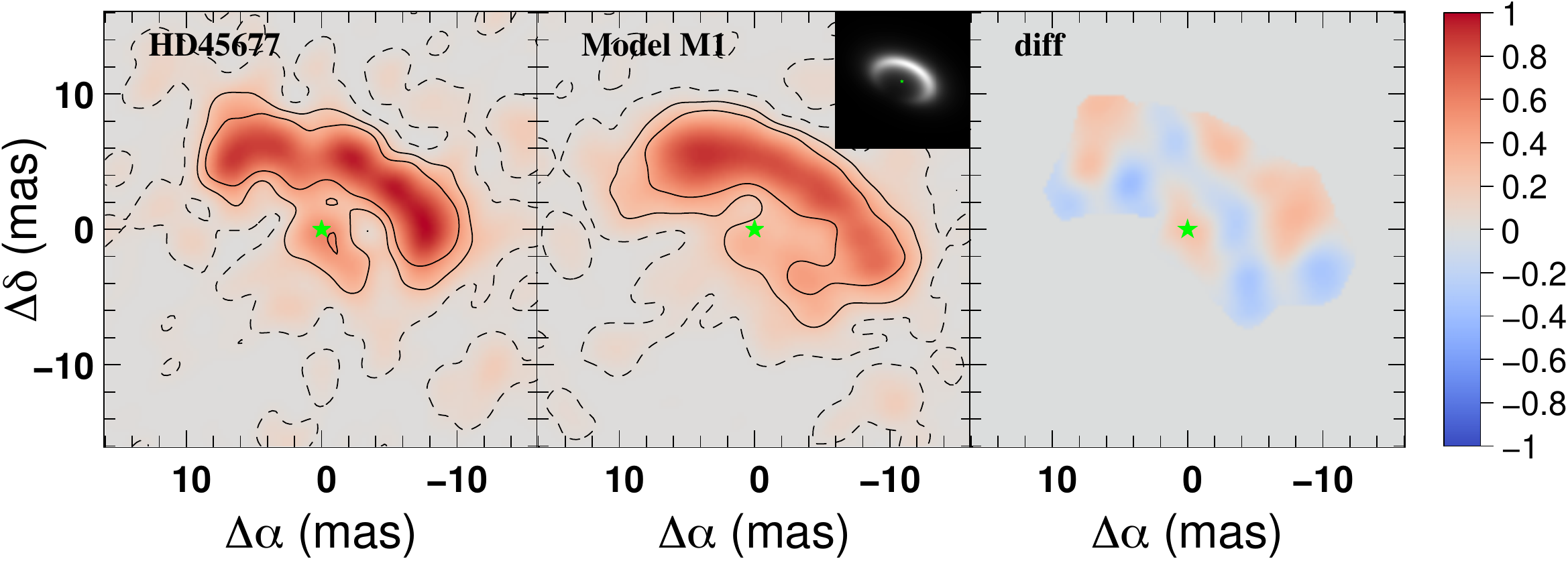}
\includegraphics[width=7cm]{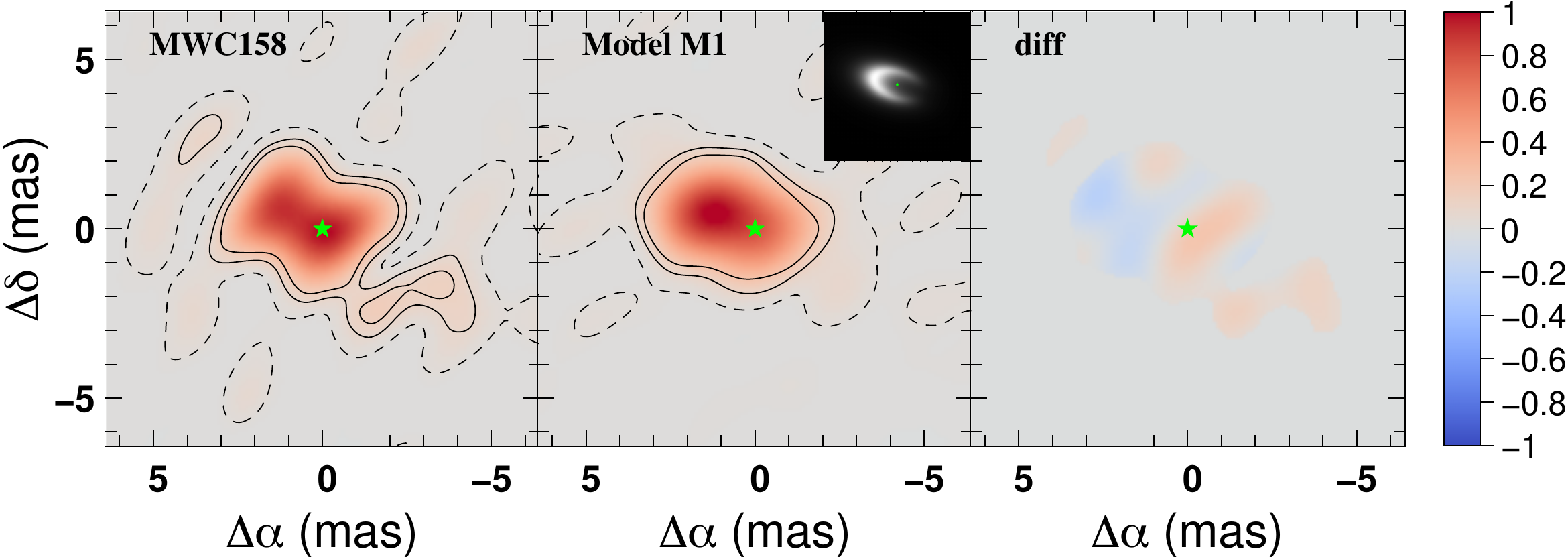}
\includegraphics[width=7cm]{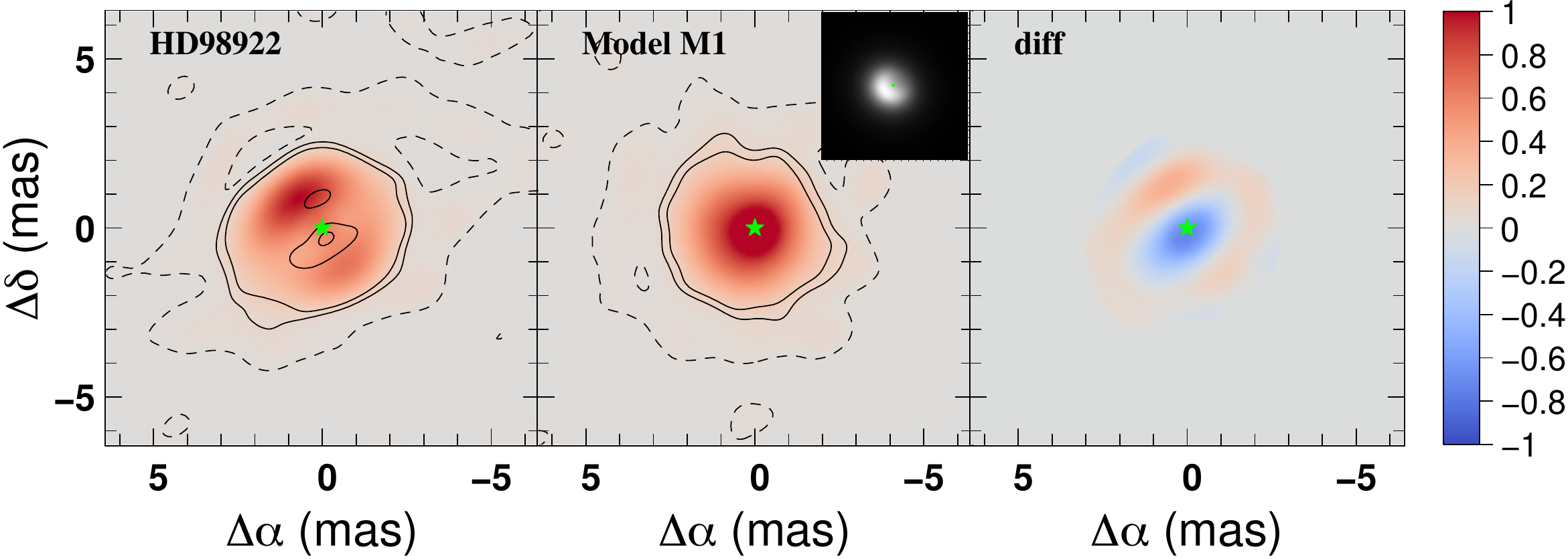}
\includegraphics[width=7cm]{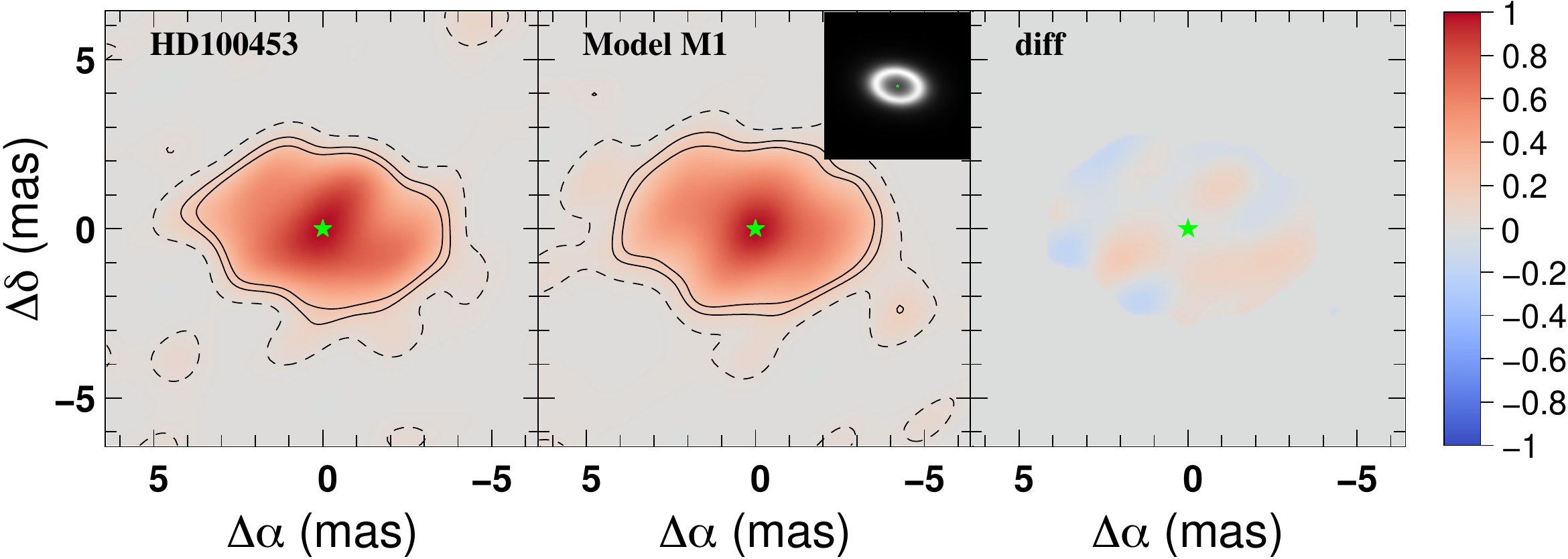}
\includegraphics[width=7cm]{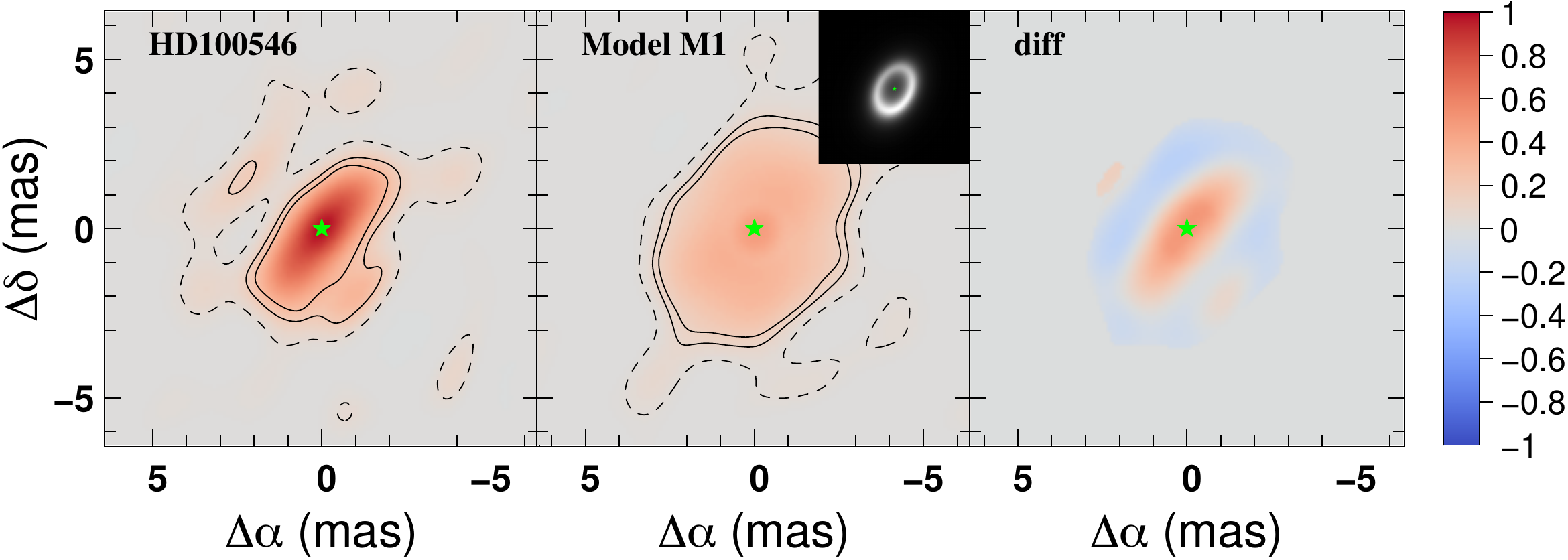}
\includegraphics[width=7cm]{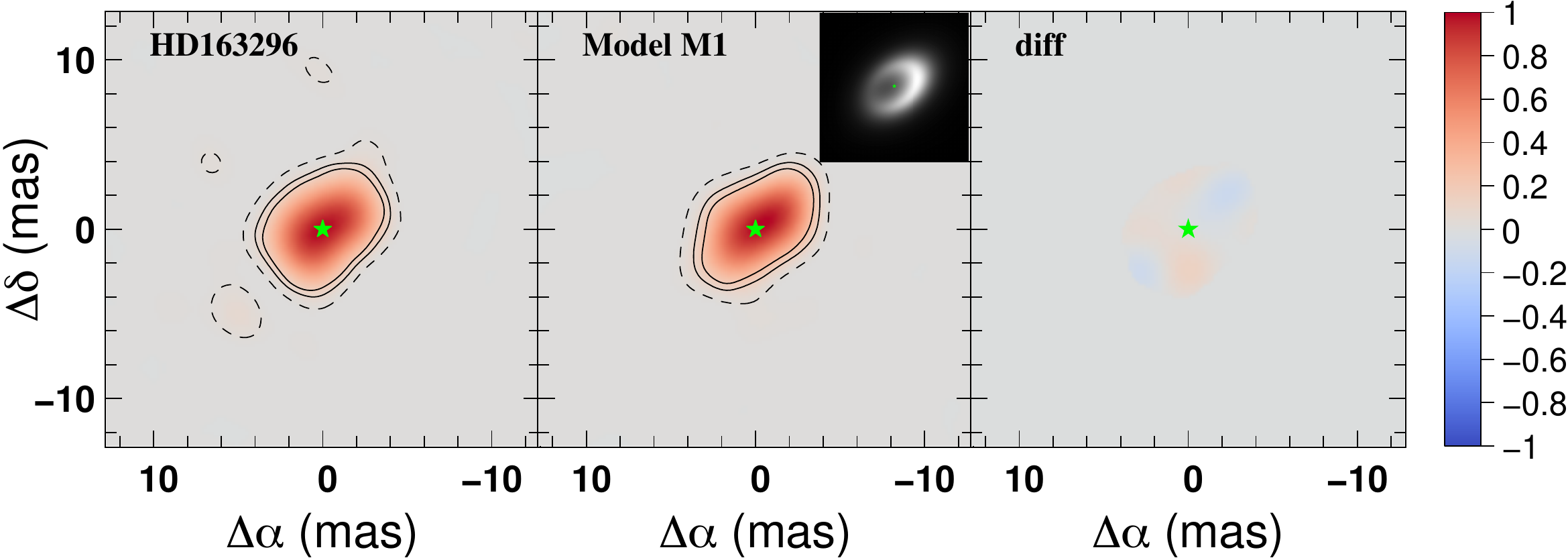}
\includegraphics[width=7cm]{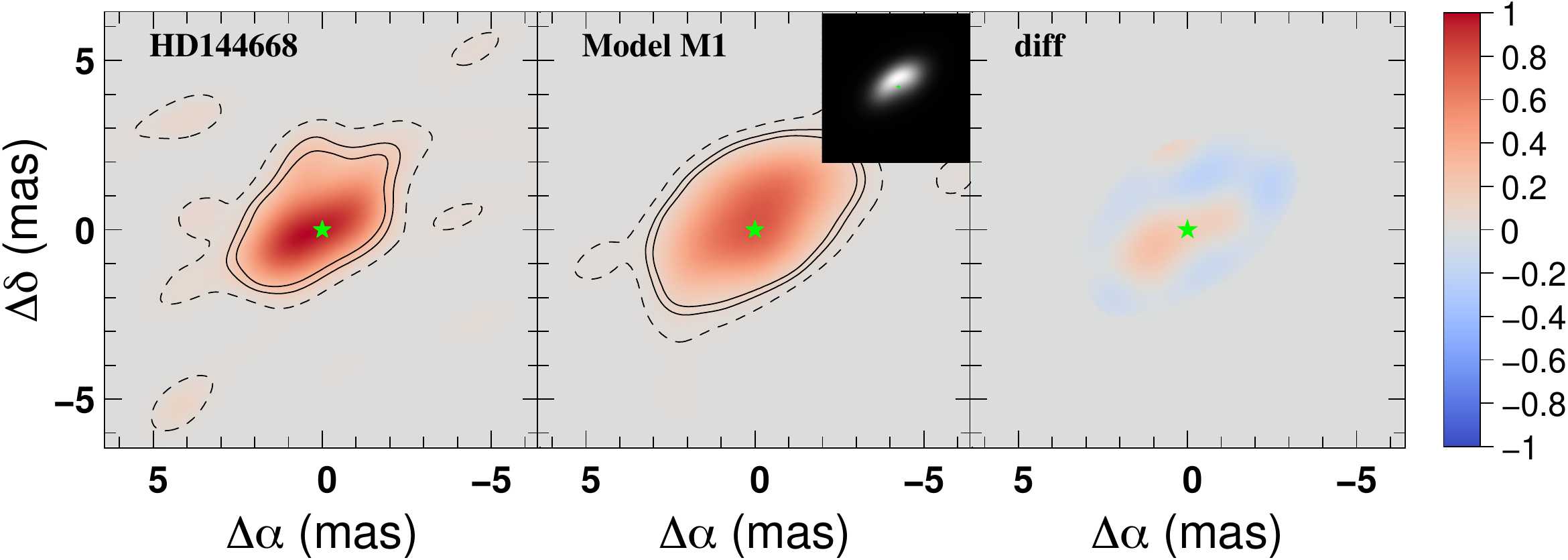}
\includegraphics[width=7cm]{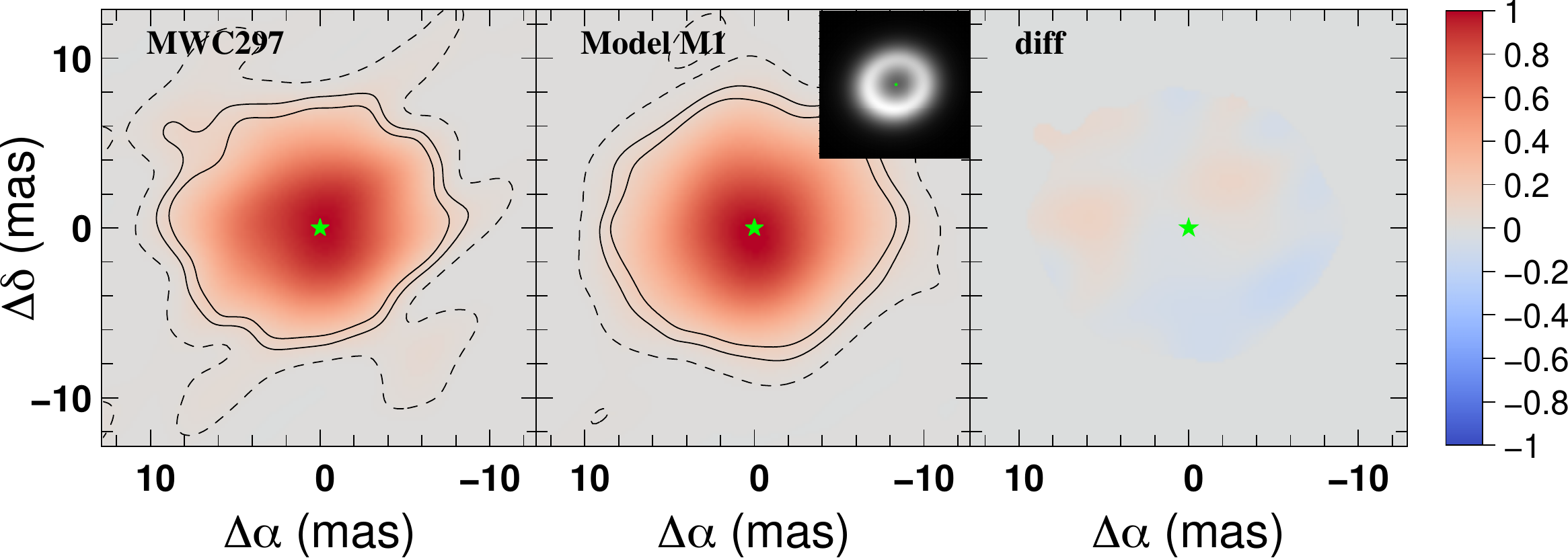}
\includegraphics[width=7cm]{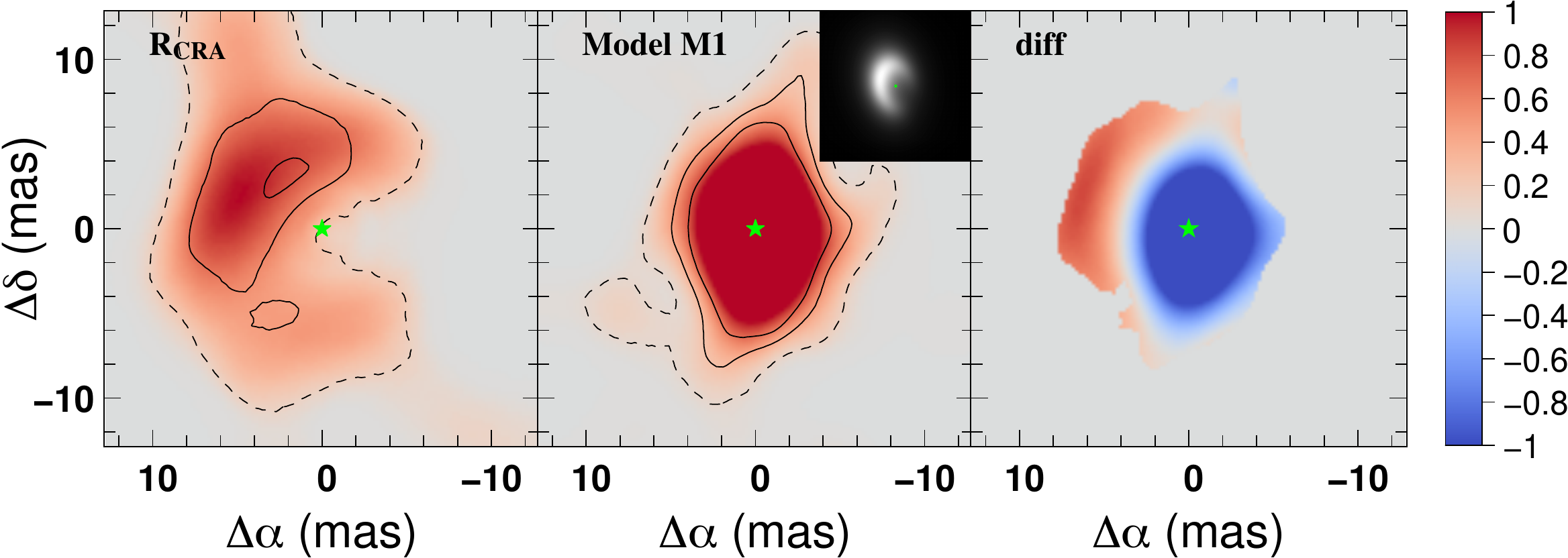}
\caption{\modif{Comparison between the image reconstructions on each target from the dataset (left panels) and the synthetic data based on the best-fit m1 model from L17 (middle panels) and the difference (right panels) in the 3-$\sigma$ significance area of the image. In the top-right corner of the model image reconstruction, there is the high resolution image of the parametric model. The green star represents the position of the central star.}}
\label{fig:m1m2comp}
\end{figure*}

\modif{Whereas the least resolved targets do not show specific morphology features, some of the most resolved ones display features that are not included in the parametric models of L17.
To make a fair comparison, we performed image reconstruction on datasets that were created from the best fits of L17 of ring models, including one so-called m1 azimuthal mode that corresponds to sinusoidal azimuthal modulation with a 2$\pi$-period.
The model dataset has the same \uv-coverage as the real data on a given target and a Gaussian noise based on the error bars of the real dataset.
}

\modif{In Fig.\ref{fig:m1m2comp}, we compare the image reconstructions on the real dataset with the image reconstructions on the synthetic dataset on best fit m1-models from L17 for the nine best resolved targets (HD45677, R CrA, HD100453, HD100546, HD163296, MWC297, MWC158, HD144668, and HD98922).
The m1-models can reproduce inclined inner rims of protoplanetary disks and especially any inclination effects.
The orientations of the targets are similar to the ones of the images from the models.
HD98922 is the only target for which the orientation is not correct.
This is due to the complex azimuthal structure of the extended emission where two maxima are present.
A model with second order modulation retrieves an orientation of the disk that is compatible with what we find in this work (see Fig.\,\ref{fig:incPA}).}

\modif{
We can see that for HD100453, HD163296, and MWC297 the images are similar.
For HD45677, the overall structure is similar.
However, because the disk rim is well resolved, we can discuss its details.
The emission from the rim shows a dip in the north direction that was not \modiff{recovered} in the image from the model that \modiff{has a smooth azimuthal brightness distribution}.
\modiff{The dip is detected at 3-$\sigma$ significance.}
Also, the sides of the rim are sharper than in the model. 
This is not surprising as the azimuthal modulation in the m1-model is done by a sinusoidal azimuthal modulation and may not reproduce azimuthal distributions of actual inner rims.
Moreover, the flux close to the central star is not as strong as in the real image.
The inner rim is therefore perturbed (for example by density or scale-height perturbations) and has a specific geometric profile that makes the closest part of the disk sharply self-shadowed.
A hypothesis to test would be the presence of a close companion, which contributes to the marginally resolved flux close to the star, that carves and perturbs the disk rim, which is located three times farther than the theoretical sublimation radius (see Sect.\,\ref{sec:Rsub} and Table\,\ref{tab:param2}).
The actual origin of those features are worth investigating in future work dedicated to this target.}

\modif{
The R CrA model image does not show an inner cavity, whereas it is clearly present in the image on the real dataset.
Recently, a binary inside the disk inner rim of R CrA was deduced from photometric times-series \citep{Sissa2019}. 
However, we do not see signs of disk truncation as the inner rim is compatible with dust sublimation.
Also, the asymmetry is compatible with inclination effects which do not display perturbations due to the inner binary.
For HD100546, the image shows a smaller emission for the real dataset.
There are also more structures in the image on the real dataset.
HD100546 is a transition disk and shows perturbation in the outer parts of the disk. It is further discussed in Sect.\,\ref{sec:RTcompare}.
MWC158 has an environment which is asymmetric and the position of the maximum is along the major axis.
This asymmetry is reproduced by the model. However, in the real images there are structures that are not present in the smoothly changing azimuthal profile of the ring model.
Finally, for HD144668, we see that the sinusoidal azimuthal variation of the ring model produces emission coming from the disk that is too extended.
}

\modif{From this qualitative analysis, we can see that the m1 ring models reproduce the global target morphology. However, the sinusoidal azimuthal modulation and the presence of a single ring are not sufficient in reproducing the morphology of the several resolved targets.
The structure of the inner rim is more complex and can be more easily reproduced by the image reconstruction technique.
It is therefore likely that an additional physical process \citep[e.g., disk instability or the presence of a companion;][]{Flock2017} produces this kind of asymmetry.
\modiff{A first improvement to the study would be to ensure that observations of a given object on all telescope configurations are all carried out well within the expected Keplerian orbital time at the inner rim. This would allow a possible smoothing out of the asymmetric emission to be lifted. A second improvement would be to push for a higher angular resolution at VLTI using the longest baseline ($\approx 200$\,m). This implementation is currently under study at the European Southern Observatory (ESO). Finally we note that some objects are sufficiently to the north so that they could be observed with VLTI and the Center for High Angular Resolution Astronomy (CHARA), which would provide a remarkable (u,v) coverage and angular resolution.}
}

\subsection{\modif{Comparing the images to the theoretical dust sublimation radius}}
\label{sec:Rsub}

\begin{figure}
\centering
\includegraphics[width=9cm,angle=0]{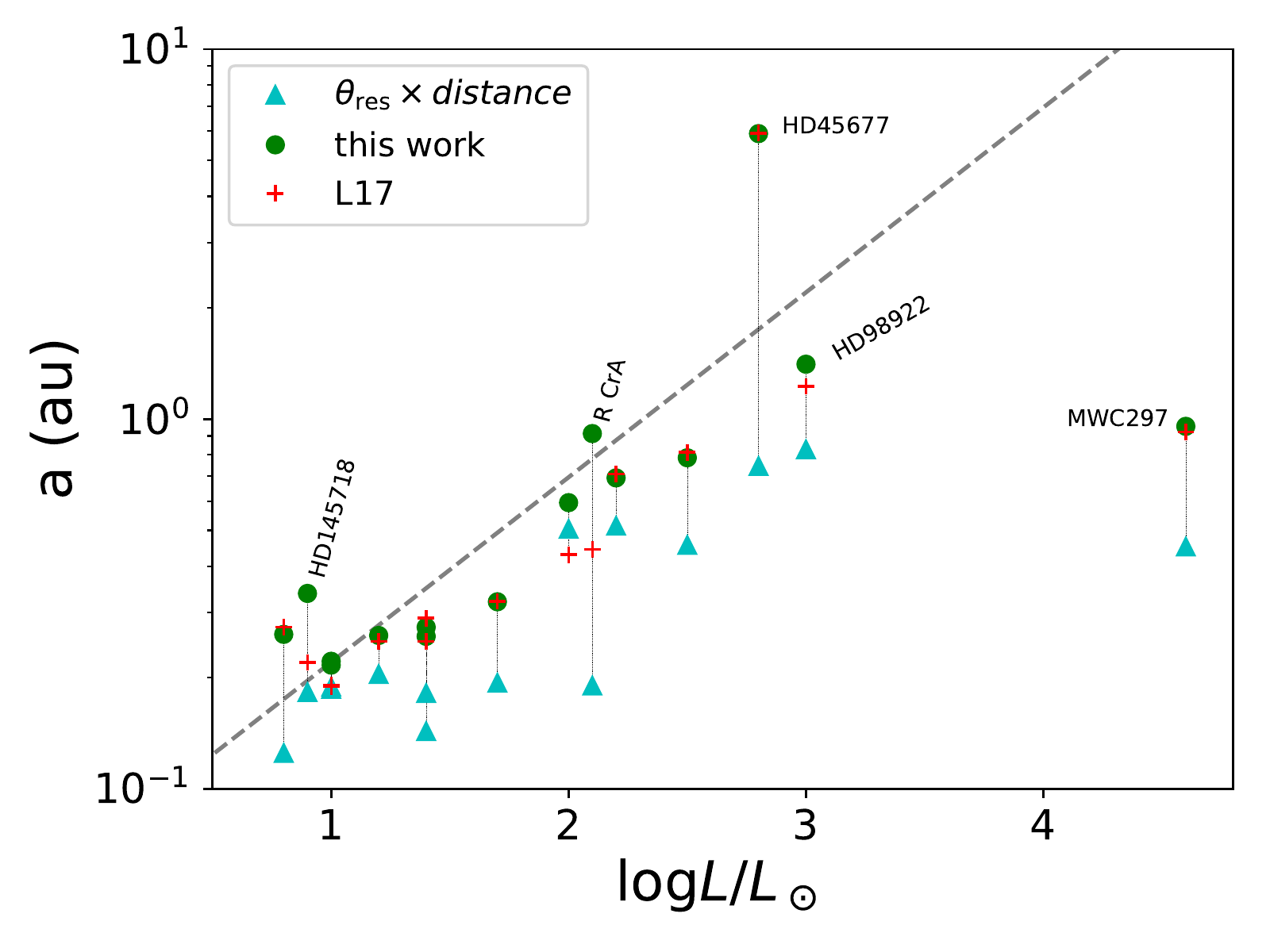}
\caption{\modif{Size-luminosity relation. The green circles are the sizes measured from images, red crosses are the sizes from parametric models as determined in L17, \modiff{and cyan triangles are the physical spatial resolution reached by our observations}. \modiff{T}he two size estimations \modiff{and the spatial resolution} are linked by a vertical dotted line. The dashed gray line represents the theoretical sublimation radius.}}
\label{fig:SizeLum}
\end{figure}

\modif{
We can measure the sizes of circumstellar emission from the images by taking the half-light radius corrected for the disk orientation (see Appendix\,\ref{app:incPA}).
We assessed the accuracy of the size measurement against the true sizes from radiative transfer models in Appendix\,\ref{sec:lessonsRT} by reconstructing images on the synthetic data generated from those models.
The recovered sizes mat\modiff{c}h the true sizes of the models with a sharp inner rim as long as the inner rim is resolved enough (i.e., $\theta_\mathrm{1/2}/\theta_\mathrm{res}\geq1.2$).
However, we find a bias for sizes from extended inner rim models \modiff{(see Appendix\,\ref{sec:lessonsRT})}.
We tend to recover sizes that are smaller than the true sizes for very resolved targets (i.e., $\theta_\mathrm{1/2}/\theta_\mathrm{res}\geq2.5$). }

\modif{
Most of the targets (12/15) are relatively well resolved (i.e., $\theta_\mathrm{1/2}/\theta_\mathrm{res}\geq1.3$).
In Fig.\,\ref{fig:SizeLum}, we plotted the sizes of the targets against their luminosity for both this work and L17. 
For the imaged object, we see good agreement between both size determinations, that is, using parametric models and image reconstruction. 
For some targets (HD145718 and R CrA), the difference is significant (ratio larger than 1.5 between the two sizes) and this is likely due to model bias (as the parameteric model for R CrA was an ellipsoid instead of a ring and the stellar-to-total flux ratios are different for HD145718). 
We note that the size difference is in the opposite direction of the size bias from image reconstruction that we identified in Appendix\,\ref{sec:lessonsRT}.
}

\modif{
In order to have a reference to interpret the sizes of the circumstellar emission from the reconstructed images, we computed the theoretical dust sublimation radii using:
\begin{equation}
\label{eq:sizeLum}
R_\mathrm{sub} = 1.1 (C_\mathrm{bw}/\epsilon)^\frac{1}{2} (L_* / 1000L_\odot)^{\frac{1}{2}} (T_\mathrm{sub}/1500K)^{-2},
\end{equation}
\citep[see, e.g.,][L17]{Monnier2002,Kama2009}, where $C_\mathrm{bw}$ is the back-warming coefficient that we fixed to four, $\epsilon$ is the cooling efficiency dust grain that we set to one, $L_*$ is the stellar luminosity that we took from L17, and $T_\mathrm{sub}$ is the sublimation temperature that we set to 1500\,K.
Since the physical size and the square root of the stellar luminosity are proportional to the distance, we can estimate the angular radius of sublimation without being sensitive to the distance error for a given target.
The angular sublimation radius for each target is defined in Table\,\ref{tab:param2} and is shown on the reconstructed images (Fig.\,\ref{fig:images}).
However, this theoretical sublimation radius can only be used as a reference since the different parameters involved ($C_\mathrm{bw}$, $\epsilon,$ or $T_\mathrm{sub}$) are not absolutely constrained and they may vary from object to object.
}

\modif{
While most of the targets are reasonably aligned with the reference size-luminosity relation defined by Eq.\,\ref{eq:sizeLum}, two targets are outliers: HD45677 and MWC297.
HD45677 has a resolved cavity and has a size that is more than three times larger than the theoretical dust sublimation radius.
This could be due to an inner binary truncating the disk, for example, even though no evidence for such a companion has been found in literature.
For the other two targets that have a resolved inner cavity (HD98922 and R CrA), the dust sublimation radius matches the radius of the centrally depressed emission well.}
\modif{MWC297 shows that its emission is more compact than the typical dust sublimation radius.
It is possible that for this target, which is a B1.5 star \citep{Fairlamb2015}, the emission is not coming from the dust sublimation but from another source, for example, free-free emission as it is the case for Be stars.
For all the other targets, the size-luminosity diagram confirms the findings of L17, where the targets near-infrared circumstellar sizes are proportional to the square root of stellar luminosity.
}

\begin{table}[!t]
\caption{Sizes of the stellar environments: theoretical sublimation radius ($\theta_\mathrm{sub}$) and the measured half-flux radius ($\theta_\mathrm{1/2}$ ).
}
\begin{center}
\begin{tabular}{c c c c c}
 \hline
  \hline
 Object & $\theta_\mathrm{sub}$  & $\theta_\mathrm{1/2}$ & $\theta_\mathrm{res}$ & $\theta_\mathrm{1/2}$/$\theta_\mathrm{res}$   \\
  & [mas]& [mas] & [mas]\\
 \hline
\object{HD37806} & 3.1 & 1.65$^{+0.05}_{-0.10}$ & 1.2 & 1.4 \\ [0.1cm]
\object{HD45677}  &  3.1 & 9.6$^{+1.8}_{-1.0}$ & 1.2 &  8 \\ [0.1cm]
\object{MWC158}  & 3.1 & $2.1^{+0.2}_{-0.1}$ & 1.2 & 1.7 \\ [0.1cm]
\object{HD98922} & 3.5 & $2.1^{+0.2}_{-0.2}$ & 1.2 & 1.7 \\ [0.1cm]
\object{HD100453}  & 1.7 & $2.6^{+0.1}_{-0.1}$ & 1.2 & 2.1 \\ [0.1cm]
\object{HD100546}  & 3.2 & $2.4^{+0.3}_{-0.2}$ & 1.3 & 1.8 \\ [0.1cm]
\object{HD142527}  & 1.4 & $1.4^{+0.1}_{-0.1}$ & 1.2 & 1.1 \\ [0.1cm]
\object{HD144432}  & 1.4 & $1.5^{+0.1}_{-0.1}$ & 1.2 & 1.2 \\ [0.1cm]
\object{HD144668} & 3.3 & $2.1^{+0.1}_{-0.2}$ & 1.2 & 1.7  \\[0.1cm]
\object{HD145718}  & 1.3 & $2.3^{+3.3}_{-0.2}$ & 1.2 & 1.9 \\[0.1cm]
\object{HD150193} & 2.3 & $1.9^{+0.1}_{-0.1}$ & 1.2 & 1.5 \\[0.1cm]
\object{HD163296}  & 2.8 & $2.5^{+0.2}_{-0.1}$ & 2.0 & 1.2 \\ [0.1cm]
\object{MWC297}  & 38.6 & $2.6^{+0.1}_{-0.1}$ & 1.2 & 2.1 \\ [0.1cm]
\object{VV Ser} & 1.6 & $1.5^{+0.1}_{-0.1}$ & 1.2 & 1.3 \\[0.1cm]
\object{R CrA}  &7.3 & $9.3^{+2.0}_{-1.0}$ & 2.0 & 4.7 
\end{tabular}
\end{center}
\label{tab:param2}
\end{table}%

\subsection{Asymmetry orientation compared with axisymmetric models}
\label{sec:fasym}
\begin{figure}
\centering
\includegraphics[width=9cm,angle=0]{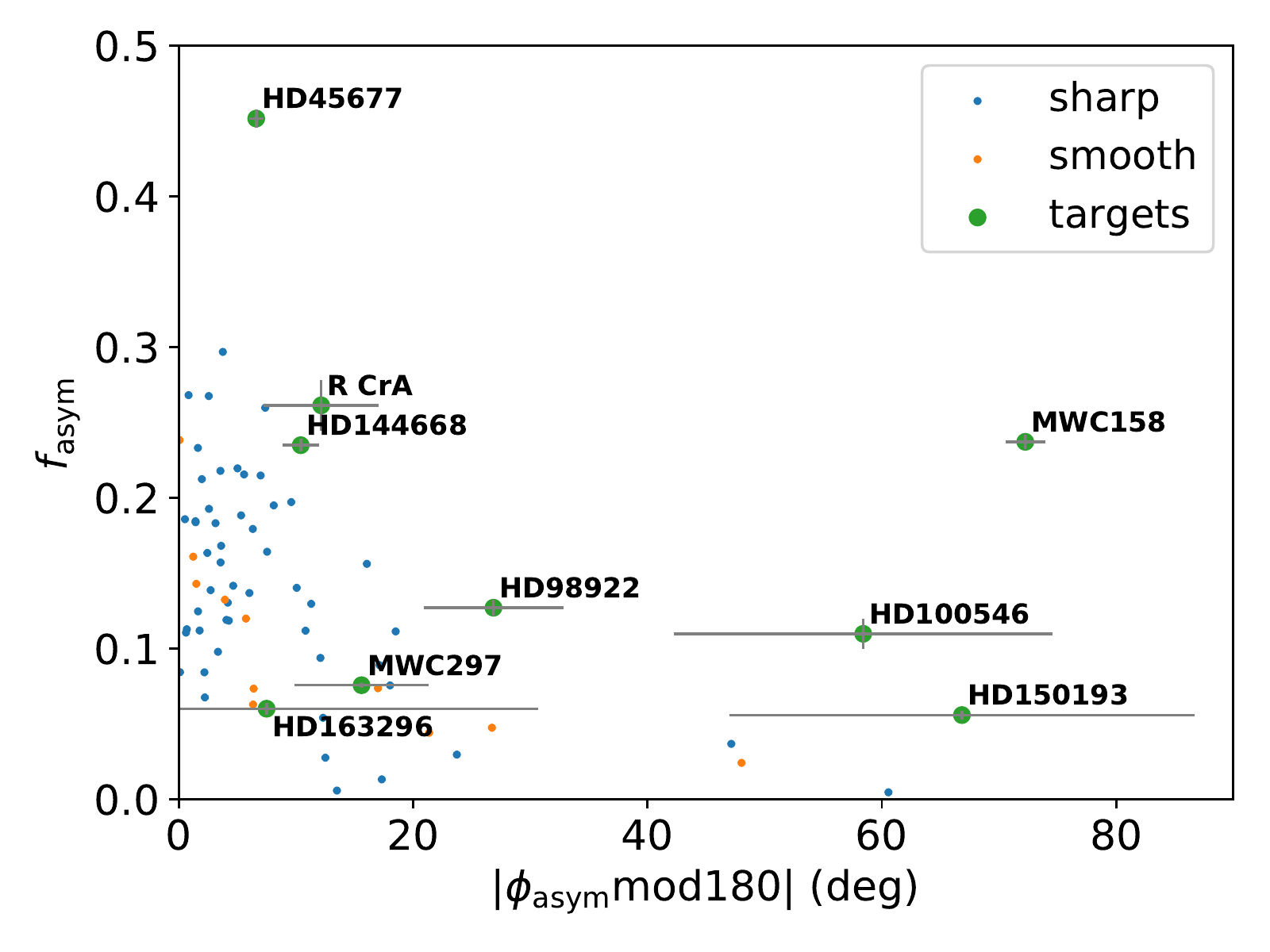}
\caption{$f_\mathrm{asym}$ versus \modiff{$|\arcsin {(\sin{(\phi_\mathrm{asym})})}|$} for observed (green) and model (blue and orange) objects. }
\label{fig:fskewmod}
\end{figure}

We have investigated the asymmetry that we could f\modif{i}nd by analyzing images that were reconstructed on synthetic datasets.
We used the synthetic datasets generated from MCFOST radiative transfer models used in L17.
\modif{Those models are described in Appendix\,\ref{sec:lessonsRT}.
}


We selected data that have a significant closure phase signal.
We then applied the same image reconstruction approach to obtain images for each of the model inclinations and distances.
\modif{In Fig.\,\ref{fig:fskewmod}, we present the asymmetry factors found from the models.
\modiff{Instead of plotting $\phi_\mathrm{asym}$, we plotted $|\arcsin {(\sin{(\phi_\mathrm{asym})})}|$, which is the absolute angle between the minor-axis and the asymmetry, independent of its absolute orientation.}
We can see that \modiff{$|\arcsin {(\sin{(\phi_\mathrm{asym})})}|$}  is below 20$^\circ $ for most
of the models.
The sharp models are more represented as they have a higher closure phase signal. They also have a larger $f_\mathrm{asym}$.}

\modif{We also placed the observed asymmetric objects on this diagram for comparison purposes.
First, we can see that the objects that have an inclination-like asymmetry are in the left part of the diagram and are in the area that is populated with the axisymmetric models.
While \object{MWC~158} cannot be reproduced by the models, \object{HD~100546} and \object{HD~150193} are located outside of the parameter space reproduced by the models \modif{and they are therefore \modifff{classified} as asymmetric in our analysis}. 
However, they have a large error bar.
\modif{HD98922 has an asymmetry angle ($\phi_\mathrm{asym}$) of 27$^\circ\pm$6$^\circ$ and is located outside the range span by the axisymmetric models.
This is due to the asymmetry located at the southeast of the star.
This is a hint for disk asymmetry that needs to be followed up by further observations.}
Further investigation of these targets would help to clearly identify if those stars are surrounded by nonaxisymmetric disks,}
\modiff{especially since we note that six out of ten objects that are the most resolved ($\theta_\mathrm{1/2}/\theta_\mathrm{res}\geq1.5$) have an $f_\mathrm{asym}>0.1$, meaning that potentially about half of the unresolved targets could have an asymmetric morphology.
This is in contrast to the five less-resolved targets that all have $f_\mathrm{asym}\leq0.11$.
}

\subsection{Origin of the noncentrosymmetry}
\label{sec:RTcompare}

\modif{
An obvious cause for the noncentrosymmetry of the stellar environment is the disk inclination.
This seems to be the case for five targets (HD45677, HD98922, HD163296, MWC297, and HD144668).
However, in this section, we further investigate the origin of environment noncentrosymmetry that could not be due to inclination effects.
}

\modif{
The disk dust sublimation rim can be perturbed by instabilities due to irradiation from the central star \citep{Fung2014} or by Rossby wave instability occurring at the inner edge of the dead zone \citep{Flock2017}.
Any perturbation of the inner disk affecting its vertical structure should generate a detectable departure from central-symmetry that would add to the natural one arising due to the inclination effect. It's orbital motion around the central star should leave a time-dependent signature.
Such a bright spot on the inner disk rim produces an asymmetric image but  also orbits the central star.}

\modif{The three objects with irregular departure from axial symmetry} discussed in Sect.\,\ref{sec:classify} \modif{can therefore trace dynamical phenomena in their disk, which induce temporal variability}. A clear and strong change in the disk morphology of \object{MWC~158} was previously noted \citep{Kluska2016}. 

\object{HD100546} is a well known transition disk, with evidence for asymmetric features, such as spiral arms in the outer disk (seen in scattered light), and for a possible protoplanet candidate \citep[e.g.,][]{Quanz2013,Quanz2015,Currie2015,Garufi2016,Follette2017,Rameau2017} that could propagate to the inner regions.
\object{HD163296} is known to host a jet that launches a knot every 16\,yrs \citep{Sitko2008,Ellerbroek2014}.
The origin to this regular periodicity can be a perturbation by a companion \citep{Terquem1999,Whelan2010,Estalella2012} or a warped inner disk \citep{Lai2003}.
More recently, a slight misalignment of the inner disk was suggested by modeling the direct imaging observations \citep{Muro2018}.
We also detect a disk misalignment for \object{HD150193} (see Sect.\,\ref{sec:misal}), suggesting a link with the perturbations we detect in the inner disk.

\begin{table}[!t]
\caption{Estimation of the number of sampled orbits between the first and the last point of our observations (time span) of asymmetric objects with the assumption of Keplerian motion. }
\begin{center}
\begin{tabular}{l l c c c c c}
 \hline
  \hline
 Object & radius & period & obs. time-span & sampled   \\
 &[au] & [days] & [days]  & orbits \\
 \hline
\object{HD45677} & \modif{6.0} & \modiff{2476} & \modif{64} & \modiff{0.03}  \\ [0.1cm]
\object{MWC158} &  \modif{0.8} & \modiff{128} & \modif{63} & \modiff{0.49} \\ [0.1cm]
\object{HD98922} &  \modif{1.4} &   \modiff{243} & \modif{62} & \modiff{0.26}  \\ [0.1cm]
\object{HD100546}&  \modif{0.3} &  \modiff{41} & \modif{62}  & \modiff{0.72} \\ [0.1cm]
\object{HD144668} & \modif{0.3} &  \modiff{39} & \modif{30} & \modiff{0.72}  \\ [0.1cm]
\object{HD150193} & \modif{0.3} &  \modiff{44} & \modif{28}  & \modiff{0.42} \\[0.1cm]
\object{HD163296} & \modif{0.3} & \modiff{38} & \modif{18} &\modiff{1.66}  \\ [0.1cm]
\object{MWC297} &  \modif{1.0} &  \modiff{227} & \modif{32} & \modiff{0.14} \\ [0.1cm]
\object{R CrA} &  \modif{0.9} & \modiff{102} & \modif{29} & \modiff{0.29}
\end{tabular}
\end{center}
\label{tab:orbit}
\end{table}%

\modif{\modiff{W}e discuss how an asymmetry located at the detected circumstellar \modiff{structure} would move during the observing time span.}
Because several different configuration\modif{s} on the VLTI are required to cover the \uv-plane, interferometric measurements can be collected during a period of a few weeks.
We therefore compare the temporal span of our observation with the Keplerian period at the position of the detected asymmetric circumstellar \modiff{structure} (Table\,\ref{tab:orbit}).
To convert the angular distances to the physical ones, we used \modiff{distance from \citet[]{Vioque2018} who used} parallaxes from the second data release of the Gaia mission \citep{Gaia2016,Gaia2018}.
Most of the objects completed at least half of their orbit.
Therefore, our observation\modif{s} sample a non-negligible part of the orbital time-scale of these objects.
\modiff{The exceptions are HD45677 and MWC297 for which the observation samples less than 15\% of the orbital period.}
The irregular features we observe can be orbiting the star during the time span of the observations (as it was the case for \object{MWC158}).
To explore this hypothesis, well-sampled interferometric observations \modiff{both in time and spatial frequencies} are necessary to follow the evolution of the disk morphology.

\subsection{Inner and outer disk misalignment}
\label{sec:misal}

\begin{table*}[!t]
\caption{Comparison between inner and outer disk orientations }
\begin{center}
\begin{tabular}{l r r r r r r r c}
 \hline
  \hline
Object & inc. (inner)   & PA (inner) &  inc. (outer) & PA (outer) & $\alpha_1$ & $\alpha_2$ & $\sigma_\alpha$ & Ref.  \\
& [$^\circ$] & [$^\circ$] & [$^\circ$] & [$^\circ$] & [$^\circ$] & [$^\circ$] & [$^\circ$] &  \\
\hline
\object{HD37806}  &  44 $\pm$ 2 &  57 $\pm$ 3 &- & - & - & - & - & - \\ 
\object{HD45677}  &   45 $\pm$ 8 &76 $\pm$ 11 & - &  - & - & - & - & -\\ 
\object{MWC158}  &  44 $\pm$ 3 &  74 $\pm$ 4 &- & - & - & - & - & -\\ 
\object{HD98922}  &  31 $\pm$ 5 & 125 $\pm$ 9 &- &  -& - & - & - & -\\ 
\object{HD100453}  &  44 $\pm$ 5 & 92 $\pm$ 8 & 38$\pm$5 & 142$\pm$5 & 32.8 & 73.2 & 9.6 & (1) \\ 
\object{HD100546}  &  56 $\pm$ 2  & 142 $\pm$ 2 & 41.34$\pm$0.03 & 145.14$\pm$0.04 & 14.6 & 83.0 & 2.6 & (2)\\ 
\object{HD142527}  &  30 $\pm$ 2  & 5 $\pm$ 5 & 27$\pm$5 & -19$\pm$5 & 11.7 & 55.1 & 6.4 & (3)\\ 
\object{HD144432}  &  13 $\pm$ 6 &  110 $\pm$ 29 & - & - & - & - & - & -\\ 
\object{HD144668}  &  56 $\pm$ 1 &125 $\pm$ 1 & - &  - & - & - & - & -\\ 
\object{HD145718}  &   48 $\pm$ 3 &  -3 $\pm$ 4 & - & - & - & - & - & -\\ 
\object{HD150193}  &  40 $\pm$ 2  & 131 $\pm$ 4 & 38$\pm$9 & 358$\pm$6 & 28.9 & 70.0  & 10.3 & (4)\\ 
\object{HD163296}  &  46 $\pm$ 1 & 131 $\pm$ 1 & 46$\pm$5 & 135$\pm$11 & 2.8 & 87.8 & 9.4 & (5)\\ 
\object{MWC297}  &  38 $\pm$ 5 & 122 $\pm$ 8 & - & - & - & - & - & -\\ 
\object{VV Ser}  &  51 $\pm$ 1 &  0 $\pm$ 1 & - & - & - & - & - & -\\ 
\object{R CrA}  &  49 $\pm$ 6 &  165 $\pm$ 8 & - & - & - & - & - & -
\end{tabular}
\tablefoot{inc.: inclination; PA: position angle.
References: (1) \citet{Benisty2017}, (2) \citet[]{Pineda2014}, (3) \citet[]{Soon2017}, (4) \citet[]{Fukagawa2003}, (5) \citet{Muro2018} \\
Where no error was provided, we applied an error of 5$^\circ$.}
\end{center}
\label{tab:incPA}
\end{table*}

Several disks around young stars display shadows that are attributed to misalignment between the inner and outer parts of the disk \citep[e.g.,][]{Marino2015,Benisty2017,Casassus2018}.
These disks are mainly transition disks, which have a gap\modif{p}ed \modif{disk} structure.
In these gap\modif{p}ed disks, the observed dark spots on the outer disk are interpreted as shadows, which are being cast by the inner disk on the inner edge of the outer disk. 
The origin of such a misalignment is still under debate, but the main invoked cause is the presence of a companion \modif{in the gap} (see e.g., \citet{Nealon2019}).

\modif{A direct} way of probing the disk misalignment is to compare the orientation of the inner disk from this work with values of the outer disk coming from near-infrared scattered light imaging or millimeter interferometric observations.
We computed the angle difference ($\alpha$) between the normal vector of the inner disk we determined and the one from the outer disk that we found in literature \citep[see Table\,\ref{tab:incPA};][]{Marino2015}.
\modif{
There is an angle degeneracy with the inclination as one can only measure the length ratio between the major and minor axis and not the inclination that can be negative or positive.}
We therefore computed both angles ($\alpha_1$ and $\alpha_2$).
In order to conclude on a misalignment between the inner and the outer disks, we propagated the errors on inclination and position angles to have an error on the angle difference between the two disks axes ($\sigma_\alpha$).
When the errors on the inclination and position angles were not known, we assumed an error of 5$^\circ$ \citep[][]{Min2017}.

We find orientations of the outer disk for five targets.
Four out of five of these targets show significant misalignment between the outer and the inner disk.
Disk misalignment was already observed for \object{HD100453} \citep{Benisty2017,Long2017,Min2017} and \object{HD142527} \citep{Marino2015}.
In the following, we discuss the misalignments per target.

\subsubsection{\modif{HD100453}}
For HD100453, \modif{a misalignment angle ($\alpha$)} of $\sim$72$^\circ$ was deduced \citep{Benisty2017}.
This angle is compatible with one of our solutions ($\alpha_2$=73.2$\pm$9.6).

\subsubsection{\modif{HD142527}}
From the shadow modeling of \object{HD142527,} a misalignment angle of 70$^\circ\pm$5$^\circ$ was suggested \citep{Marino2015}.
Our measurement does not confirm this value.
A modeling of the whole system, including orientations of the inner disk from this work, is needed to assess if the shadows can be due to the inner disk misalignment that we detect \citep[see, e.g.,][]{Nealon2019}.\\
\subsubsection{\modif{HD100546}}
\object{HD100546} is a well studied transition disk with evidence of a complex environment in the inner disk.
F\modif{ro}m polarimetric direct imaging observations in the visible with SPHERE, \citet{Mendigutia2017} observed bars across the gap.
ALMA observations of CO display a strong residual signal once a Keplerian model was fit \citep[]{Walsh2017}, suggesting out-of plane kinematic gas motion occurring in the inner parts of the disk.
A possible warping of the disk was previously suggested \citep{Pineda2014}.
We find misalignment angles of 14.6$^\circ$ and 83.0$^\circ\pm2.6^\circ$.
For the latter value, we would expect to detect sharp shadows but no evidence for them has been found so far \citep{Sissa2018}.
Nevertheless, the orientation\modif{s} of the inner disk that we derived are not compatible with the inner feature seen with ALMA or the bar morphology seen with SPHERE.
\\
\subsubsection{\modif{HD150193}}
\object{HD150193} is also such an object with suspected nonaxisymmetric inner disk morphology.
The misalignment is about 28.9$^\circ$ or 70.0$^\circ\pm10.3^\circ$.
Unlike the other targets showing misaligned disks, \object{HD150193} is not a known transition disk \citep[e.g.,][]{Banzatti2018}.
However, a stellar companion was detected outside the disk \citep{Fukagawa2003,Carmona2007,Garufi2014,Monnier2017}, which is able to dynamically perturb it as in the case of \object{HD100453}.\\

\subsubsection{\modif{HD163296}}
Finally, we do not detect a significant misalignment of the disk of \object{HD163296}.
A misalignment of $\sim$3$^\circ$ was suggested by a recent modeling work to reproduce both direct imaging SPHERE and millimetric ALMA observations \citep{Muro2018}.
Our observations are not \modif{precise} enough ($\sigma_\alpha$=9.4) to confirm this interpretation.
Moreover, this object can have a nonaxisymmetric morphology at the inner rim (see Section\,\ref{sec:RTcompare}).




\section{Summary and conclusions}
\label{sec:conlusion}

\modifff{We summarize here our most important findings from this work:}
\begin{enumerate}
\item We probed the inner astronomical units of the circumstellar environments of Herbig Ae/Be stars from the VLTI/PIONIER survey by applying the SPARCO image reconstruction method on 1\modif{5} objects.
It is the first imaging survey of the inner parts of protoplanetary disks through near-infrared interferometry. This imaging method \modif{is motivated by a need to investigate the morphologies of the disk inner regions and it is complementary to model fitting (L17) but it is less dependent on a strong prior hypothesis }.
    \item \modif{We note that 20\% (3/15)} of our targets show a ring-like morphology with a detected inner cavity.
The rest of the objects (8\modif{0}\%) are consistent with a \modif{centrally peaked morphology at the observed angular resolution.}
    \item We classified the objects into different categories based on their radial and azimuthal morphology.
We notice that, as expected, the \modif{less} resolved objects \modif{and the ones where near-infrared flux is dominated by emission from the star} do not show strong closure phase signals. The objects showing ring-like morphologies are dominated by inclination-like asymmetric emission.
    \item \modif{Our comparison of the most resolved targets with their geometric models from L17 indicates a more complex morphology for seven targets.}
    \item \modif{We ruled out structures outside the very inner disk emission. We worked out an \modiff{average} upper limit of ~1\modiff{0}\% of the flux at 1.65$\mu$m that could be located between 5 and 1\modiff{5}\,mas from the star.}
    \item We then analyzed the retrieved images by producing asymmetry maps and radial profiles.
We note that \modif{60}\% (9/1\modif{5}) of our targets show noncentrosymmetric emission, and \modif{40}\% of them \modif{(6/15)} are consistent with inclination-like noncentrosymmetry.
The other \modif{$\sim$20}\% (3/1\modif{5}) display irregular features that are linked with nonaxisymmetric and possibly time-variable morphologies. \modif{Those morphologies could be due to a disk warp or instabilities at the disk inner rim, such as Rossby wave instability, creating a vortex or the presence of a companion. }
    \item We probe misalignments between the inner and the outer disk regions.
We confirm the misalignments of \object{HD100453} and \object{HD142527}, even though, for the latter, the orientation of the inner disk is different from what was suggested from the shadows cast on the outer disk.
We confirm a potential disk warp in \object{HD100546}.
We also suggest that there is a disk misalignment in \object{HD150193}.
    \item Finally, we raise the question of the non-negligible dynamical evolution of the objects during interferometric observations that can disrupt the process of an image reconstruction if the time span of observation is comparable to the dynamical time-scale.

\end{enumerate}

\modiff{The simultaneous availability of PIONIER, GRAVITY, and MATISSE at VLTI offers the opportunity of multiwavelength studies covering the $H$, $K$, $L$, $M,$ and $N$ bands. This allows the observer to probe different regions in the disk, both vertically and radially. One of the first outcomes of such a study would be to determine the degree of flaring of the inner rim as predicted by Flock et al. 2017, for example, in relation with various processes at play (sublimation, wind, magnetic field). Moreover, the combination of visibility and SED fitting from 1.5 to 13 microns is a powerful tool to determine dust composition, size, and mineralogy. It would be particularly interesting to confirm L17’s hint that a significant fraction of dust responsible for near-infrared emission might be of a refractory nature. Exploiting the 200\,m available at VLTI will allow one to locate the disk inner rim and compare it to the dust composition. Additionally, GRAVITY’s polarimetric capability should be used to try to map the scattered light and further constrain the nature of the dust grains. It might even be possible to use polarization to enhance the contrast of the environment with respect to the star and better reveal the perturbed disk. We also note that combining time-resolved observations in the continuum, in the Brackett Gamma line together with spectro-polarimetry should be a powerful tool to test the magnetospheric accretion scenario. In the case of Herbig AeBe stars, as for T Tauri stars, one might expect that the inner dust is lifted into the magnetospheric funnels, thereby offering the possibility to relate continuum asymmetry with the exquisite line spectro-astrometry that could be related to possible accretion impacts and magnetic field topology. Finally, a dedicated search for putative inner companions responsible for disk perturbations should be carried out by pushing the dynamic range of the interferometric observations (better precision, many observing points) in combination with the high resolution infrared spectrometers that will  be available soon.}



\begin{acknowledgements}
\modifff{We want to acknowledge the referee for his/her comments.}
This work is supported by the French ANR POLCA project (Processing of pOLychromatic interferometriC data for Astrophysics, ANR-10-BLAN-0511).
JK acknowledges support from an Marie Curie Career Integration Grant (SH-06192, PI: Stefan Kraus) and a Philip Leverhulme Prize (PLP-2013-110).
JK acknowledges support from the Research Council of the KU Leuven under grant number C14/17/082.
FB acknowledges funding from National Science Foundation awards number 1210972 and 1616483.
SK acknowledges support from an ERC Starting Grant (Grant Agreement No.\ 639889) 
FMe acknowledges funding from the EU FP7-2011 under Grant Agreement No 284405.
FMe acknowledges funding from ANR of France under contract number ANR-16-CE31-0013.
C.P. acknowledges funding from the Australian Research Council via FT170100040, and DP180104235.
PIONIER is funded by the Universit\'e Joseph Fourier (UJF, Grenoble) through its Poles TUNES and SMING, the Institut de Plan\'etologie et d'Astrophysique de Grenoble, the ``Agence Nationale  pour la Recherche'' with the program ANR EXOZODI, and the Institut National des Science de l'Univers (INSU) via the ``Programme National de Physique Stellaire'' and ``Programme National de Plan\'etologie''. 
The authors want to warmly thank all the people involved in the VLTI project. 
This work is based on observations made with the ESO telescopes. 
It made use of the Smithsonian NASA Astrophysics Data System (ADS) and of the Centre de Donn\'ees astronomiques de Strasbourg (CDS). 
All calculations and graphics were performed with the open source software \texttt{Yorick}. 
This research has made use of the Jean-Marie Mariotti Center ASPRO2 and SearchCal services  co-developed by CRAL, IPAG and FIZEAU. 
\end{acknowledgements}


\bibliographystyle{bibtex/aa}
\bibliography{bibtex/biblio}

\begin{thebibliography}{84}
\expandafter\ifx\csname natexlab\endcsname\relax\def\natexlab#1{#1}\fi

\bibitem[{{Alibert} {et~al.}(2011){Alibert}, {Mordasini}, \&
  {Benz}}]{Alibert2011}
{Alibert}, Y., {Mordasini}, C., \& {Benz}, W. 2011, \aap, 526, A63

\bibitem[{{Andrews} {et~al.}(2018){Andrews}, {Huang}, {P{\'e}rez}, {Isella},
  {Dullemond}, {Kurtovic}, {Guzm{\'a}n}, {Carpenter}, {Wilner}, {Zhang}, {Zhu},
  {Birnstiel}, {Bai}, {Benisty}, {Hughes}, {{\"O}berg}, \&
  {Ricci}}]{Andrews2018}
{Andrews}, S.~M., {Huang}, J., {P{\'e}rez}, L.~M., {et~al.} 2018, \apjl, 869,
  L41

\bibitem[{{Anthonioz} {et~al.}(2015){Anthonioz}, {M{\'e}nard}, {Pinte}, {Le
  Bouquin}, {Benisty}, {Thi}, {Absil}, {Duch{\^e}ne}, {Augereau}, {Berger},
  {Casassus}, {Duvert}, {Lazareff}, {Malbet}, {Millan-Gabet}, {Schreiber},
  {Traub}, \& {Zins}}]{Anthonioz2015}
{Anthonioz}, F., {M{\'e}nard}, F., {Pinte}, C., {et~al.} 2015, \aap, 574, A41

\bibitem[{{Avenhaus} {et~al.}(2018){Avenhaus}, {Quanz}, {Garufi}, {Perez},
  {Casassus}, {Pinte}, {Bertrang}, {Caceres}, {Benisty}, \&
  {Dominik}}]{Avenhaus2018}
{Avenhaus}, H., {Quanz}, S.~P., {Garufi}, A., {et~al.} 2018, \apj, 863, 44

\bibitem[{{Bans} \& {K{\"o}nigl}(2012)}]{Bans2012}
{Bans}, A. \& {K{\"o}nigl}, A. 2012, \apj, 758, 100

\bibitem[{{Banzatti} {et~al.}(2018){Banzatti}, {Garufi}, {Kama}, {Benisty},
  {Brittain}, {Pontoppidan}, \& {Rayner}}]{Banzatti2018}
{Banzatti}, A., {Garufi}, A., {Kama}, M., {et~al.} 2018, \aap, 609, L2

\bibitem[{{Benisty} {et~al.}(2015){Benisty}, {Juhasz}, {Boccaletti},
  {Avenhaus}, {Milli}, {Thalmann}, {Dominik}, {Pinilla}, {Buenzli}, {Pohl},
  {Beuzit}, {Birnstiel}, {de Boer}, {Bonnefoy}, {Chauvin}, {Christiaens},
  {Garufi}, {Grady}, {Henning}, {Huelamo}, {Isella}, {Langlois}, {M{\'e}nard},
  {Mouillet}, {Olofsson}, {Pantin}, {Pinte}, \& {Pueyo}}]{Benisty2015}
{Benisty}, M., {Juhasz}, A., {Boccaletti}, A., {et~al.} 2015, \aap, 578, L6

\bibitem[{{Benisty} {et~al.}(2018){Benisty}, {Juh{\'a}sz}, {Facchini},
  {Pinilla}, {de Boer}, {P{\'e}rez}, {Keppler}, {Muro-Arena}, {Villenave},
  {Andrews}, {Dominik}, {Dullemond}, {Gallenne}, {Garufi}, {Ginski}, \&
  {Isella}}]{Benisty2018}
{Benisty}, M., {Juh{\'a}sz}, A., {Facchini}, S., {et~al.} 2018, \aap, 619, A171

\bibitem[{{Benisty} {et~al.}(2011){Benisty}, {Renard}, {Natta}, {Berger},
  {Massi}, {Malbet}, {Garcia}, {Isella}, {M{\'e}rand}, {Monin}, {Testi},
  {Thi{\'e}baut}, {Vannier}, \& {Weigelt}}]{Benisty2011}
{Benisty}, M., {Renard}, S., {Natta}, A., {et~al.} 2011, \aap, 531, A84

\bibitem[{{Benisty} {et~al.}(2017){Benisty}, {Stolker}, {Pohl}, {de Boer},
  {Lesur}, {Dominik}, {Dullemond}, {Langlois}, {Min}, {Wagner}, {Henning},
  {Juhasz}, {Pinilla}, {Facchini}, {Apai}, {van Boekel}, {Garufi}, {Ginski},
  {M{\'e}nard}, {Pinte}, {Quanz}, {Zurlo}, {Boccaletti}, {Bonnefoy}, {Beuzit},
  {Chauvin}, {Cudel}, {Desidera}, {Feldt}, {Fontanive}, {Gratton}, {Kasper},
  {Lagrange}, {LeCoroller}, {Mouillet}, {Mesa}, {Sissa}, {Vigan}, {Antichi},
  {Buey}, {Fusco}, {Gisler}, {Llored}, {Magnard}, {Moeller-Nilsson}, {Pragt},
  {Roelfsema}, {Sauvage}, \& {Wildi}}]{Benisty2017}
{Benisty}, M., {Stolker}, T., {Pohl}, A., {et~al.} 2017, \aap, 597, A42

\bibitem[{{Benisty} {et~al.}(2010){Benisty}, {Tatulli}, {M{\'e}nard}, \&
  {Swain}}]{Benisty2010}
{Benisty}, M., {Tatulli}, E., {M{\'e}nard}, F., \& {Swain}, M.~R. 2010, \aap,
  511, A75

\bibitem[{{Carmona} {et~al.}(2007){Carmona}, {van den Ancker}, \&
  {Henning}}]{Carmona2007}
{Carmona}, A., {van den Ancker}, M.~E., \& {Henning}, T. 2007, \aap, 464, 687

\bibitem[{{Casassus} {et~al.}(2018){Casassus}, {Avenhaus}, {P{\'e}rez},
  {Navarro}, {C{\'a}rcamo}, {Marino}, {Cieza}, {Quanz}, {Alarc{\'o}n}, {Zurlo},
  {Osses}, {Rannou}, {Rom{\'a}n}, \& {Barraza}}]{Casassus2018}
{Casassus}, S., {Avenhaus}, H., {P{\'e}rez}, S., {et~al.} 2018, \mnras, 477,
  5104

\bibitem[{{Colavita}(1999)}]{Colavita1999}
{Colavita}, M.~M. 1999, \pasp, 111, 111

\bibitem[{{Currie} {et~al.}(2015){Currie}, {Cloutier}, {Brittain}, {Grady},
  {Burrows}, {Muto}, {Kenyon}, \& {Kuchner}}]{Currie2015}
{Currie}, T., {Cloutier}, R., {Brittain}, S., {et~al.} 2015, \apjl, 814, L27

\bibitem[{{Debes} {et~al.}(2017){Debes}, {Poteet}, {Jang-Condell}, {Gaspar},
  {Hines}, {Kastner}, {Pueyo}, {Rapson}, {Roberge}, {Schneider}, \&
  {Weinberger}}]{Debes2017}
{Debes}, J.~H., {Poteet}, C.~A., {Jang-Condell}, H., {et~al.} 2017, \apj, 835,
  205

\bibitem[{{Ellerbroek} {et~al.}(2014){Ellerbroek}, {Podio}, {Dougados},
  {Cabrit}, {Sitko}, {Sana}, {Kaper}, {de Koter}, {Klaassen}, {Mulders},
  {Mendigut{\'{\i}}a}, {Grady}, {Grankin}, {van Winckel}, {Bacciotti},
  {Russell}, {Lynch}, {Hammel}, {Beerman}, {Day}, {Huelsman}, {Werren},
  {Henden}, \& {Grindlay}}]{Ellerbroek2014}
{Ellerbroek}, L.~E., {Podio}, L., {Dougados}, C., {et~al.} 2014, \aap, 563, A87

\bibitem[{{Espaillat} {et~al.}(2011){Espaillat}, {Furlan}, {D'Alessio},
  {Sargent}, {Nagel}, {Calvet}, {Watson}, \& {Muzerolle}}]{Espaillat2011}
{Espaillat}, C., {Furlan}, E., {D'Alessio}, P., {et~al.} 2011, \apj, 728, 49

\bibitem[{{Estalella} {et~al.}(2012){Estalella}, {L{\'o}pez}, {Anglada},
  {G{\'o}mez}, {Riera}, \& {Carrasco- Gonz{\'a}lez}}]{Estalella2012}
{Estalella}, R., {L{\'o}pez}, R., {Anglada}, G., {et~al.} 2012, \aj, 144, 61

\bibitem[{{Fairlamb} {et~al.}(2015){Fairlamb}, {Oudmaijer}, {Mendigut{\'\i}a},
  {Ilee}, \& {van den Ancker}}]{Fairlamb2015}
{Fairlamb}, J.~R., {Oudmaijer}, R.~D., {Mendigut{\'\i}a}, I., {Ilee}, J.~D., \&
  {van den Ancker}, M.~E. 2015, \mnras, 453, 976

\bibitem[{{Faure} {et~al.}(2015){Faure}, {Fromang}, {Latter}, \&
  {Meheut}}]{Faure2015}
{Faure}, J., {Fromang}, S., {Latter}, H., \& {Meheut}, H. 2015, \aap, 573, A132

\bibitem[{{Faure} \& {Nelson}(2016)}]{Faure2016}
{Faure}, J. \& {Nelson}, R.~P. 2016, \aap, 586, A105

\bibitem[{{Flaherty} {et~al.}(2016){Flaherty}, {DeMarchi}, {Muzerolle},
  {Balog}, {Herbst}, {Megeath}, {Furlan}, \& {Gutermuth}}]{Flaherty2016}
{Flaherty}, K.~M., {DeMarchi}, L., {Muzerolle}, J., {et~al.} 2016, \apj, 833,
  104

\bibitem[{{Flock} {et~al.}(2016){Flock}, {Fromang}, {Turner}, \&
  {Benisty}}]{Flock2016}
{Flock}, M., {Fromang}, S., {Turner}, N.~J., \& {Benisty}, M. 2016, \apj, 827,
  144

\bibitem[{{Flock} {et~al.}(2017){Flock}, {Fromang}, {Turner}, \&
  {Benisty}}]{Flock2017}
{Flock}, M., {Fromang}, S., {Turner}, N.~J., \& {Benisty}, M. 2017, \apj, 835,
  230

\bibitem[{{Follette} {et~al.}(2017){Follette}, {Rameau}, {Dong}, {Pueyo},
  {Close}, {Duch{\^e}ne}, {Fung}, {Leonard}, {Macintosh}, {Males}, {Marois},
  {Millar-Blanchaer}, {Morzinski}, {Mullen}, {Perrin}, {Spiro}, {Wang},
  {Ammons}, {Bailey}, {Barman}, {Bulger}, {Chilcote}, {Cotten}, {De Rosa},
  {Doyon}, {Fitzgerald}, {Goodsell}, {Graham}, {Greenbaum}, {Hibon}, {Hung},
  {Ingraham}, {Kalas}, {Konopacky}, {Larkin}, {Maire}, {Marchis}, {Metchev},
  {Nielsen}, {Oppenheimer}, {Palmer}, {Patience}, {Poyneer}, {Rajan},
  {Rantakyr{\"o}}, {Savransky}, {Schneider}, {Sivaramakrishnan}, {Song},
  {Soummer}, {Thomas}, {Vega}, {Wallace}, {Ward-Duong}, {Wiktorowicz}, \&
  {Wolff}}]{Follette2017}
{Follette}, K.~B., {Rameau}, J., {Dong}, R., {et~al.} 2017, \aj, 153, 264

\bibitem[{{Fukagawa} {et~al.}(2003){Fukagawa}, {Tamura}, {Itoh}, {Hayashi}, \&
  {Oasa}}]{Fukagawa2003}
{Fukagawa}, M., {Tamura}, M., {Itoh}, Y., {Hayashi}, S.~S., \& {Oasa}, Y. 2003,
  \apj, 590, L49

\bibitem[{{Fung} \& {Artymowicz}(2014)}]{Fung2014}
{Fung}, J. \& {Artymowicz}, P. 2014, \apj, 790, 78

\bibitem[{{Gaia Collaboration} {et~al.}(2018){Gaia Collaboration}, {Brown},
  {Vallenari}, {Prusti}, {de Bruijne}, {Babusiaux}, \&
  {Bailer-Jones}}]{Gaia2018}
{Gaia Collaboration}, {Brown}, A.~G.~A., {Vallenari}, A., {et~al.} 2018, ArXiv
  e-prints, arXiv:1804.09365

\bibitem[{{Gaia Collaboration} {et~al.}(2016){Gaia Collaboration}, {Prusti},
  {de Bruijne}, {Brown}, {Vallenari}, {Babusiaux}, {Bailer-Jones}, {Bastian},
  {Biermann}, {Evans}, {Eyer}, {Jansen}, {Jordi}, {Klioner}, {Lammers},
  {Lindegren}, {Luri}, {Mignard}, {Milligan}, {Panem}, {Poinsignon},
  {Pourbaix}, {Randich}, {Sarri}, {Sartoretti}, {Siddiqui}, {Soubiran},
  {Valette}, {van Leeuwen}, {Walton}, {Aerts}, {Arenou}, {Cropper}, {Drimmel},
  {H{\o}g}, {Katz}, {Lattanzi}, {O'Mullane}, {Grebel}, {Holland}, {Huc},
  {Passot}, {Bramante}, {Cacciari}, {Casta{\~n}eda}, {Chaoul}, {Cheek}, {De
  Angeli}, {Fabricius}, {Guerra}, {Hern{\'a}ndez}, {Jean-Antoine-Piccolo},
  {Masana}, {Messineo}, {Mowlavi}, {Nienartowicz}, {Ord{\'o}{\~n}ez- Blanco},
  {Panuzzo}, {Portell}, {Richards}, {Riello}, {Seabroke}, {Tanga},
  {Th{\'e}venin}, {Torra}, {Els}, {Gracia- Abril}, {Comoretto},
  {Garcia-Reinaldos}, {Lock}, {Mercier}, {Altmann}, {Andrae}, {Astraatmadja},
  {Bellas-Velidis}, {Benson}, {Berthier}, {Blomme}, {Busso}, {Carry},
  {Cellino}, {Clementini}, {Cowell}, {Creevey}, {Cuypers}, {Davidson}, {De
  Ridder}, {de Torres}, {Delchambre}, {Dell'Oro}, {Ducourant}, {Fr{\'e}mat},
  {Garc{\'\i}a-Torres}, {Gosset}, {Halbwachs}, {Hambly}, {Harrison}, {Hauser},
  {Hestroffer}, {Hodgkin}, {Huckle}, {Hutton}, {Jasniewicz}, {Jordan},
  {Kontizas}, {Korn}, {Lanzafame}, {Manteiga}, {Moitinho}, {Muinonen},
  {Osinde}, {Pancino}, {Pauwels}, {Petit}, {Recio-Blanco}, {Robin}, {Sarro},
  {Siopis}, {Smith}, {Smith}, {Sozzetti}, {Thuillot}, {van Reeven}, {Viala},
  {Abbas}, {Abreu Aramburu}, {Accart}, {Aguado}, {Allan}, {Allasia},
  {Altavilla}, {{\'A}lvarez}, {Alves}, {Anderson}, {Andrei}, {Anglada Varela},
  {Antiche}, {Antoja}, {Ant{\'o}n}, {Arcay}, {Atzei}, {Ayache}, {Bach},
  {Baker}, {Balaguer-N{\'u}{\~n}ez}, {Barache}, {Barata}, {Barbier}, {Barblan},
  {Baroni}, {Barrado y Navascu{\'e}s}, {Barros}, {Barstow}, {Becciani},
  {Bellazzini}, {Bellei}, {Bello Garc{\'\i}a}, {Belokurov}, {Bendjoya},
  {Berihuete}, {Bianchi}, {Bienaym{\'e}}, {Billebaud}, {Blagorodnova},
  {Blanco-Cuaresma}, {Boch}, {Bombrun}, {Borrachero}, {Bouquillon}, {Bourda},
  {Bouy}, {Bragaglia}, {Breddels}, {Brouillet}, {Br{\"u}semeister},
  {Bucciarelli}, {Budnik}, {Burgess}, {Burgon}, {Burlacu}, {Busonero}, {Buzzi},
  {Caffau}, {Cambras}, {Campbell}, {Cancelliere}, {Cantat-Gaudin}, {Carlucci},
  {Carrasco}, {Castellani}, {Charlot}, {Charnas}, {Charvet}, {Chassat},
  {Chiavassa}, {Clotet}, {Cocozza}, {Collins}, {Collins}, {Costigan}, {Crifo},
  {Cross}, {Crosta}, {Crowley}, {Dafonte}, {Damerdji}, {Dapergolas}, {David},
  {David}, {De Cat}, {de Felice}, {de Laverny}, {De Luise}, {De March}, {de
  Martino}, {de Souza}, {Debosscher}, {del Pozo}, {Delbo}, {Delgado},
  {Delgado}, {di Marco}, {Di Matteo}, {Diakite}, {Distefano}, {Dolding}, {Dos
  Anjos}, {Drazinos}, {Dur{\'a}n}, {Dzigan}, {Ecale}, {Edvardsson}, {Enke},
  {Erdmann}, {Escolar}, {Espina}, {Evans}, {Eynard Bontemps}, {Fabre},
  {Fabrizio}, {Faigler}, {Falc{\~a}o}, {Farr{\`a}s Casas}, {Faye}, {Federici},
  {Fedorets}, {Fern{\'a}ndez-Hern{\'a}ndez}, {Fernique}, {Fienga}, {Figueras},
  {Filippi}, {Findeisen}, {Fonti}, {Fouesneau}, {Fraile}, {Fraser}, {Fuchs},
  {Furnell}, {Gai}, {Galleti}, {Galluccio}, {Garabato}, {Garc{\'\i}a-Sedano},
  {Gar{\'e}}, {Garofalo}, {Garralda}, {Gavras}, {Gerssen}, {Geyer}, {Gilmore},
  {Girona}, {Giuffrida}, {Gomes}, {Gonz{\'a}lez-Marcos},
  {Gonz{\'a}lez-N{\'u}{\~n}ez}, {Gonz{\'a}lez-Vidal}, {Granvik}, {Guerrier},
  {Guillout}, {Guiraud}, {G{\'u}rpide}, {Guti{\'e}rrez-S{\'a}nchez}, {Guy},
  {Haigron}, {Hatzidimitriou}, {Haywood}, {Heiter}, {Helmi}, {Hobbs},
  {Hofmann}, {Holl}, {Holland}, {Hunt}, {Hypki}, {Icardi}, {Irwin}, {Jevardat
  de Fombelle}, {Jofr{\'e}}, {Jonker}, {Jorissen}, {Julbe}, {Karampelas},
  {Kochoska}, {Kohley}, {Kolenberg}, {Kontizas}, {Koposov}, {Kordopatis},
  {Koubsky}, {Kowalczyk}, {Krone-Martins}, {Kudryashova}, {Kull}, {Bachchan},
  {Lacoste-Seris}, {Lanza}, {Lavigne}, {Le Poncin-Lafitte}, {Lebreton},
  {Lebzelter}, {Leccia}, {Leclerc}, {Lecoeur-Taibi}, {Lemaitre}, {Lenhardt},
  {Leroux}, {Liao}, {Licata}, {Lindstr{\o}m}, {Lister}, {Livanou}, {Lobel},
  {L{\"o}ffler}, {L{\'o}pez}, {Lopez-Lozano}, {Lorenz}, {Loureiro},
  {MacDonald}, {Magalh{\~a}es Fernandes}, {Managau}, {Mann}, {Mantelet},
  {Marchal}, {Marchant}, {Marconi}, {Marie}, {Marinoni}, {Marrese},
  {Marschalk{\'o}}, {Marshall}, {Mart{\'\i}n-Fleitas}, {Martino}, {Mary},
  {Matijevi{\v{c}}}, {Mazeh}, {McMillan}, {Messina}, {Mestre}, {Michalik},
  {Millar}, {Miranda}, {Molina}, {Molinaro}, {Molinaro}, {Moln{\'a}r},
  {Moniez}, {Montegriffo}, {Monteiro}, {Mor}, {Mora}, {Morbidelli}, {Morel},
  {Morgenthaler}, {Morley}, {Morris}, {Mulone}, {Muraveva}, {Musella},
  {Narbonne}, {Nelemans}, {Nicastro}, {Noval}, {Ord{\'e}novic},
  {Ordieres-Mer{\'e}}, {Osborne}, {Pagani}, {Pagano}, {Pailler}, {Palacin},
  {Palaversa}, {Parsons}, {Paulsen}, {Pecoraro}, {Pedrosa}, {Pentik{\"a}inen},
  {Pereira}, {Pichon}, {Piersimoni}, {Pineau}, {Plachy}, {Plum}, {Poujoulet},
  {Pr{\v{s}}a}, {Pulone}, {Ragaini}, {Rago}, {Rambaux}, {Ramos-Lerate},
  {Ranalli}, {Rauw}, {Read}, {Regibo}, {Renk}, {Reyl{\'e}}, {Ribeiro},
  {Rimoldini}, {Ripepi}, {Riva}, {Rixon}, {Roelens}, {Romero-G{\'o}mez},
  {Rowell}, {Royer}, {Rudolph}, {Ruiz-Dern}, {Sadowski}, {Sagrist{\`a}
  Sell{\'e}s}, {Sahlmann}, {Salgado}, {Salguero}, {Sarasso}, {Savietto},
  {Schnorhk}, {Schultheis}, {Sciacca}, {Segol}, {Segovia}, {Segransan},
  {Serpell}, {Shih}, {Smareglia}, {Smart}, {Smith}, {Solano}, {Solitro},
  {Sordo}, {Soria Nieto}, {Souchay}, {Spagna}, {Spoto}, {Stampa}, {Steele},
  {Steidelm{\"u}ller}, {Stephenson}, {Stoev}, {Suess}, {S{\"u}veges}, {Surdej},
  {Szabados}, {Szegedi-Elek}, {Tapiador}, {Taris}, {Tauran}, {Taylor},
  {Teixeira}, {Terrett}, {Tingley}, {Trager}, {Turon}, {Ulla}, {Utrilla},
  {Valentini}, {van Elteren}, {Van Hemelryck}, {van Leeuwen}, {Varadi},
  {Vecchiato}, {Veljanoski}, {Via}, {Vicente}, {Vogt}, {Voss}, {Votruba},
  {Voutsinas}, {Walmsley}, {Weiler}, {Weingrill}, {Werner}, {Wevers},
  {Whitehead}, {Wyrzykowski}, {Yoldas}, {{\v{Z}}erjal}, {Zucker}, {Zurbach},
  {Zwitter}, {Alecu}, {Allen}, {Allende Prieto}, {Amorim},
  {Anglada-Escud{\'e}}, {Arsenijevic}, {Azaz}, {Balm}, {Beck}, {Bernstein},
  {Bigot}, {Bijaoui}, {Blasco}, {Bonfigli}, {Bono}, {Boudreault}, {Bressan},
  {Brown}, {Brunet}, {Bunclark}, {Buonanno}, {Butkevich}, {Carret}, {Carrion},
  {Chemin}, {Ch{\'e}reau}, {Corcione}, {Darmigny}, {de Boer}, {de Teodoro}, {de
  Zeeuw}, {Delle Luche}, {Domingues}, {Dubath}, {Fodor}, {Fr{\'e}zouls},
  {Fries}, {Fustes}, {Fyfe}, {Gallardo}, {Gallegos}, {Gardiol}, {Gebran},
  {Gomboc}, {G{\'o}mez}, {Grux}, {Gueguen}, {Heyrovsky}, {Hoar}, {Iannicola},
  {Isasi Parache}, {Janotto}, {Joliet}, {Jonckheere}, {Keil}, {Kim},
  {Klagyivik}, {Klar}, {Knude}, {Kochukhov}, {Kolka}, {Kos}, {Kutka}, {Lainey},
  {LeBouquin}, {Liu}, {Loreggia}, {Makarov}, {Marseille}, {Martayan},
  {Martinez-Rubi}, {Massart}, {Meynadier}, {Mignot}, {Munari}, {Nguyen},
  {Nordlander}, {Ocvirk}, {O'Flaherty}, {Olias Sanz}, {Ortiz}, {Osorio},
  {Oszkiewicz}, {Ouzounis}, {Palmer}, {Park}, {Pasquato}, {Peltzer}, {Peralta},
  {P{\'e}turaud}, {Pieniluoma}, {Pigozzi}, {Poels}, {Prat}, {Prod'homme},
  {Raison}, {Rebordao}, {Risquez}, {Rocca-Volmerange}, {Rosen}, {Ruiz-Fuertes},
  {Russo}, {Sembay}, {Serraller Vizcaino}, {Short}, {Siebert}, {Silva},
  {Sinachopoulos}, {Slezak}, {Soffel}, {Sosnowska}, {Strai{\v{z}}ys}, {ter
  Linden}, {Terrell}, {Theil}, {Tiede}, {Troisi}, {Tsalmantza}, {Tur},
  {Vaccari}, {Vachier}, {Valles}, {Van Hamme}, {Veltz}, {Virtanen}, {Wallut},
  {Wichmann}, {Wilkinson}, {Ziaeepour}, \& {Zschocke}}]{Gaia2016}
{Gaia Collaboration}, {Prusti}, T., {de Bruijne}, J.~H.~J., {et~al.} 2016,
  \aap, 595, A1

\bibitem[{{Garcia Lopez} {et~al.}(2015){Garcia Lopez}, {Tambovtseva},
  {Schertl}, {Grinin}, {Hofmann}, {Weigelt}, \& {Caratti o
  Garatti}}]{Garcia2015}
{Garcia Lopez}, R., {Tambovtseva}, L.~V., {Schertl}, D., {et~al.} 2015, \aap,
  576, A84

\bibitem[{{Garufi} {et~al.}(2014){Garufi}, {Quanz}, {Schmid}, {Avenhaus},
  {Buenzli}, \& {Wolf}}]{Garufi2014}
{Garufi}, A., {Quanz}, S.~P., {Schmid}, H.~M., {et~al.} 2014, \aap, 568, A40

\bibitem[{{Garufi} {et~al.}(2016){Garufi}, {Quanz}, {Schmid}, {Mulders},
  {Avenhaus}, {Boccaletti}, {Ginski}, {Langlois}, {Stolker}, {Augereau},
  {Benisty}, {Lopez}, {Dominik}, {Gratton}, {Henning}, {Janson}, {M{\'e}nard},
  {Meyer}, {Pinte}, {Sissa}, {Vigan}, {Zurlo}, {Bazzon}, {Buenzli}, {Bonnefoy},
  {Brandner}, {Chauvin}, {Cheetham}, {Cudel}, {Desidera}, {Feldt}, {Galicher},
  {Kasper}, {Lagrange}, {Lannier}, {Maire}, {Mesa}, {Mouillet}, {Peretti},
  {Perrot}, {Salter}, \& {Wildi}}]{Garufi2016}
{Garufi}, A., {Quanz}, S.~P., {Schmid}, H.~M., {et~al.} 2016, \aap, 588, A8

\bibitem[{{GRAVITY Collaboration: Perraut} {et~al.}(2019){GRAVITY
  Collaboration: Perraut}, {Labadie}, {Lazareff}, {Klarmann}, {Segura-Cox},
  {Benisty}, {Bouvier}, {Brand ner}, {Garatti}, {Caselli}, {Dougados},
  {Garcia}, {Garcia-Lopez}, {Kendrew}, {Koutoulaki}, {Kervella}, {Lin},
  {Pineda}, {Sanchez-Bermudez}, {van Dishoeck}, {Abuter}, {Amorim}, {Berger},
  {Bonnet}, {Buron}, {Cantalloube}, {Cl{\'e}net}, {Coud{\'e} du Foresto},
  {Dexter}, {de Zeeuw}, {Duvert}, {Eckart}, {Eisenhauer}, {Eupen}, {Gao},
  {Gendron}, {Genzel}, {Gillessen}, {Gordo}, {Grellmann}, {Haubois},
  {Haussmann}, {Henning}, {Hippler}, {Horrobin}, {Hubert}, {Jocou}, {Lacour},
  {Le Bouquin}, {L{\'e}na}, {M{\'e}rand}, {Ott}, {Paumard}, {Perrin}, {Pfuhl},
  {Rabien}, {Ray}, {Rau}, {Rousset}, {Scheithauer}, {Straub}, {Straubmeier},
  {Sturm}, {Vincent}, {Waisberg}, {Wank}, {Widmann}, {Wieprecht}, {Wiest},
  {Wiezorrek}, {Woillez}, \& {Yazici}}]{Perraut2019}
{GRAVITY Collaboration: Perraut}, K., {Labadie}, L., {Lazareff}, B., {et~al.}
  2019, arXiv e-prints, arXiv:1911.00611

\bibitem[{Hansen(2001)}]{Lcurve}
Hansen, P. 2001, Computational Inverse Problems in Electrocardiology

\bibitem[{{Hillen} {et~al.}(2016){Hillen}, {Kluska}, {Le Bouquin}, {Van
  Winckel}, {Berger}, {Kamath}, \& {Bujarrabal}}]{Hillen2016}
{Hillen}, M., {Kluska}, J., {Le Bouquin}, J.-B., {et~al.} 2016, \aap, 588, L1

\bibitem[{{Isella} \& {Natta}(2005)}]{Isella2005}
{Isella}, A. \& {Natta}, A. 2005, \aap, 438, 899

\bibitem[{{Kama} {et~al.}(2009){Kama}, {Min}, \& {Dominik}}]{Kama2009}
{Kama}, M., {Min}, M., \& {Dominik}, C. 2009, \aap, 506, 1199

\bibitem[{{Klarmann} {et~al.}(2017){Klarmann}, {Benisty}, {Min}, {Dominik},
  {Berger}, {Waters}, {Kluska}, {Lazareff}, \& {Le Bouquin}}]{Klarmann2016}
{Klarmann}, L., {Benisty}, M., {Min}, M., {et~al.} 2017, \aap, 599, A80

\bibitem[{{Kluska} {et~al.}(2016){Kluska}, {Benisty}, {Soulez}, {Berger}, {Le
  Bouquin}, {Malbet}, {Lazareff}, \& {Thi{\'e}baut}}]{Kluska2016}
{Kluska}, J., {Benisty}, M., {Soulez}, F., {et~al.} 2016, \aap, 591, A82

\bibitem[{{Kluska} {et~al.}(2014){Kluska}, {Malbet}, {Berger}, {Baron},
  {Lazareff}, {Le Bouquin}, {Monnier}, {Soulez}, \&
  {Thi{\'e}baut}}]{Kluska2014}
{Kluska}, J., {Malbet}, F., {Berger}, J.-P., {et~al.} 2014, \aap, 564, A80

\bibitem[{{Kraus} {et~al.}(2008){Kraus}, {Preibisch}, \& {Ohnaka}}]{Kraus2008}
{Kraus}, S., {Preibisch}, T., \& {Ohnaka}, K. 2008, \apj, 676, 490

\bibitem[{{Kretke} {et~al.}(2009){Kretke}, {Lin}, {Garaud}, \&
  {Turner}}]{Kretke2009}
{Kretke}, K.~A., {Lin}, D.~N.~C., {Garaud}, P., \& {Turner}, N.~J. 2009, \apj,
  690, 407

\bibitem[{{Kurosawa} {et~al.}(2016){Kurosawa}, {Kreplin}, {Weigelt}, {Natta},
  {Benisty}, {Isella}, {Tatulli}, {Massi}, {Testi}, {Kraus}, {Duvert},
  {Petrov}, \& {Stee}}]{Kurosawa2016}
{Kurosawa}, R., {Kreplin}, A., {Weigelt}, G., {et~al.} 2016, \mnras, 457, 2236

\bibitem[{{Lai}(2003)}]{Lai2003}
{Lai}, D. 2003, \apj, 591, L119

\bibitem[{{Lazareff} {et~al.}(2017){Lazareff}, {Berger}, {Kluska}, {Le
  Bouquin}, {Benisty}, {Malbet}, {Koen}, {Pinte}, {Thi}, {Absil}, {Baron},
  {Delboulb{\'e}}, {Duvert}, {Isella}, {Jocou}, {Juhasz}, {Kraus}, {Lachaume},
  {M{\'e}nard}, {Millan-Gabet}, {Monnier}, {Moulin}, {Perraut}, {Rochat},
  {Soulez}, {Tallon}, {Thi{\'e}baut}, {Traub}, \& {Zins}}]{Lazareff2017}
{Lazareff}, B., {Berger}, J.-P., {Kluska}, J., {et~al.} 2017, \aap, 599, A85

\bibitem[{{Le Bouquin} {et~al.}(2011){Le Bouquin}, {Berger}, {Lazareff},
  {Zins}, {Haguenauer}, {Jocou}, {Kern}, {Millan-Gabet}, {Traub}, {Absil},
  {Augereau}, {Benisty}, {Blind}, {Bonfils}, {Bourget}, {Delboulbe},
  {Feautrier}, {Germain}, {Gitton}, {Gillier}, {Kiekebusch}, {Kluska},
  {Knudstrup}, {Labeye}, {Lizon}, {Monin}, {Magnard}, {Malbet}, {Maurel},
  {M{\'e}nard}, {Micallef}, {Michaud}, {Montagnier}, {Morel}, {Moulin},
  {Perraut}, {Popovic}, {Rabou}, {Rochat}, {Rojas}, {Roussel}, {Roux},
  {Stadler}, {Stefl}, {Tatulli}, \& {Ventura}}]{Lebouquin2011}
{Le Bouquin}, J.-B., {Berger}, J.-P., {Lazareff}, B., {et~al.} 2011, \aap, 535,
  A67

\bibitem[{{Long} {et~al.}(2017){Long}, {Fernandes}, {Sitko}, {Wagner}, {Muto},
  {Hashimoto}, {Follette}, {Grady}, {Fukagawa}, {Hasegawa}, {Kluska}, {Kraus},
  {Mayama}, {McElwain}, {Oh}, {Tamura}, {Uyama}, {Wisniewski}, \&
  {Yang}}]{Long2017}
{Long}, Z.~C., {Fernandes}, R.~B., {Sitko}, M., {et~al.} 2017, \apj, 838, 62

\bibitem[{{Malbet} {et~al.}(2007){Malbet}, {Benisty}, {de Wit}, {Kraus},
  {Meilland}, {Millour}, {Tatulli}, {Berger}, {Chesneau}, {Hofmann}, {Isella},
  {Natta}, {Petrov}, {Preibisch}, {Stee}, {Testi}, {Weigelt}, {Antonelli},
  {Beckmann}, {Bresson}, {Chelli}, {Dugu{\'e}}, {Duvert}, {Gennari},
  {Gl{\"u}ck}, {Kern}, {Lagarde}, {Le Coarer}, {Lisi}, {Perraut}, {Puget},
  {Rantakyr{\"o}}, {Robbe-Dubois}, {Roussel}, {Zins}, {Accardo}, {Acke},
  {Agabi}, {Altariba}, {Arezki}, {Aristidi}, {Baffa}, {Behrend}, {Bl{\"o}cker},
  {Bonhomme}, {Busoni}, {Cassaing}, {Clausse}, {Colin}, {Connot},
  {Delboulb{\'e}}, {Domiciano de Souza}, {Driebe}, {Feautrier}, {Ferruzzi},
  {Forveille}, {Fossat}, {Foy}, {Fraix-Burnet}, {Gallardo}, {Giani}, {Gil},
  {Glentzlin}, {Heiden}, {Heininger}, {Hernandez Utrera}, {Kamm}, {Kiekebusch},
  {Le Contel}, {Le Contel}, {Lesourd}, {Lopez}, {Lopez}, {Magnard}, {Marconi},
  {Mars}, {Martinot-Lagarde}, {Mathias}, {M{\`e}ge}, {Monin}, {Mouillet},
  {Mourard}, {Nussbaum}, {Ohnaka}, {Pacheco}, {Perrier}, {Rabbia}, {Rebattu},
  {Reynaud}, {Richichi}, {Robini}, {Sacchettini}, {Schertl}, {Sch{\"o}ller},
  {Solscheid}, {Spang}, {Stefanini}, {Tallon}, {Tallon-Bosc}, {Tasso},
  {Vakili}, {von der L{\"u}he}, {Valtier}, {Vannier}, \&
  {Ventura}}]{Malbet2007}
{Malbet}, F., {Benisty}, M., {de Wit}, W.~J., {et~al.} 2007, \aap, 464, 43

\bibitem[{{Marino} {et~al.}(2015){Marino}, {Perez}, \& {Casassus}}]{Marino2015}
{Marino}, S., {Perez}, S., \& {Casassus}, S. 2015, \apj, 798, L44

\bibitem[{{McGinnis} {et~al.}(2015){McGinnis}, {Alencar}, {Guimar{\~a}es},
  {Sousa}, {Stauffer}, {Bouvier}, {Rebull}, {Fonseca}, {Venuti}, {Hillenbrand},
  {Cody}, {Teixeira}, {Aigrain}, {Favata}, {F{\H u}r{\'e}sz}, {Vrba},
  {Flaccomio}, {Turner}, {Gameiro}, {Dougados}, {Herbst},
  {Morales-Calder{\'o}n}, \& {Micela}}]{McGinnis2015}
{McGinnis}, P.~T., {Alencar}, S.~H.~P., {Guimar{\~a}es}, M.~M., {et~al.} 2015,
  \aap, 577, A11

\bibitem[{{Mendigut{\'{\i}}a} {et~al.}(2017){Mendigut{\'{\i}}a}, {Oudmaijer},
  {Garufi}, {Lumsden}, {Hu{\'e}lamo}, {Cheetham}, {de Wit}, {Norris}, {Olguin},
  \& {Tuthill}}]{Mendigutia2017}
{Mendigut{\'{\i}}a}, I., {Oudmaijer}, R.~D., {Garufi}, A., {et~al.} 2017, \aap,
  608, A104

\bibitem[{{Min} {et~al.}(2017){Min}, {Stolker}, {Dominik}, \&
  {Benisty}}]{Min2017}
{Min}, M., {Stolker}, T., {Dominik}, C., \& {Benisty}, M. 2017, \aap, 604, L10

\bibitem[{{Monnier} {et~al.}(2006){Monnier}, {Berger}, {Millan-Gabet}, {Traub},
  {Schloerb}, {Pedretti}, {Benisty}, {Carleton}, {Haguenauer}, {Kern},
  {Labeye}, {Lacasse}, {Malbet}, {Perraut}, {Pearlman}, \&
  {Zhao}}]{Monnier2006}
{Monnier}, J.~D., {Berger}, J.-P., {Millan-Gabet}, R., {et~al.} 2006, \apj,
  647, 444

\bibitem[{{Monnier} {et~al.}(2017){Monnier}, {Harries}, {Aarnio}, {Adams},
  {Andrews}, {Calvet}, {Espaillat}, {Hartmann}, {Hinkley}, {Kraus}, {McClure},
  {Oppenheimer}, {Perrin}, \& {Wilner}}]{Monnier2017}
{Monnier}, J.~D., {Harries}, T.~J., {Aarnio}, A., {et~al.} 2017, \apj, 838, 20

\bibitem[{{Monnier} \& {Millan-Gabet}(2002)}]{Monnier2002}
{Monnier}, J.~D. \& {Millan-Gabet}, R. 2002, \apj, 579, 694

\bibitem[{{Mordasini} {et~al.}(2009{\natexlab{a}}){Mordasini}, {Alibert}, \&
  {Benz}}]{Mordasini2009a}
{Mordasini}, C., {Alibert}, Y., \& {Benz}, W. 2009{\natexlab{a}}, \aap, 501,
  1139

\bibitem[{{Mordasini} {et~al.}(2009{\natexlab{b}}){Mordasini}, {Alibert},
  {Benz}, \& {Naef}}]{Mordasini2009b}
{Mordasini}, C., {Alibert}, Y., {Benz}, W., \& {Naef}, D. 2009{\natexlab{b}},
  \aap, 501, 1161

\bibitem[{{Muro-Arena} {et~al.}(2018){Muro-Arena}, {Dominik}, {Waters}, {Min},
  {Klarmann}, {Ginski}, {Isella}, {Benisty}, {Pohl}, {Garufi}, {Hagelberg},
  {Langlois}, {Menard}, {Pinte}, {Sezestre}, {van der Plas}, {Villenave},
  {Delboulb{\'e}}, {Magnard}, {M{\"o}ller- Nilsson}, {Pragt}, {Rabou}, \&
  {Roelfsema}}]{Muro2018}
{Muro-Arena}, G.~A., {Dominik}, C., {Waters}, L.~B.~F.~M., {et~al.} 2018, \aap,
  614, A24

\bibitem[{{Nealon} {et~al.}(2019){Nealon}, {Pinte}, {Alexander}, {Mentiplay},
  \& {Dipierro}}]{Nealon2019}
{Nealon}, R., {Pinte}, C., {Alexander}, R., {Mentiplay}, D., \& {Dipierro}, G.
  2019, \mnras, 484, 4951

\bibitem[{{P{\'e}rez} {et~al.}(2014){P{\'e}rez}, {Isella}, {Carpenter}, \&
  {Chandler}}]{Perez2014}
{P{\'e}rez}, L.~M., {Isella}, A., {Carpenter}, J.~M., \& {Chandler}, C.~J.
  2014, \apjl, 783, L13

\bibitem[{{Pineda} {et~al.}(2014){Pineda}, {Quanz}, {Meru}, {Mulders}, {Meyer},
  {Pani{\'c}}, \& {Avenhaus}}]{Pineda2014}
{Pineda}, J.~E., {Quanz}, S.~P., {Meru}, F., {et~al.} 2014, \apj, 788, L34

\bibitem[{{Pinte} {et~al.}(2009){Pinte}, {Harries}, {Min}, {Watson},
  {Dullemond}, {Woitke}, {M{\'e}nard}, \& {Dur{\'a}n-Rojas}}]{Pinte2009}
{Pinte}, C., {Harries}, T.~J., {Min}, M., {et~al.} 2009, \aap, 498, 967

\bibitem[{{Pinte} {et~al.}(2008){Pinte}, {M{\'e}nard}, {Berger}, {Benisty}, \&
  {Malbet}}]{Pinte2008}
{Pinte}, C., {M{\'e}nard}, F., {Berger}, J.~P., {Benisty}, M., \& {Malbet}, F.
  2008, \apjl, 673, L63

\bibitem[{{Pinte} {et~al.}(2006){Pinte}, {M{\'e}nard}, {Duch{\^e}ne}, \&
  {Bastien}}]{Pinte2006}
{Pinte}, C., {M{\'e}nard}, F., {Duch{\^e}ne}, G., \& {Bastien}, P. 2006, \aap,
  459, 797

\bibitem[{{Poppenhaeger} {et~al.}(2015){Poppenhaeger}, {Cody}, {Covey},
  {G{\"u}nther}, {Hillenbrand}, {Plavchan}, {Rebull}, {Stauffer}, {Wolk},
  {Espaillat}, {Forbrich}, {Gutermuth}, {Hora}, {Morales-Calder{\'o}n}, \&
  {Song}}]{Poppenhaeger2015}
{Poppenhaeger}, K., {Cody}, A.~M., {Covey}, K.~R., {et~al.} 2015, \aj, 150, 118

\bibitem[{{Quanz} {et~al.}(2015){Quanz}, {Amara}, {Meyer}, {Girard},
  {Kenworthy}, \& {Kasper}}]{Quanz2015}
{Quanz}, S.~P., {Amara}, A., {Meyer}, M.~R., {et~al.} 2015, \apj, 807, 64

\bibitem[{{Quanz} {et~al.}(2013){Quanz}, {Amara}, {Meyer}, {Kenworthy},
  {Kasper}, \& {Girard}}]{Quanz2013}
{Quanz}, S.~P., {Amara}, A., {Meyer}, M.~R., {et~al.} 2013, \apjl, 766, L1

\bibitem[{{Rameau} {et~al.}(2017){Rameau}, {Follette}, {Pueyo}, {Marois},
  {Macintosh}, {Millar- Blanchaer}, {Wang}, {Vega}, {Doyon}, {Lafreni{\`e}re},
  {Nielsen}, {Bailey}, {Chilcote}, {Close}, {Esposito}, {Males}, {Metchev},
  {Morzinski}, {Ruffio}, {Wolff}, {Ammons}, {Barman}, {Bulger}, {Cotten}, {De
  Rosa}, {Duchene}, {Fitzgerald}, {Goodsell}, {Graham}, {Greenbaum}, {Hibon},
  {Hung}, {Ingraham}, {Kalas}, {Konopacky}, {Larkin}, {Maire}, {Marchis},
  {Oppenheimer}, {Palmer}, {Patience}, {Perrin}, {Poyneer}, {Rajan},
  {Rantakyr{\"o}}, {Marley}, {Savransky}, {Schneider}, {Sivaramakrishnan},
  {Song}, {Soummer}, {Thomas}, {Wallace}, {Ward-Duong}, \&
  {Wiktorowicz}}]{Rameau2017}
{Rameau}, J., {Follette}, K.~B., {Pueyo}, L., {et~al.} 2017, \aj, 153, 244

\bibitem[{{Renard} {et~al.}(2011){Renard}, {Thi{\'e}baut}, \&
  {Malbet}}]{Renard2011}
{Renard}, S., {Thi{\'e}baut}, E., \& {Malbet}, F. 2011, \aap, 533, A64

\bibitem[{{Sissa} {et~al.}(2019){Sissa}, {Gratton}, {Alcal{\`a}}, {Desidera},
  {Messina}, {Mesa}, {D'Orazi}, \& {Rigliaco}}]{Sissa2019}
{Sissa}, E., {Gratton}, R., {Alcal{\`a}}, J.~M., {et~al.} 2019, \aap, 630, A132

\bibitem[{{Sissa} {et~al.}(2018){Sissa}, {Gratton}, {Garufi}, {Rigliaco},
  {Zurlo}, {Mesa}, {Langlois}, {de Boer}, {Desidera}, {Ginski}, {Lagrange},
  {Maire}, {Vigan}, {Dima}, {Antichi}, {Baruffolo}, {Bazzon}, {Benisty},
  {Beuzit}, {Biller}, {Boccaletti}, {Bonavita}, {Bonnefoy}, {Brandner},
  {Bruno}, {Buenzli}, {Cascone}, {Chauvin}, {Cheetham}, {Claudi}, {Cudel}, {De
  Caprio}, {Dominik}, {Fantinel}, {Farisato}, {Feldt}, {Fontanive}, {Galicher},
  {Giro}, {Hagelberg}, {Incorvaia}, {Janson}, {Kasper}, {Keppler}, {Kopytova},
  {Lagadec}, {Lannier}, {Lazzoni}, {LeCoroller}, {Lessio}, {Ligi}, {Marzari},
  {Menard}, {Meyer}, {Mouillet}, {Peretti}, {Perrot}, {Potiron}, {Rouan},
  {Salasnich}, {Salter}, {Samland}, {Schmidt}, {Scuderi}, \&
  {Wildi}}]{Sissa2018}
{Sissa}, E., {Gratton}, R., {Garufi}, A., {et~al.} 2018, \aap, 619, A160

\bibitem[{{Sitko} {et~al.}(2008){Sitko}, {Carpenter}, {Kimes}, {Wilde},
  {Lynch}, {Russell}, {Rudy}, {Mazuk}, {Venturini}, {Puetter}, {Grady},
  {Polomski}, {Wisnewski}, {Brafford}, {Hammel}, \& {Perry}}]{Sitko2008}
{Sitko}, M.~L., {Carpenter}, W.~J., {Kimes}, R.~L., {et~al.} 2008, \apj, 678,
  1070

\bibitem[{{Soon} {et~al.}(2017){Soon}, {Hanawa}, {Muto}, {Tsukagoshi}, \&
  {Momose}}]{Soon2017}
{Soon}, K.-L., {Hanawa}, T., {Muto}, T., {Tsukagoshi}, T., \& {Momose}, M.
  2017, Publications of the Astronomical Society of Japan, 69, 34

\bibitem[{{Stolker} {et~al.}(2017){Stolker}, {Sitko}, {Lazareff}, {Benisty},
  {Dominik}, {Waters}, {Min}, {Perez}, {Milli}, {Garufi}, {de Boer}, {Ginski},
  {Kraus}, {Berger}, \& {Avenhaus}}]{Stolker2017}
{Stolker}, T., {Sitko}, M., {Lazareff}, B., {et~al.} 2017, \apj, 849, 143

\bibitem[{{Tannirkulam} {et~al.}(2007){Tannirkulam}, {Harries}, \&
  {Monnier}}]{Tannirkulam2007}
{Tannirkulam}, A., {Harries}, T.~J., \& {Monnier}, J.~D. 2007, \apj, 661, 374

\bibitem[{{Tannirkulam} {et~al.}(2008){Tannirkulam}, {Monnier}, {Harries},
  {Millan-Gabet}, {Zhu}, {Pedretti}, {Ireland}, {Tuthill}, {ten Brummelaar},
  {McAlister}, {Farrington}, {Goldfinger}, {Sturmann}, {Sturmann}, \&
  {Turner}}]{Tannirkulam2008}
{Tannirkulam}, A., {Monnier}, J.~D., {Harries}, T.~J., {et~al.} 2008, \apj,
  689, 513

\bibitem[{{Tatulli} {et~al.}(2011){Tatulli}, {Benisty}, {M{\'e}nard},
  {Varni{\`e}re}, {Martin-Za{\"\i}di}, {Thi}, {Pinte}, {Massi}, {Weigelt},
  {Hofmann}, \& {Petrov}}]{Tatulli2011}
{Tatulli}, E., {Benisty}, M., {M{\'e}nard}, F., {et~al.} 2011, \aap, 531, A1

\bibitem[{{Terquem} {et~al.}(1999){Terquem}, {Eisl{\"o}ffel}, {Papaloizou}, \&
  {Nelson}}]{Terquem1999}
{Terquem}, C., {Eisl{\"o}ffel}, J., {Papaloizou}, J.~C.~B., \& {Nelson}, R.~P.
  1999, \apj, 512, L131

\bibitem[{{Thi{\'e}baut}(2008)}]{Thiebaut2008}
{Thi{\'e}baut}, E. 2008, in Society of Photo-Optical Instrumentation Engineers
  (SPIE) Conference Series, Vol. 7013, Society of Photo-Optical Instrumentation
  Engineers (SPIE) Conference Series, 1

\bibitem[{{Ueda} {et~al.}(2019){Ueda}, {Flock}, \& {Okuzumi}}]{Ueda2019}
{Ueda}, T., {Flock}, M., \& {Okuzumi}, S. 2019, \apj, 871, 10

\bibitem[{{Vioque} {et~al.}(2018){Vioque}, {Oudmaijer}, {Baines},
  {Mendigut{\'\i}a}, \& {P{\'e}rez-Mart{\'\i}nez}}]{Vioque2018}
{Vioque}, M., {Oudmaijer}, R.~D., {Baines}, D., {Mendigut{\'\i}a}, I., \&
  {P{\'e}rez-Mart{\'\i}nez}, R. 2018, \aap, 620, A128

\bibitem[{{Walsh} {et~al.}(2017){Walsh}, {Daley}, {Facchini}, \&
  {Juh{\'a}sz}}]{Walsh2017}
{Walsh}, C., {Daley}, C., {Facchini}, S., \& {Juh{\'a}sz}, A. 2017, \aap, 607,
  A114

\bibitem[{{Whelan} {et~al.}(2010){Whelan}, {Dougados}, {Perrin}, {Bonnefoy},
  {Bains}, {Redman}, {Ray}, {Bouy}, {Benisty}, {Bouvier}, {Chauvin}, {Garcia},
  {Grankvin}, \& {Malbet}}]{Whelan2010}
{Whelan}, E.~T., {Dougados}, C., {Perrin}, M.~D., {et~al.} 2010, \apj, 720,
  L119

\end{thebibliography}

\begin{appendix}


\section{Determination of the chromatic parameters} \label{app:fsde}

\begin{figure*}
   \centering
   \includegraphics[width=14cm]{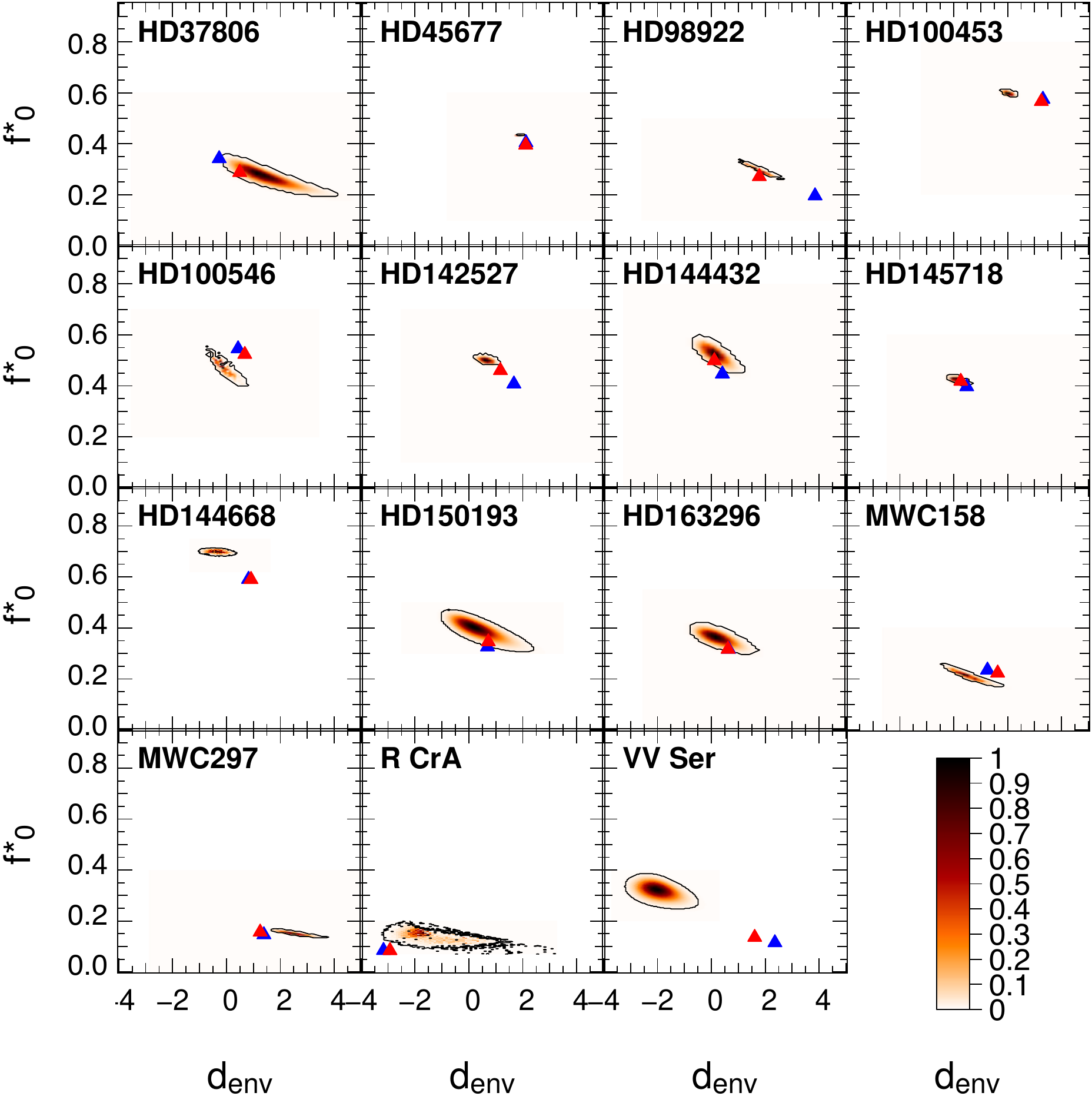}
      \caption{Significance maps for the chromatic parameters of each object. Contours indicate 5-$\sigma$ likelihood of the parameters for the actual dataset (black). Blue and red triangles represent the chromatic parameters from L17 for ring models with first order and second order modulations, respectively.}
         \label{fig:chrom}
   \end{figure*}

The determination of the chromatic parameters ($\fso$ and $\denv$) of each object was obtained by a selection a posteriori of the reconstructed images: a $100 \times 100$ grid was built using image reconstructions with different chromatic parameters.
For each couple of chromatic parameters, the value of the total function $f$ was minimized by the image reconstruction algorithm.
The probability density function $P=\exp (-\frac{f}{2})$ was then normalized to one because the \modif{densities} are negligible at the edges.
The grids are displayed in Fig.~\ref{fig:chrom}.

One bias was identified with the SPARCO method: When the environment is marginally resolved, the parameter that corresponds to the unresolved contribution can carry the flux of the star and a part of the flux from the unresolved environment.
Therefore, it yields an overestimation of the stellar-to-total flux ratio (\fso) and of the environment spectral index (\denv) since the assumption that all the flux in the parametric model is in the Rayleigh-Jeans regime is no longer correct.
It leads to a whole range of acceptable values \citep[see Fig.\,\ref{fig:chrom};][]{Kluska2014,Kluska2016}.
An interesting case is \object{HD\,100453}, where the parametric models from L17 predict a ring-like structure, but the reconstructed images from these models do not show such a feature. 
It is important to note that (1) if the data do not resolve the inner cavity, it is not seen in the image, as illustrated in the image reconstruction of the model shown in Sec.\,\ref{sec:lessonsRT} and that (2) the reconstruction is not based on a simultaneous photometric measurement. 
However, trying an image reconstruction using the chromatic parameters from L17 did not result in an image with an inner cavity.
This object is very similar to the case of a model treated in Sect.\,\ref{sec:lessonsRT} that has a transition radial profile between clearly resolved inner cavities and unresolved ones.

We also noti\modif{c}ed a change in the allowed values of the chromatic parameters with the strength of the regularization.
\modif{An increase in $\mu$  favors a smooth image and quenching noise, but it potentially fills a central cavity in the circumstellar brightness distribution. 
A side effect is that the extra flux assigned to the center of the circumstellar distribution is taken from the stellar flux, that is, a decrease of \fso. 
Furthermore, because of the degeneracy in the (\fso, \denv) parameter pair, which is visible in the oblique likelihood contours of Fig.\,\ref{fig:chrom}, the decrease of \fso is coupled with an increase of \denv. 
We show, however, below and in Fig.\,\ref{fig:Lcurve2}, that the actual impact of these effects is negligible.}

However most of the chromatic parameters are well constrained and they are in agreement with a photometric fit of the parameters. Unfortunately, the photometric observations were not usually simultaneously carried out with our observations; therefore, it is difficult to state whether the determination of the parameters from the interferometric data is wrong or if the object has varied.
In the paper, we keep the chromatic parameters determined from the interferometric data alone.

\section{\modif{Regularization weight} determination} \label{app:lcurve}
  
\begin{figure*}
   \centering
   \includegraphics[width=16cm]{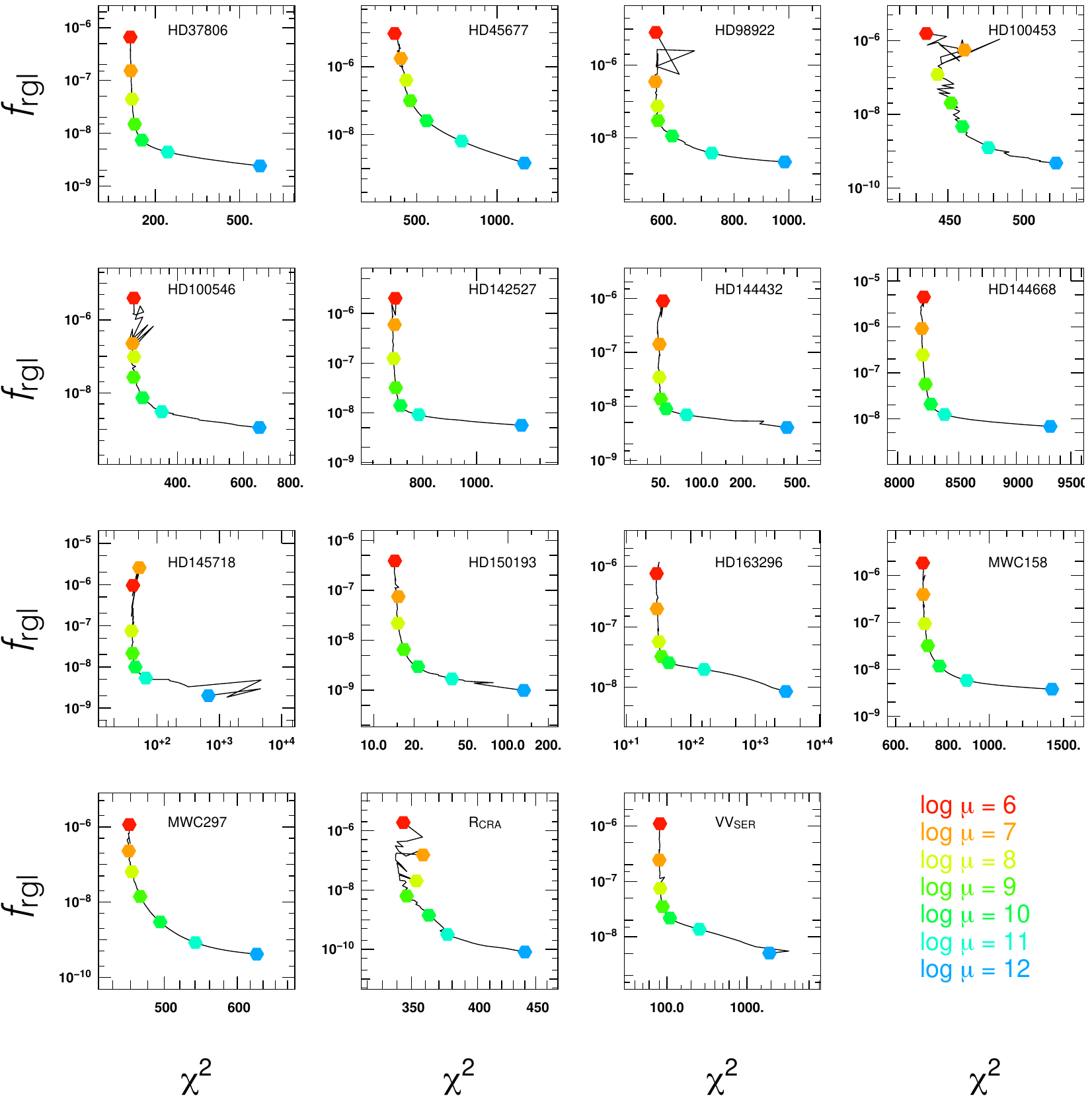}
      \caption{L-curves for all objets. The bend of the curve corresponds to the optimal regularization weight.}
         \label{fig:Lcurve}
   \end{figure*}
The optimal regularization weight depends on the \uv-coverage and the complexity of the shape of the object.
It can be determined using the L-curve method \citep{Renard2011,Kluska2016}.
Once the chromatic parameters were fixed (see previous section), image reconstructions were performed to build a grid of the regularization weight $\mu$.
In the plot representing $f_\mathrm{rgl}$ in function of $f_\mathrm{data}$, two regimes appear to create asymptotes that give an ``L" shape to the plot. The first regime is dominated by the likelihood and the second by the regularization.
The optimal \modif{regularization} weight corresponds to the one located at the bend of the L-curve.
The L-curves for each object are displayed in Fig.~\ref{fig:Lcurve}.
If the found \modif{regularization} weight differs significantly from the one used to determine the chromatic parameters, then we have iterated by redoing the chromatic parameters determination with the new \modif{regularization} weight and the L-curve was checked again.
All the parameters used for the image reconstruction including the chromatic parameters and the \modif{regularization} weights used for each object are summarized in Table~\ref{tab:param}.

\begin{figure*}
   \centering
   \includegraphics[width=15cm]{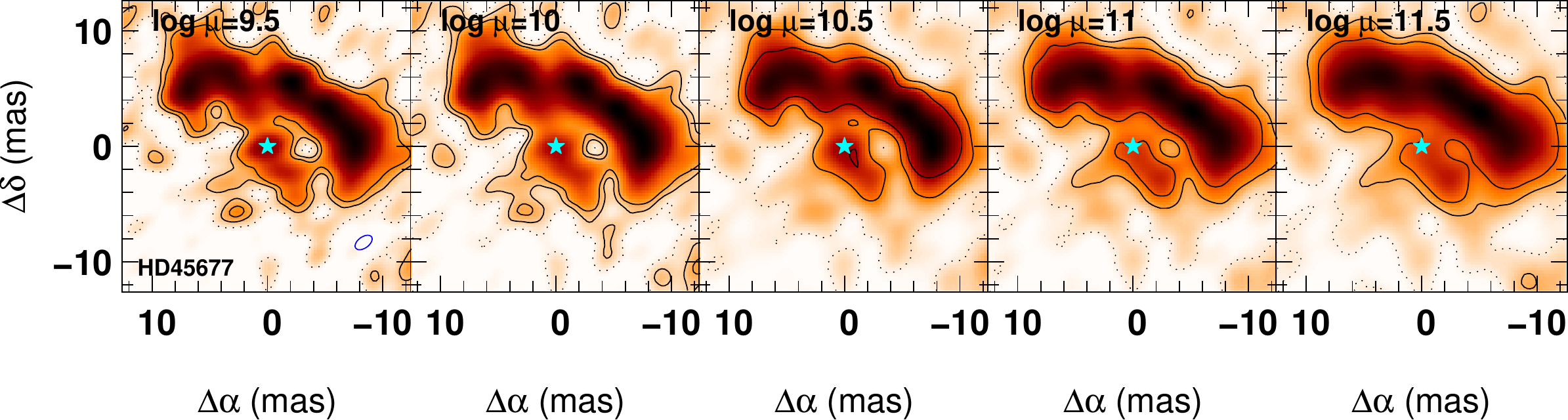}

   \includegraphics[width=15cm]{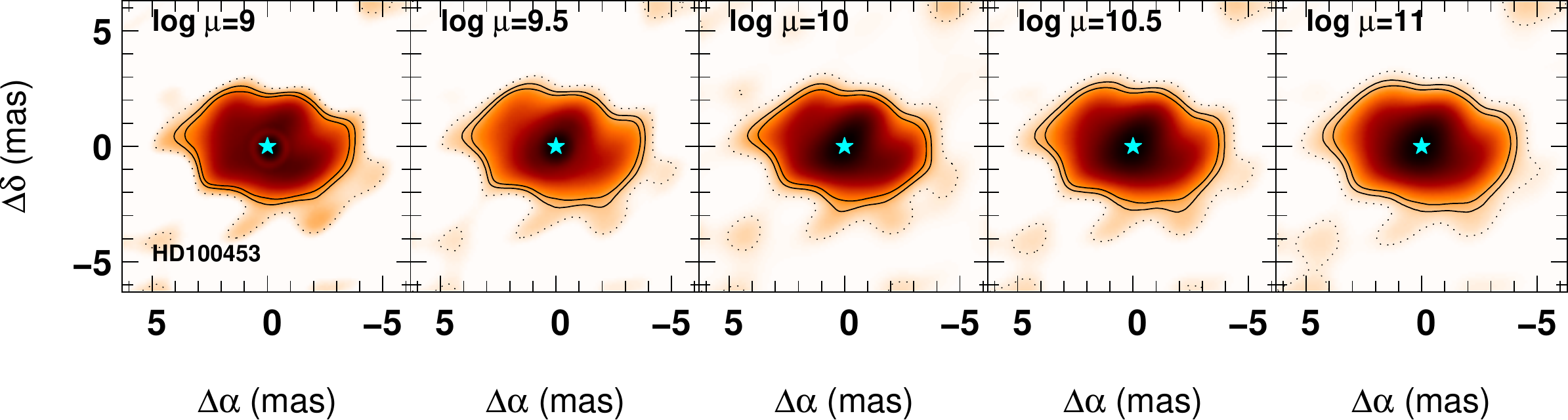} 
      \caption{Illustration of the effect of a slight variation of the regularization weight ($\mu$) around the chosen value for HD45677 (top) and HD100453 (bottom).}
         \label{fig:Lcurve2}
   \end{figure*}

\modif{We also performed image reconstructions with slightly different parameters (see Fig.\,\ref{fig:Lcurve2}).
There are some slight changes in the morphology of the images as the less regularized images appear less smooth that the most regularized once, which is as expected.
The sizes of the image do not change for the most resolved targets (HD45677 and HD100453).
This is the degeneracy between the size, the stellar-to-total flux ratio, and the spectral index that was also seen for geometrical models (L17).
The details we see for HD45677 (dip of flux in the northern part of the ring and sharp turns at the extrema of the circumstellar emission) stay in the image even if they appear dimmer in the most regularized case as the image becomes over-regularized.
}

\section{Image pixel bootstrap } 
\label{app:boot}

We estimated the significance of each pixel in the image using a bootstrap method:
500 datasets were generated from the original one by drawing the baselines and the triangles one by one.
Each baseline (or triangle for the closure phases) can be redrawn several times.
We ended this process when we had drawn as many baselines and triangle as are in the original data.
Image reconstruction was then performed for all of these datasets.

The average flux value of the pixel was divided by its \modif{standard deviation} in order to give an estimation of its significance. 
The average images are less sensitive to the \uv coverage and the rest of this work is based on them.
The difference between the $\chi^2_\mathrm{r}$ of the averaged images from the bootstrap and the reconstructed images directly from the dataset is negligible.


The flux of the extended structures, even if it is not significant pixel by pixel, is required to correctly fit the data because it contributes to the short baselines. The shapes of these extended structures vary depending on the \uv coverage and it is therefore not reliable.




  \section{Determining the inclination and position angles from the reconstructed images}
\label{app:incPA}
  
\begin{figure*}
  \centering
  \includegraphics[width=6cm]{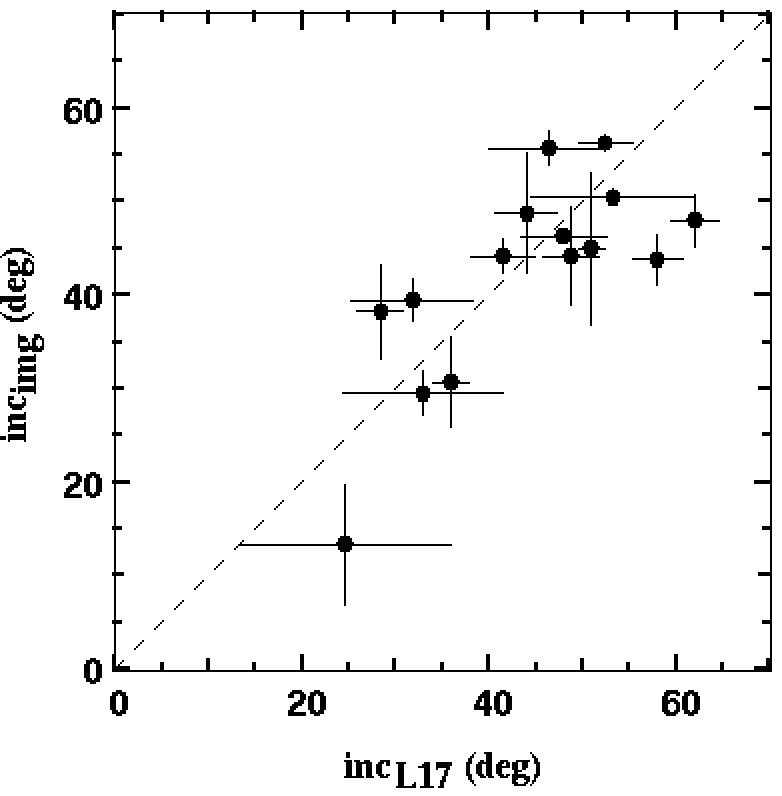} 
  \includegraphics[width=6cm]{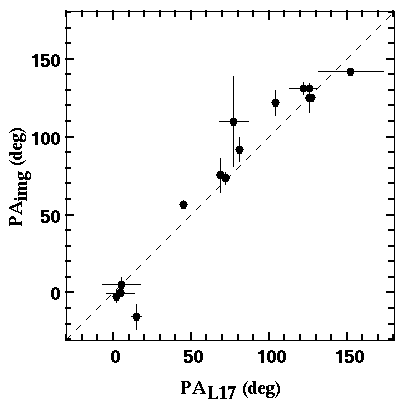}
  \caption{Comparison between inclinations (left) and position angles (right) obtained from the image and parameteric fitting in L17. The dashed line represents the one-to-one line.}
   \label{fig:incPA}
\end{figure*}

To determine the inclinations and position angle directly from the images, we fit a two-dimensionnal Gaussian to the amplitude of the Fourier transform of the image.
This method is less model-dependent than fitting a model to the dataset itself since, depending on the object, different models can be fitted and account for the object morphology (e.g., a Gaussian, a ring, ring modulations, shift between the star, and the environment).
As the central part of the image is similar if it is a ring or a Gaussian, the obtained orientations (inclination and position angle) are not dependent on the exact type of morphology.
Whereas the inclination of the Gaussian is directly the \modifff{inclination} of the image, the position angle has to be shifted by 90$^\circ$.
We compare the values of the inclination and position angle \modif{with} the those from ring models with first order modulation from L17.
The inclinations obtained from the image are similar to those obtained by parametric fitting in L17 (see Fig.\,\ref{fig:incPA}).


\section{Generic tools to analyze the images: Asymmetry maps, azimuthal and radial profiles, and asymmetry factor}
\label{sec:tools}

To analyze the reconstructed image\modif{s}, we developed several tools that allow us to quantify the morphology displayed by the image. \modifff{These tools are described below.}

\subsection{Radial profile:}
The radial profile is the azimuthally averaged profile of the image, which takes the disk orientation on the plane of sky into account.
In order to do so, we defined concentric ellipses, which were oriented using the inclination and position angle of the disk model.
In the case of the observational data, we fit the inclination and position angles in the Fourier space of the reconstructed images (see Appendix\,\ref{app:incPA}).
More specifically, the radial profiles were obtained by azimuthally averaging the flux of \modif{150} consecutive annuli ranging \modif{from the first pixel to the outer edge of the image}. 

\subsection{Asymmetry map:}
Asymmetry maps consist in subtracting the 180\,degrees-rotated image to the image itself and taking the positive part \citep[for more information see][]{Kluska2016}.
The resulting image shows all the flux that is not point-symmetric and that contributes to a nonzero closure phase signal.


\begin{figure}[t]
\centering
\includegraphics[width=6cm]{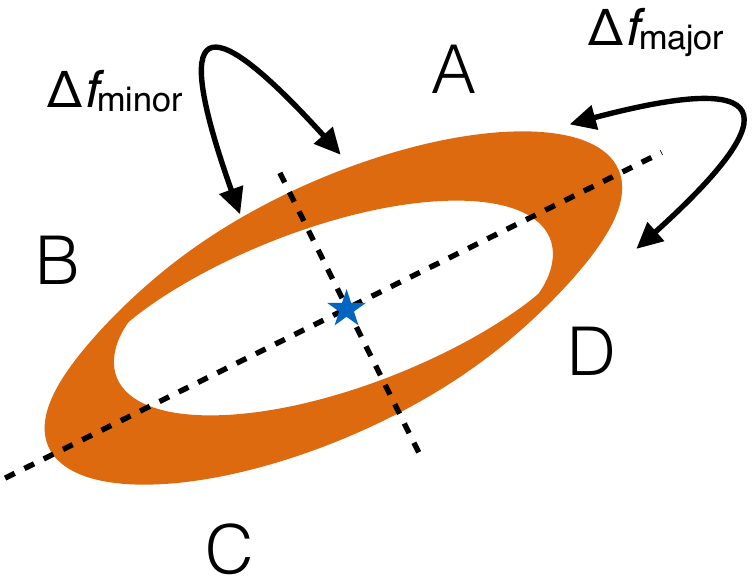}
\caption{Sketch explaining the asymmetry factor computation. We note that $\Delta f_\mathrm{major}$ (and $\Delta f_\mathrm{minor}$, respectively) is the difference of flux between the two sides of the major-axis (the minor-axis respectively). }
\label{fig:fskew}
\end{figure}

\subsection{Asymmetry factors ($f_\mathrm{asym}$ and $\phi_\mathrm{asym}$):}
\modif{To quantify the asymmetry in both amplitude and orientation, we used the asymmetry factors ($f_\mathrm{asym}$ and $\phi_\mathrm{asym}$) that we define as follows \citep[as inspired by][]{Colavita1999}:}
\begin{eqnarray}
f_\mathrm{asym} &=& \frac{\sqrt{(A-C)^2 + (B-D)^2}}{A+B+C+D}\\
\phi_\mathrm{asym} &=& \arctan\Big( \frac{B-D}{A-C} \Big) - \frac{\pi}{4} \\
\end{eqnarray}
\modif{where A, B, C, and D are the different quadrants as indicated in Fig.\,\ref{fig:fskew}.
We note that $\phi_\mathrm{asym}$ is defined so that $\phi_\mathrm{asym}$=0$^\circ$ \modiff{or $\pm$180$^\circ$} indicates asymmetry along the minor axis and $\phi_\mathrm{asym}$=90$^\circ$ \modiff{or -90$^\circ$} indicates an asymmetry along the major axis \modiff{in the anticlockwise and clockwise direction, respectively}.}
\modif{If $f_\mathrm{asym}$ is close to unity, the amplitude of the asymmetry has a maximum of double the average brightness and a minimum of zero.  
On the other hand, if $f_\mathrm{asym}$ is low, the asymmetry amplitude is not strong or is more complex to be captured by these factors. 
}

\section{\modif{Test on synthetic data from a radiative transfer model of a disk}}
\label{sec:lessonsRT}

\begin{figure}
\centering
\includegraphics[width=4.4cm]{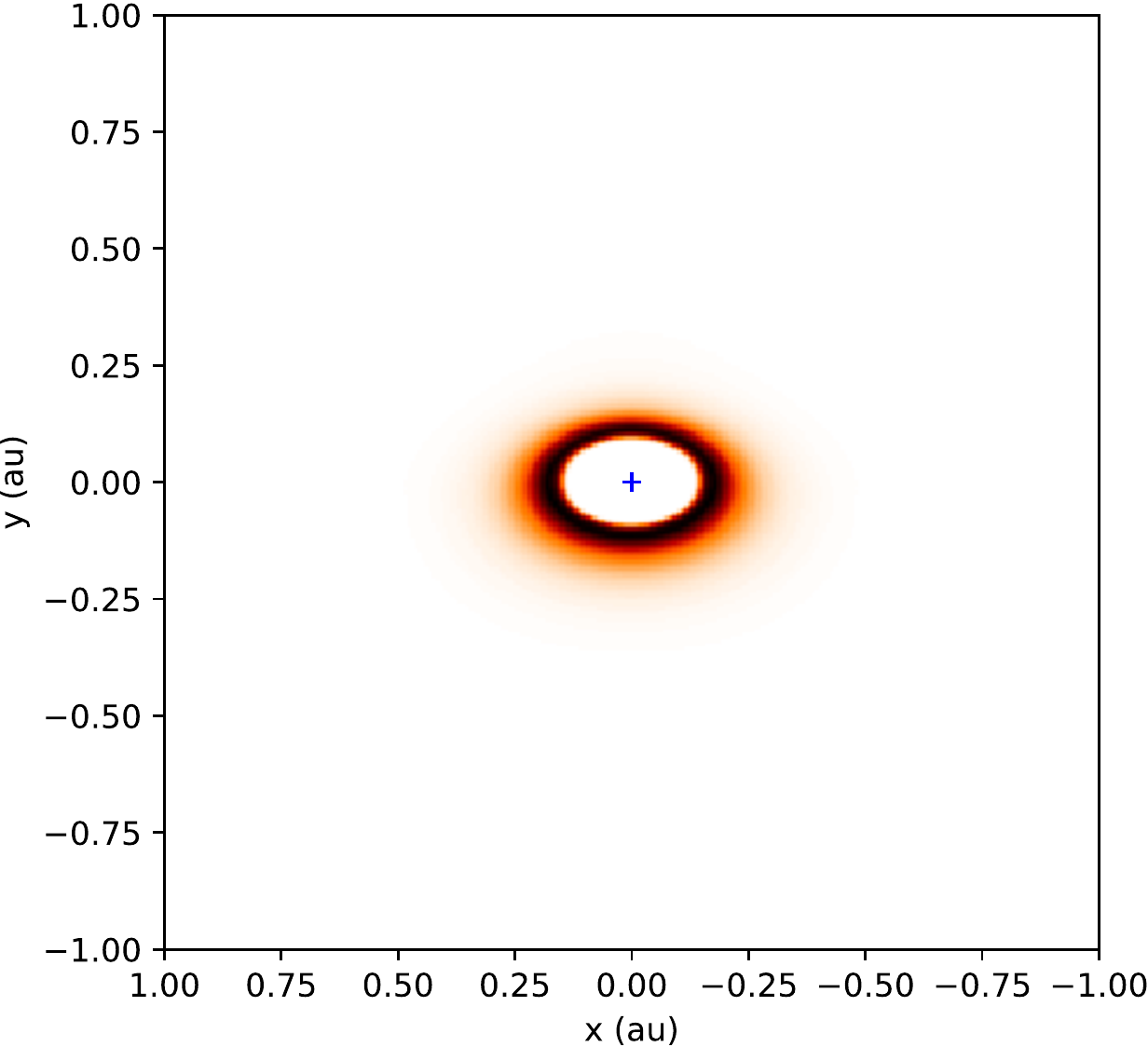} 
\includegraphics[width=4.4cm]{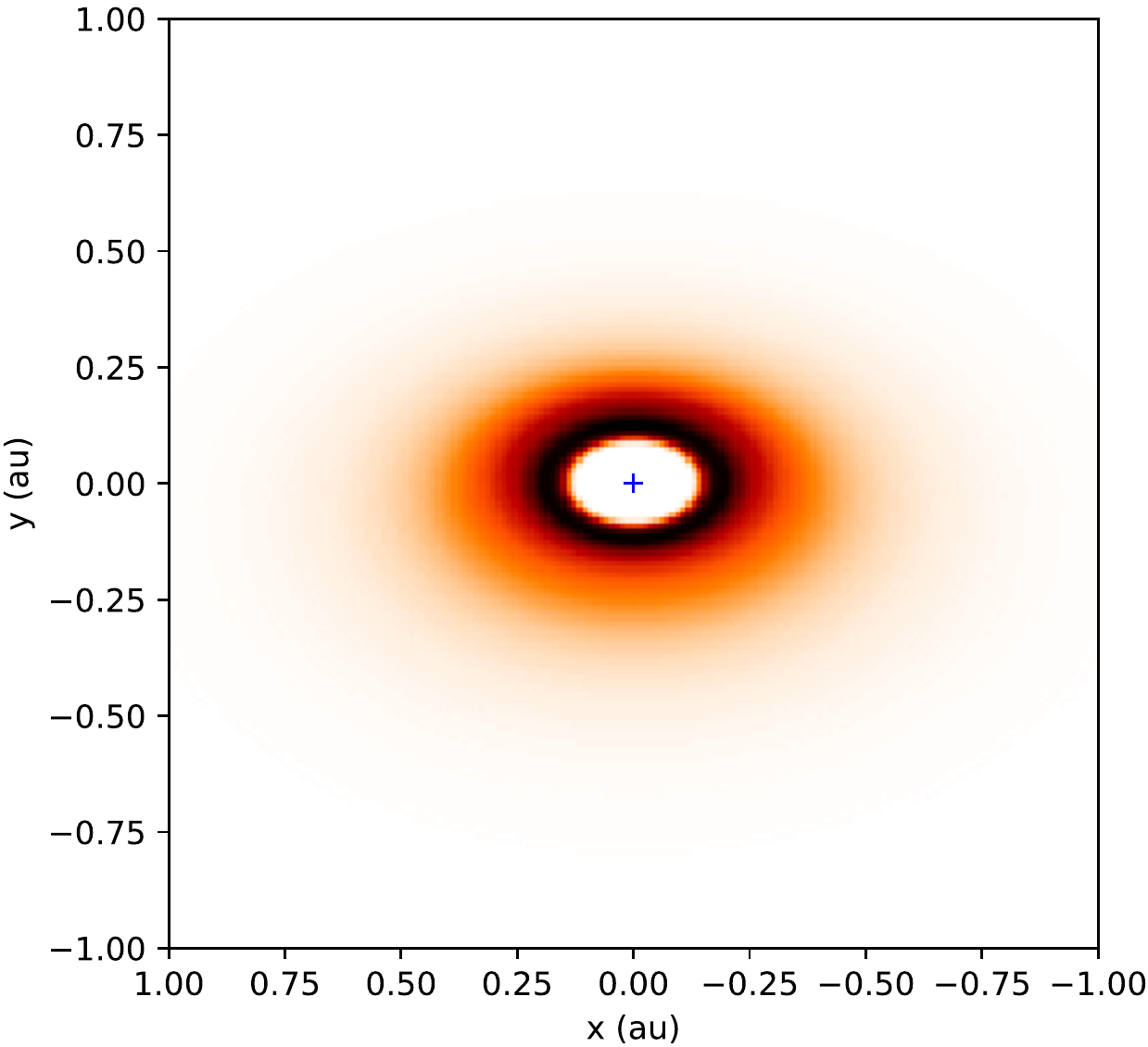} 
\caption{Images of the radiative transfer models at 1.65$\mu$m \modiff{at an inclination of 49\,degrees}. Top: the sharp rim model, bottom: the smooth rim model. The position of the star, which was subtracted from the image, is represented by the blue cross.}
\label{fig:RTmodel}
\end{figure}

\modif{As an example of the tools we have defined, here, we show their application on radiative transfer models.
We used the following two radiative transfer models \citep[computed with the radiative transfer code MCFOST;][]{Pinte2006,Pinte2009}: one with a sharp inner rim \citep[][]{Isella2005} and one with a radially extended inner rim (smooth rim) as defined by \citet[]{Tannirkulam2007}.
Those models are described in L17.
Their goal is to produce models of the disk inner rim with a radially sharp or extended near-infrared emission from the inner edge and not necessary to claim that they are representing the reality.
Both disk models are axisymmetric and the inner disk rim is set by the radiative transfer model with an inner rim at 0.1\,au.
The difference between the two models is the scale height (0.037\,au for sharp rim and 0.05\,au for smooth rim at 1\,au) and the dust size distribution with a single size of 1.2$\mu$m for the sharp rim model and sizes between 0.18 and 2.23$\mu$m for the smooth rim model.
We can see the differences between the two models in Fig.\,\ref{fig:RTmodel}, which shows the images at 1.65$\mu$m at two different inclinations.
The smooth rim models clearly have more extended emission from the inner rim.
}

\begin{figure}
\centering
\includegraphics[width=7.5cm]{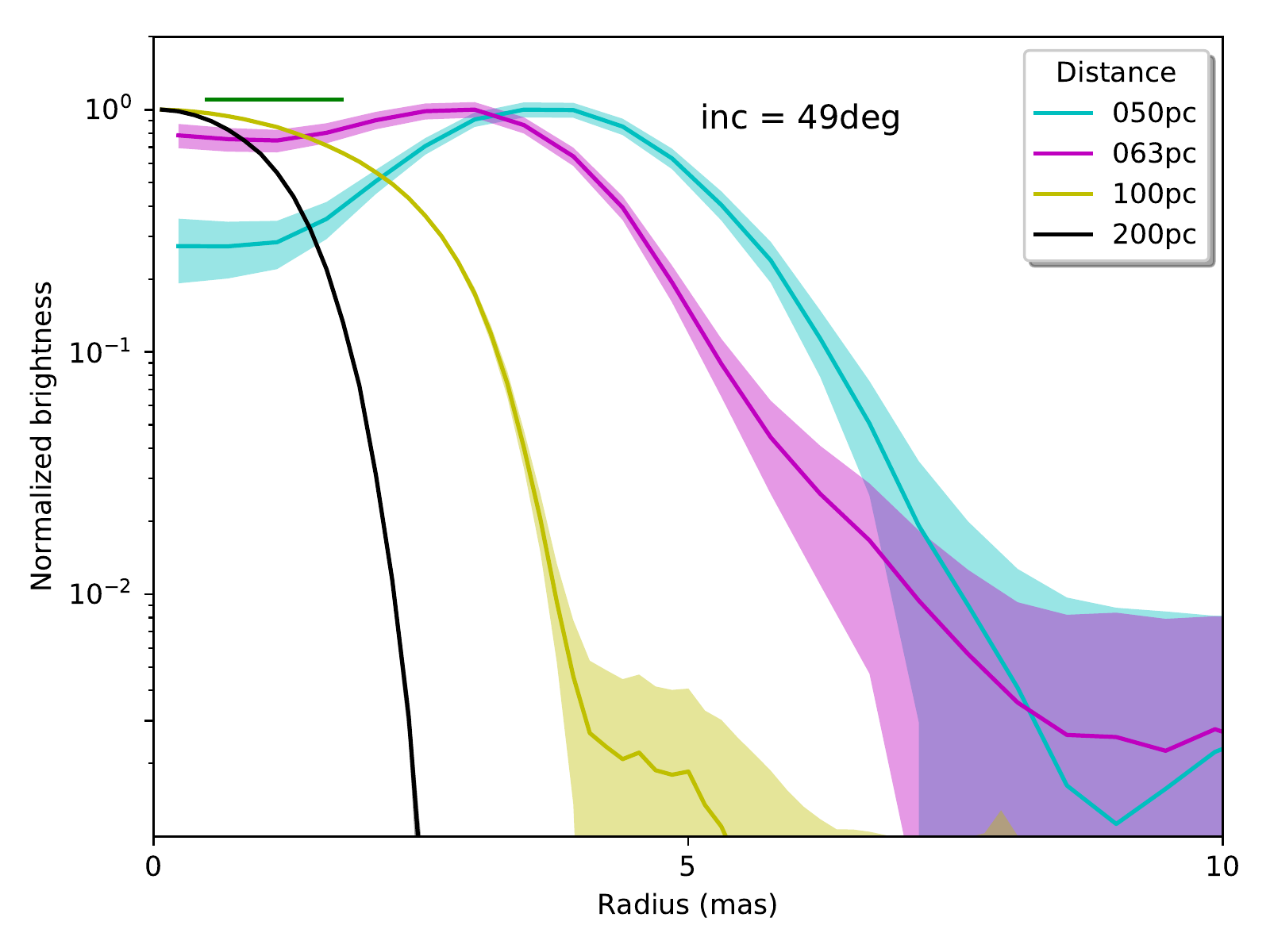} 
\caption{Radial profiles from image reconstructions on the sharp rim models \modiff{for} high inclination (49$^\circ$). The solid lines represent the radial profiles for models at different distances, which are color-coded. The green horizontal line above the profiles represents the beam size. }
\label{fig:RTmodelrad}
\end{figure}

\subsection{Inner cavity detection and radial profile} 
We performed image reconstructions on data generated from the radiative transfer models using the same methodology as for the observed targets.
We used the  (u, v)-coverage obtained on HD144668 and simulated noise on the complex visibility.
We reconstructed images of the model taking different distances and inclinations.
We then investigated the presence of a central cavity, the radial profile, and sizes.
The results of the detection of the central hole are presented in Fig.\,\ref{fig:InnerHole}.
The radial profiles (Fig.\,\ref{fig:RTmodelrad}) were used for this inner hole detection.
When the radial profile is not monotonically decreasing, then the inner hole is considered to be detected.
We can see that the distance, and therefore the angular size of the source, influence the radial profile and the detection of the inner hole.
We also see that for higher inclination, it is more difficult to detect the presence of an inner hole since the minor axis of the projection of the inner rim on the plane of the sky is too small to be resolved, whereas the major axis is resolved.

\subsection{Size determination}
\label{sec:innerhole}

\begin{figure}[t]
   \centering
   \includegraphics[width=9cm]{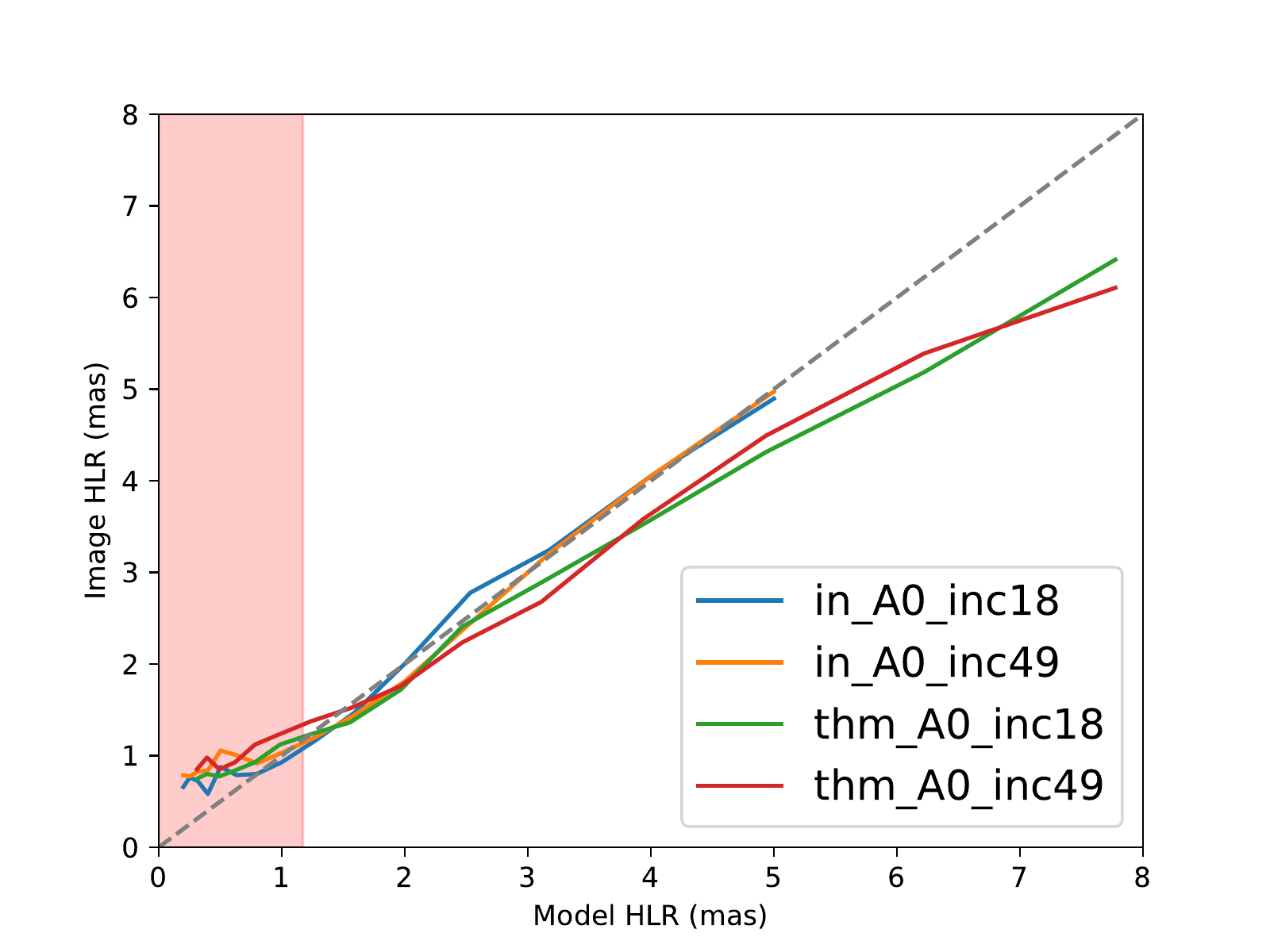}
   \caption{Comparison of the emission size measured in the image with the theoretical size. The red area indicates where the angular resolution given by the (u, v)-coverage is not sufficient in resolving the emission. The dashed line represents the 1:1 relation between the theoretical size and the size determined from the image.}
   \label{fig:sizeModel}
   \end{figure}

\modif{We wanted to test the size determination from the images (described in Sect.\,\ref{app:incPA}) and test if it is reliable to determine the emission size.
We compared the determined sizes from the reconstructed images of the models with different inclinations that are representative of our targets (a low inclination with $inc=18^\circ$ and moderate with $inc=49^\circ$) and distances in Fig.\,\ref{fig:sizeModel}.
We did this with the two models.
We can see that for the sharp rim models, the size determined from the image matches the model size well for both inclinations as long as the size is resolved by the interferometer. 
For the smooth rim model, however, the sizes determined from the images match for relatively small sizes ($\theta_\mathrm{1//2}$<2.5\,mas), but they are then biased towards smaller sizes from $\theta_\mathrm{1//2}$>2.5\,mas.
The investigation of the origin for such a bias is beyond the scope of this paper, but it is useful to keep it in mind. }

\subsection{Asymmetry map}

\begin{figure}
\centering
\includegraphics[width=7.5cm]{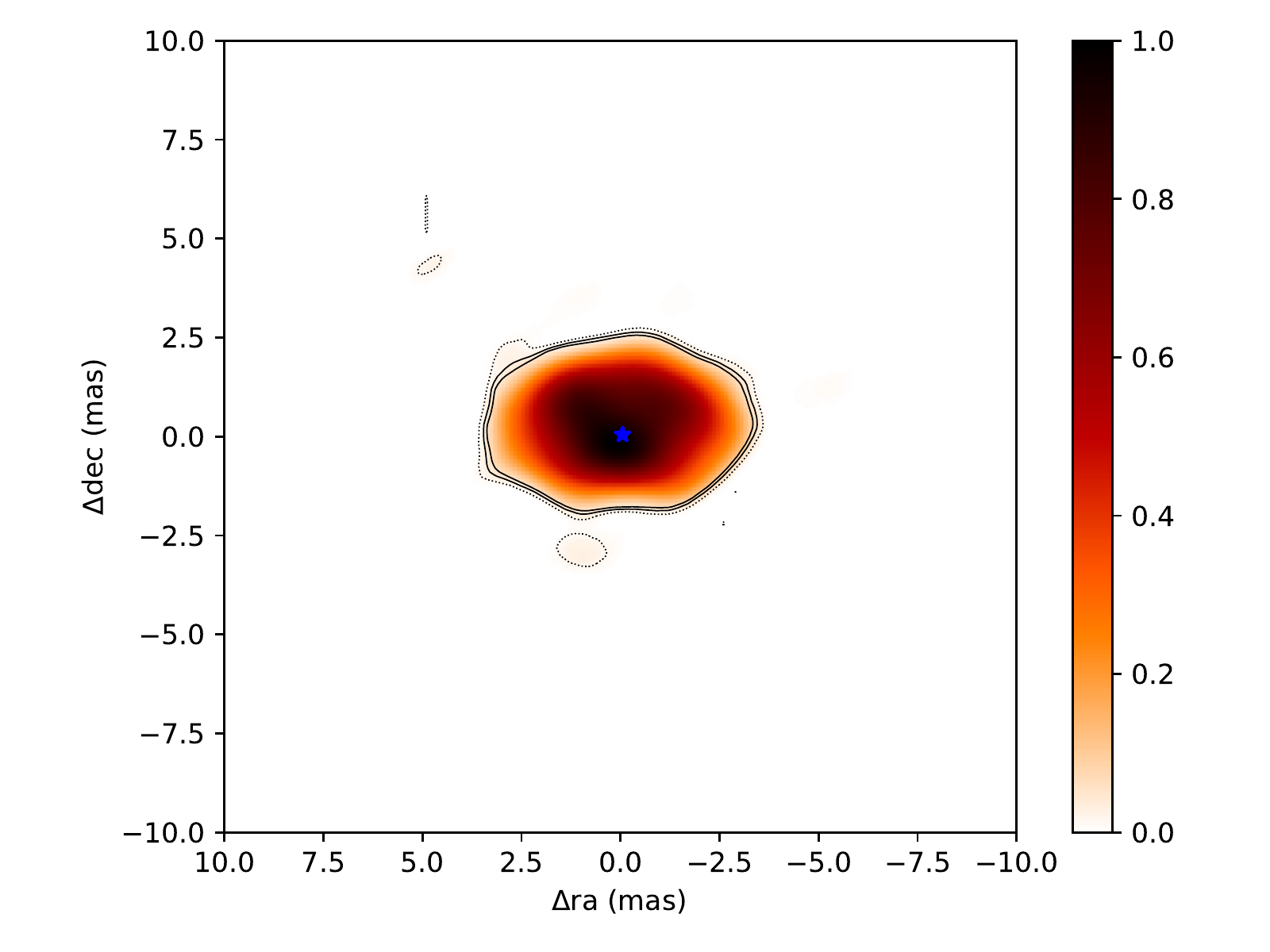} 
\includegraphics[width=7.5cm]{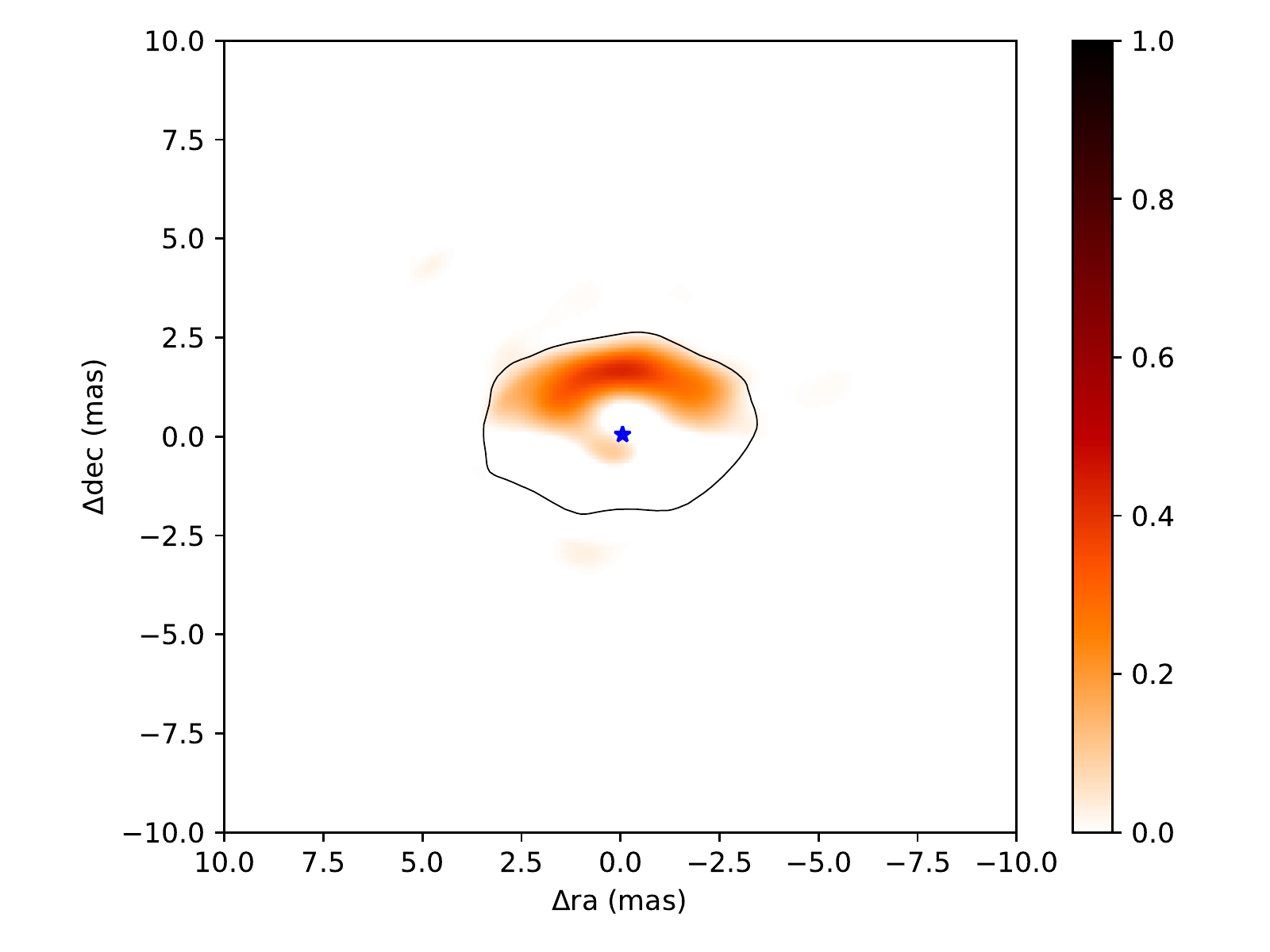} 
\caption{Top: Image reconstruction of a sharp model with an inclination of 49$^\circ$ degrees and located at 100\,pc. The blue star represents the position of the star. The contours are 1 (dotted), 2, and 3-sigma pixel significance levels. Bottom: Corresponding asymmetry map with the contour representing the 2-$\sigma$ significance level.}
\label{fig:asymmap}
\end{figure}

\modif{In Fig.\,\ref{fig:asymmap}, we present an example of an asymmetry map from a reconstructed image (with a $\chi^2_\mathrm{red}$ of 1.2) of a radiative transfer model that has a sharp rim and an inclination for 49$^\circ$ and at a distance of 100\,pc.
The asymmetry factors for this model are $f_\mathrm{asym}$= 0.14$\pm$0.01 and $\phi_\mathrm{asym}$=5$^\circ\pm$1$^\circ$.
}

 \begin{figure*}
   \centering
   \includegraphics[width=4.5cm]{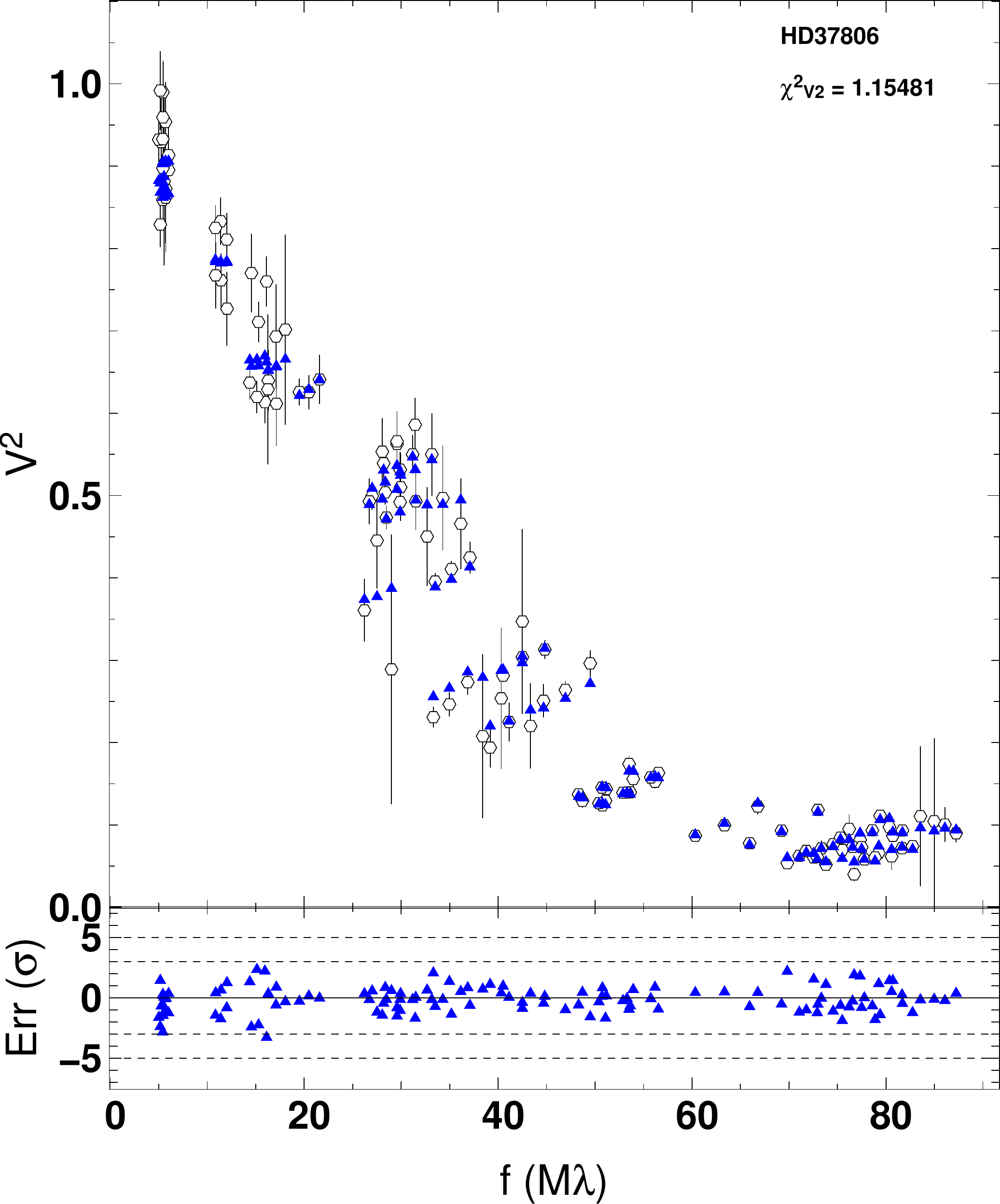} 
\includegraphics[width=4.5cm]{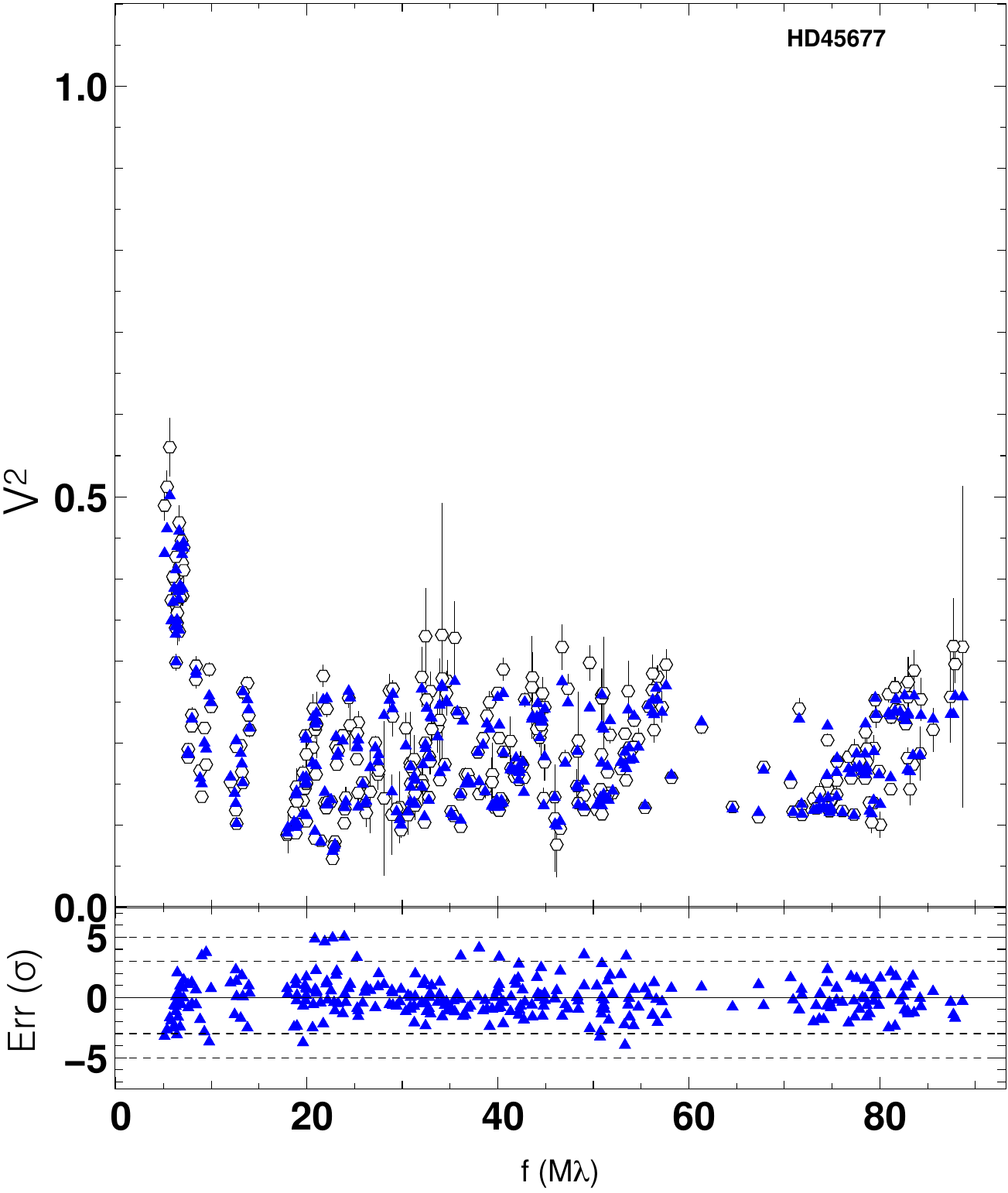} 
\includegraphics[width=4.5cm]{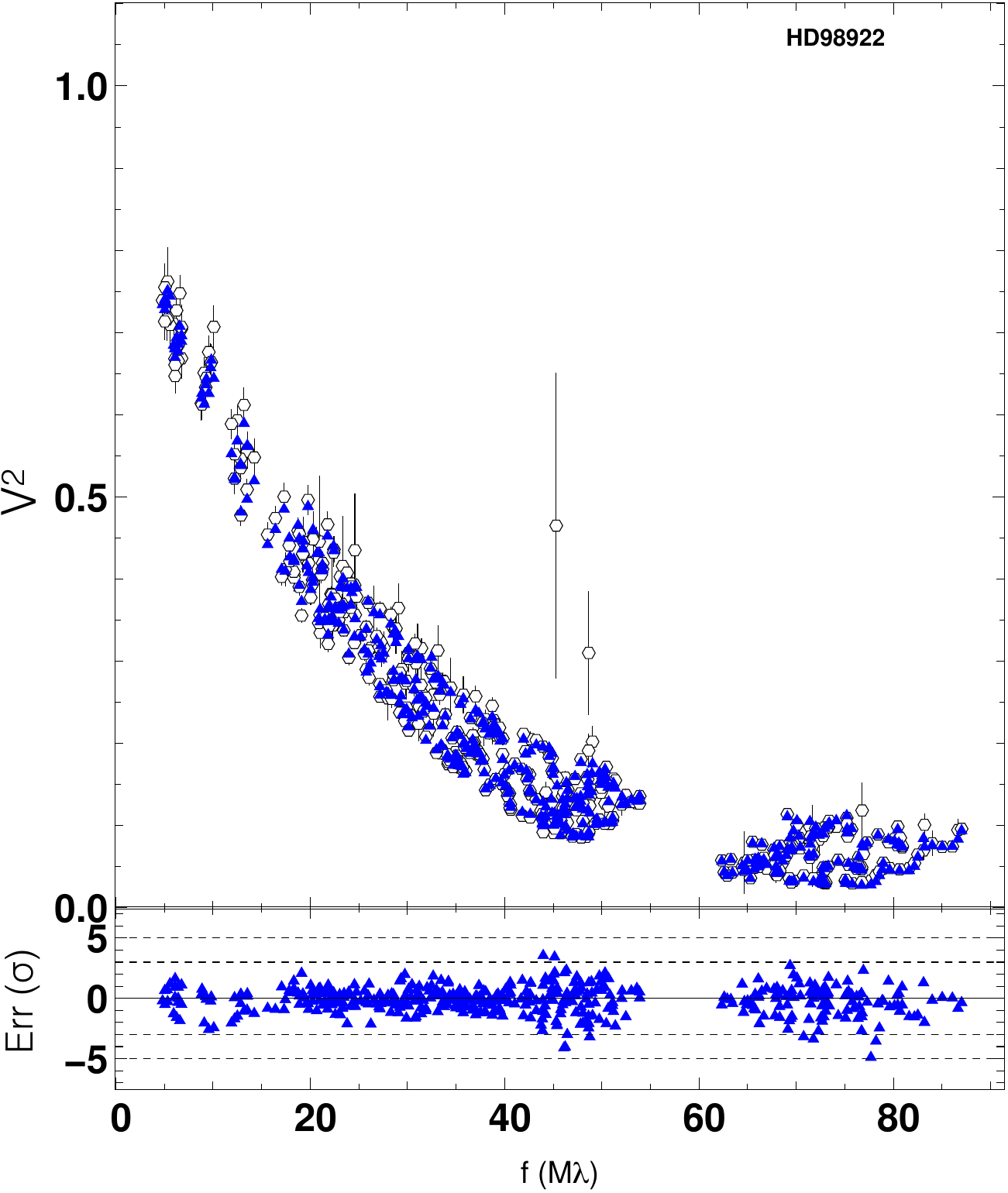} 
\includegraphics[width=4.5cm]{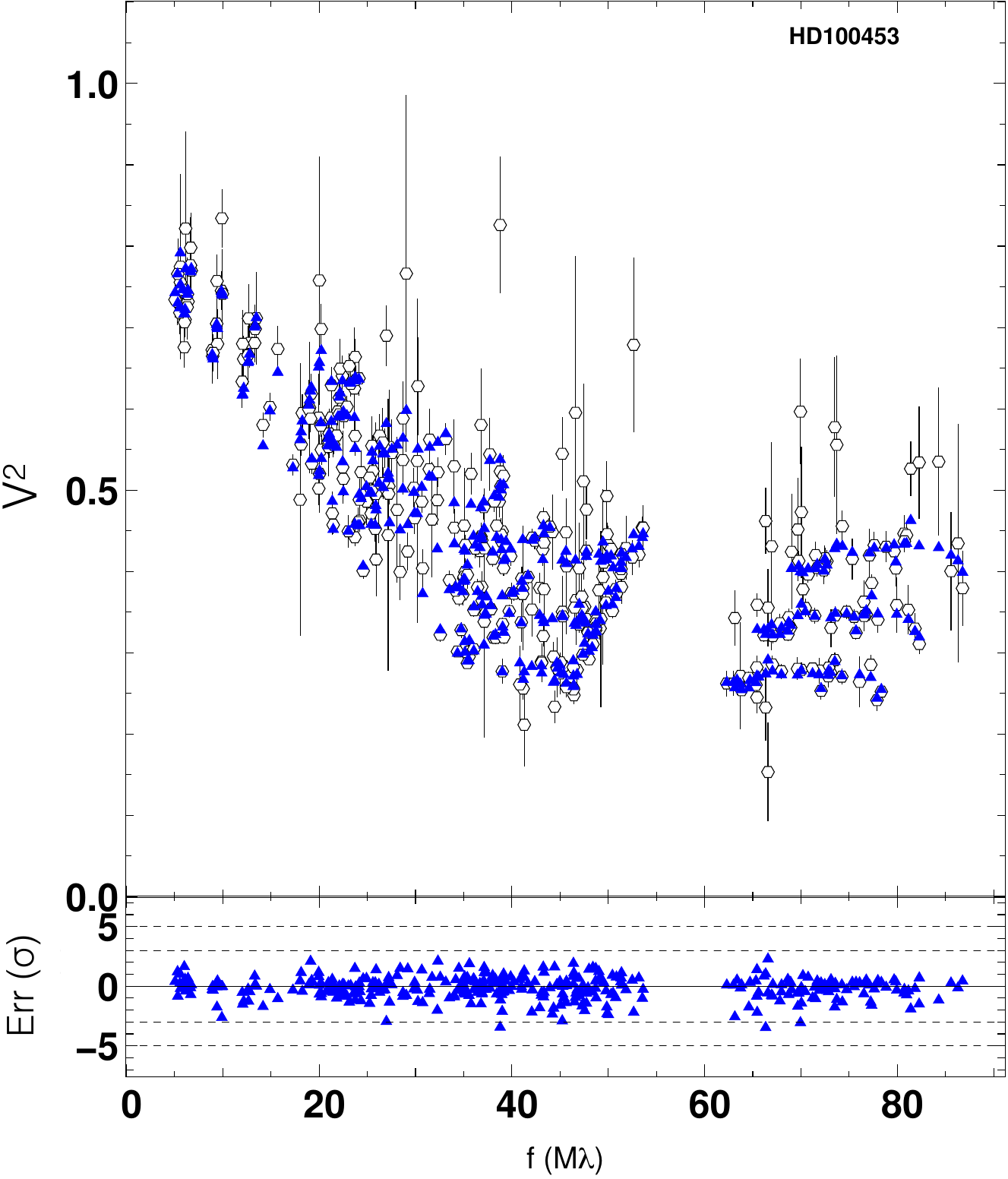} 
\includegraphics[width=4.5cm]{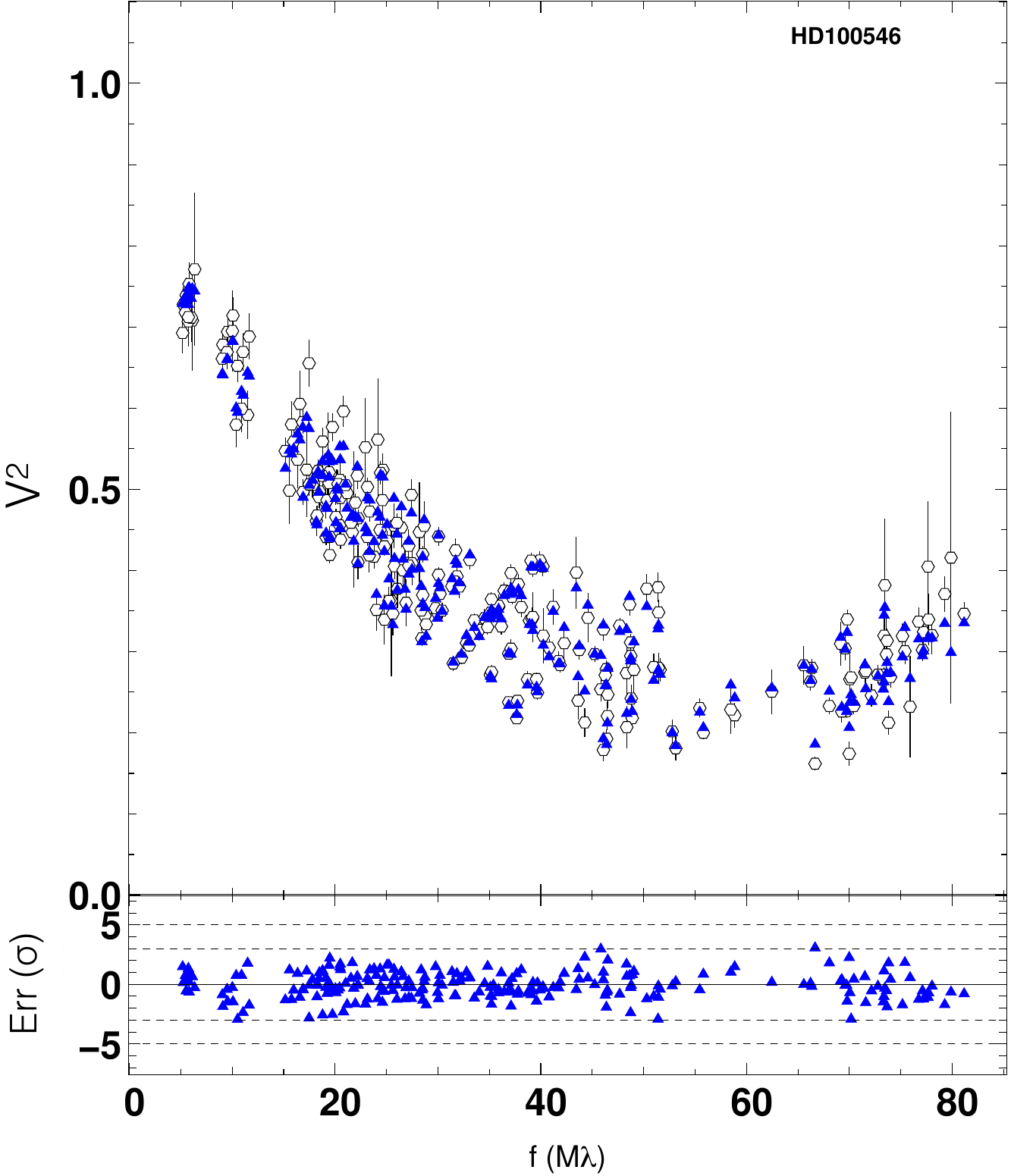} 
\includegraphics[width=4.5cm]{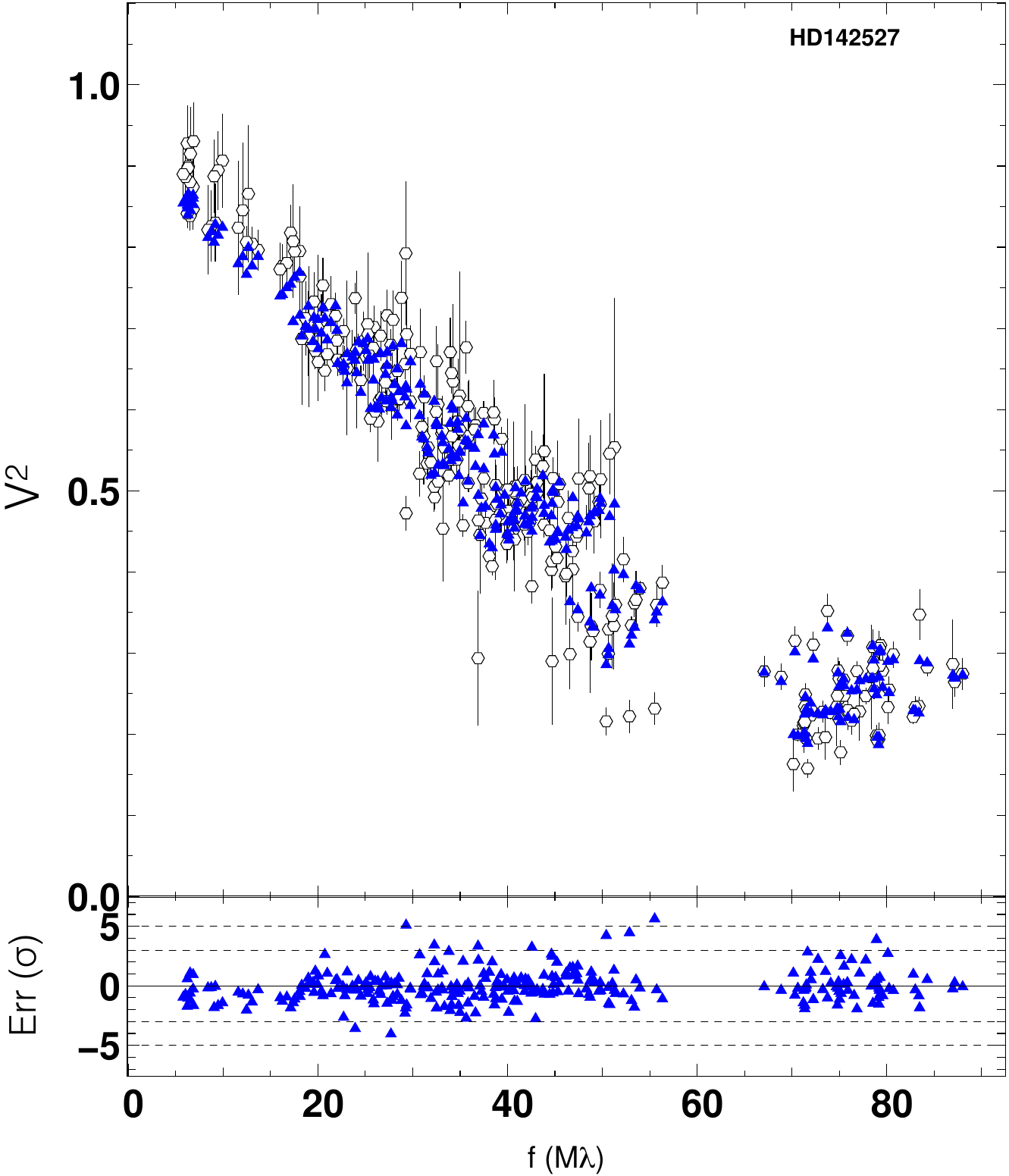} 
\includegraphics[width=4.5cm]{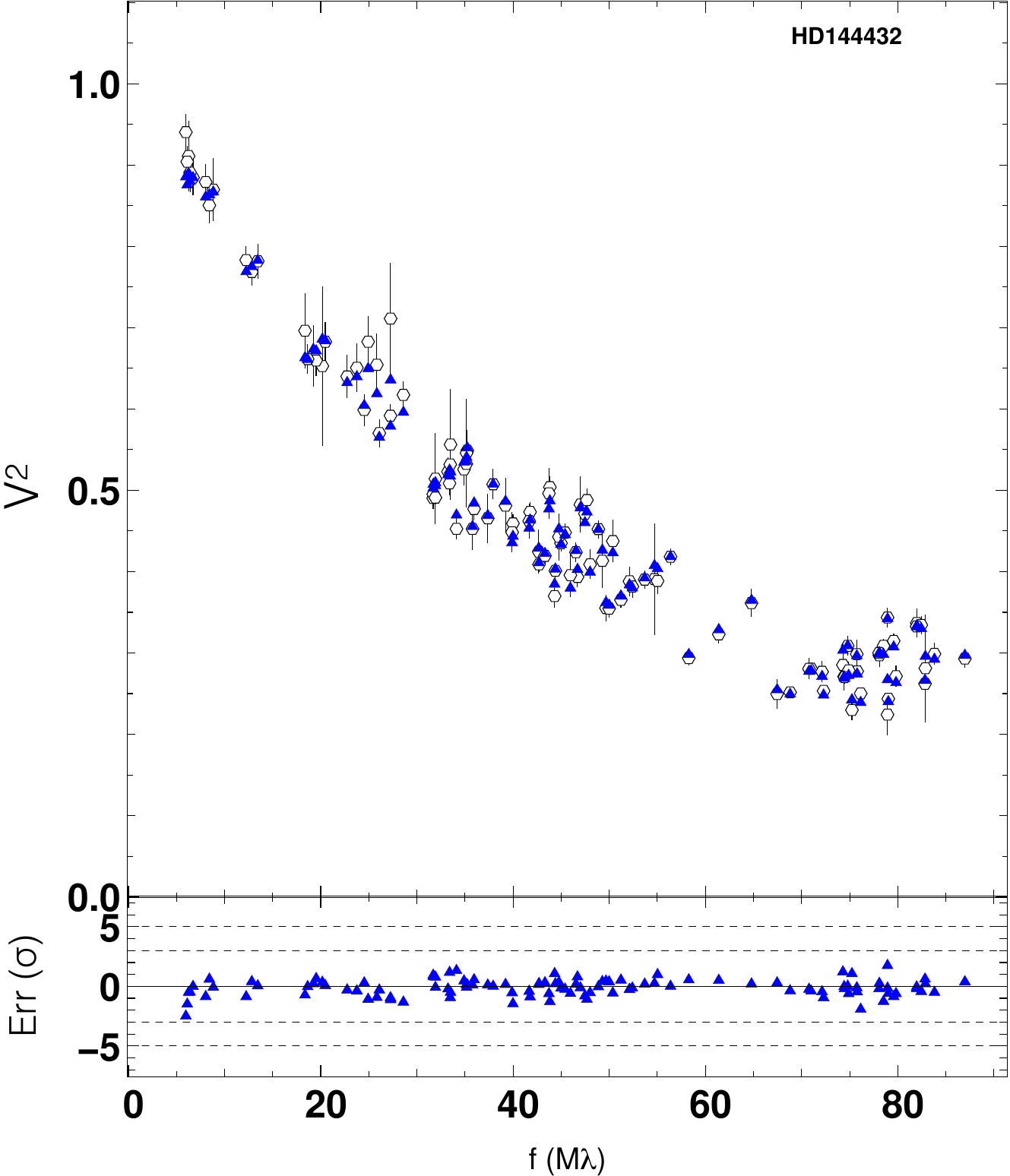} 
\includegraphics[width=4.5cm]{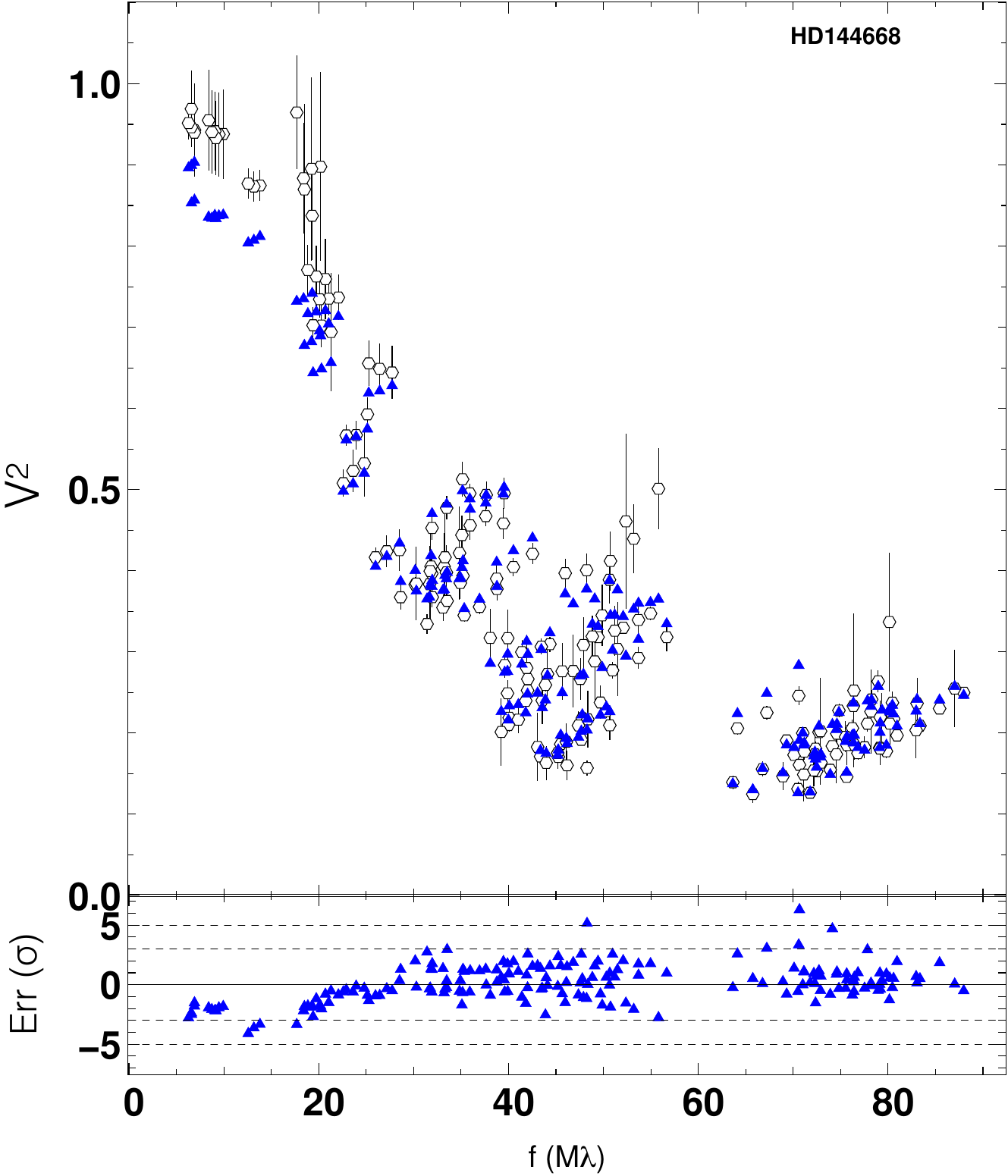} 
\includegraphics[width=4.5cm]{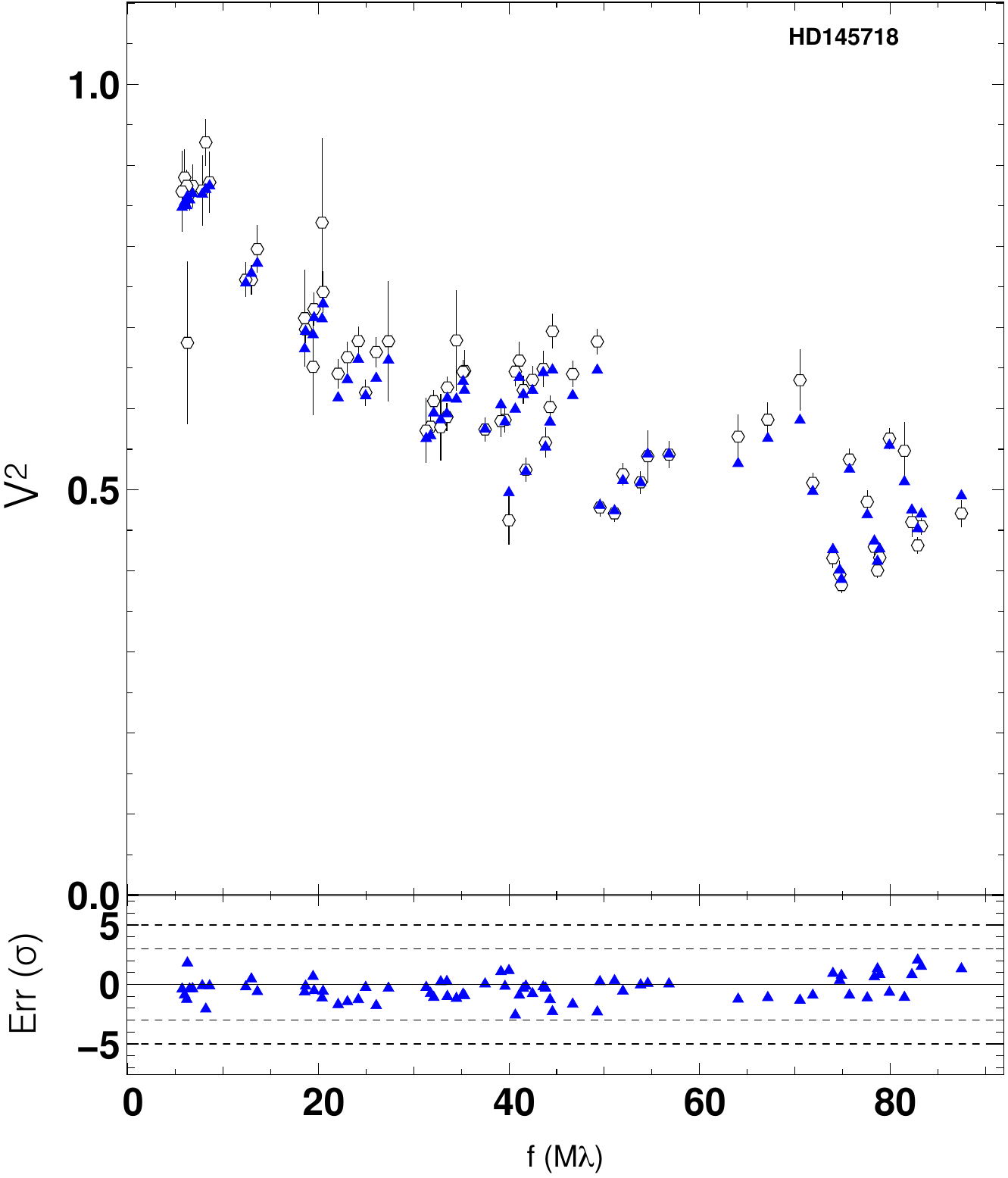} 
\includegraphics[width=4.5cm]{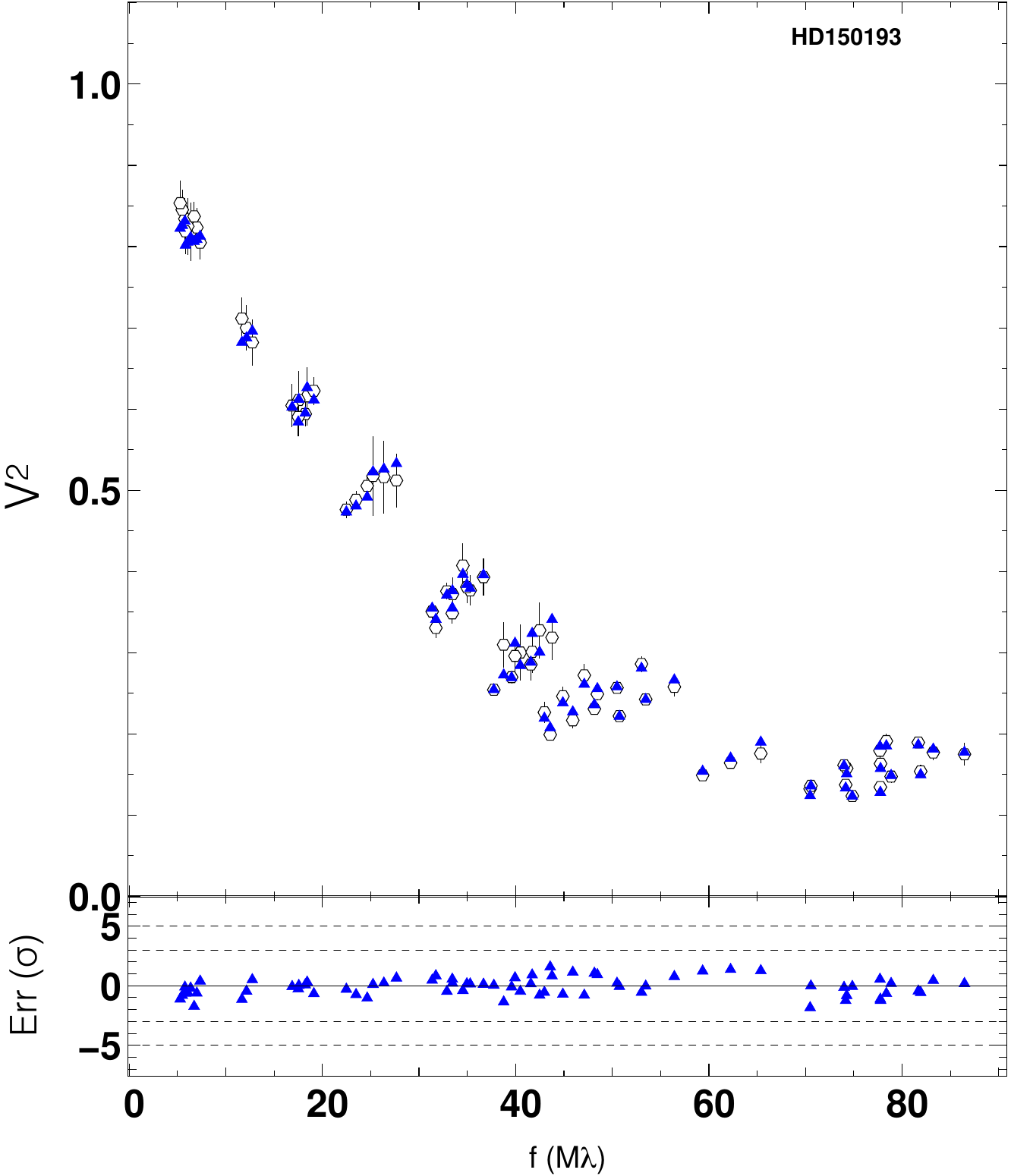} 
\includegraphics[width=4.5cm]{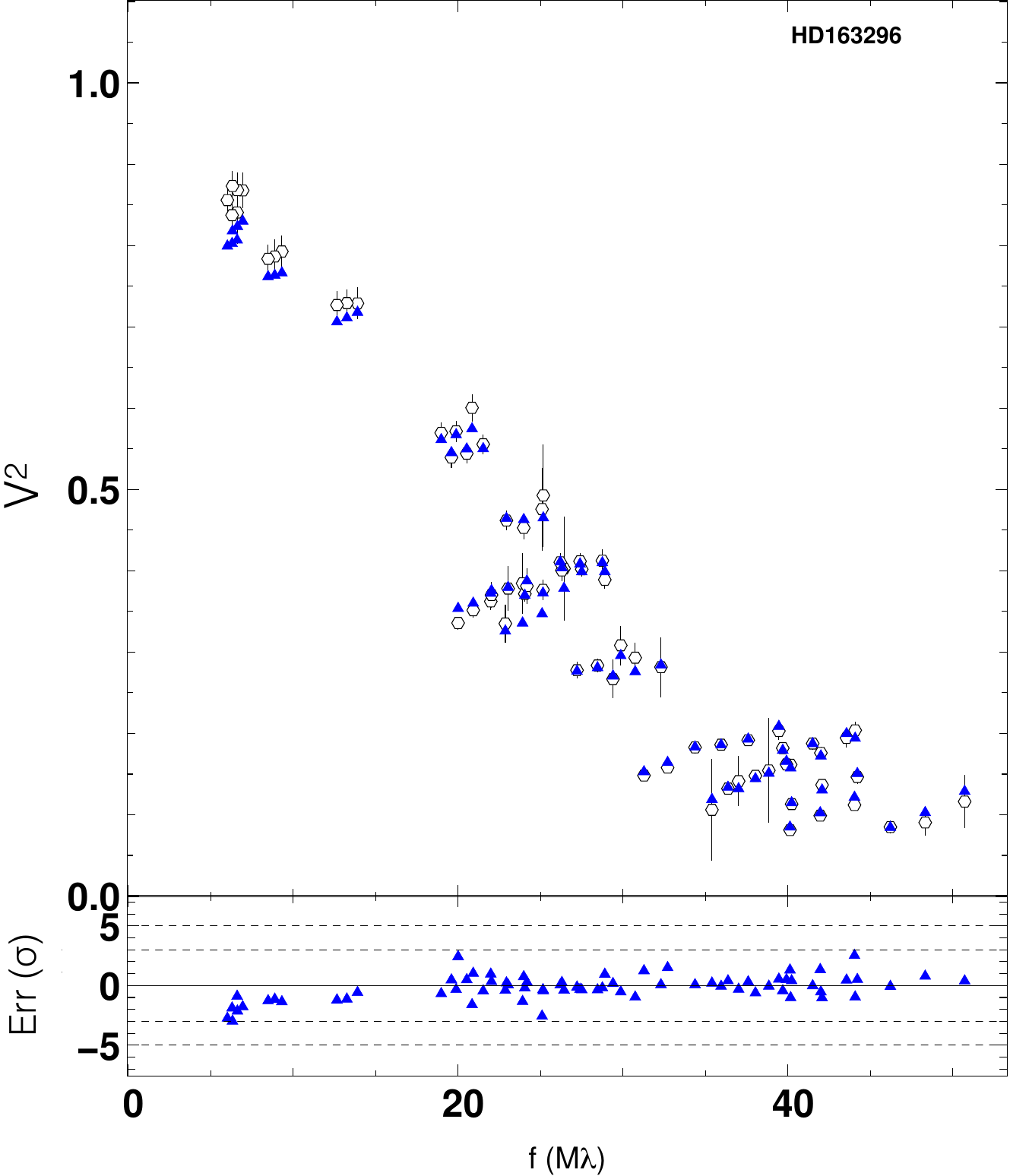} 
\includegraphics[width=4.5cm]{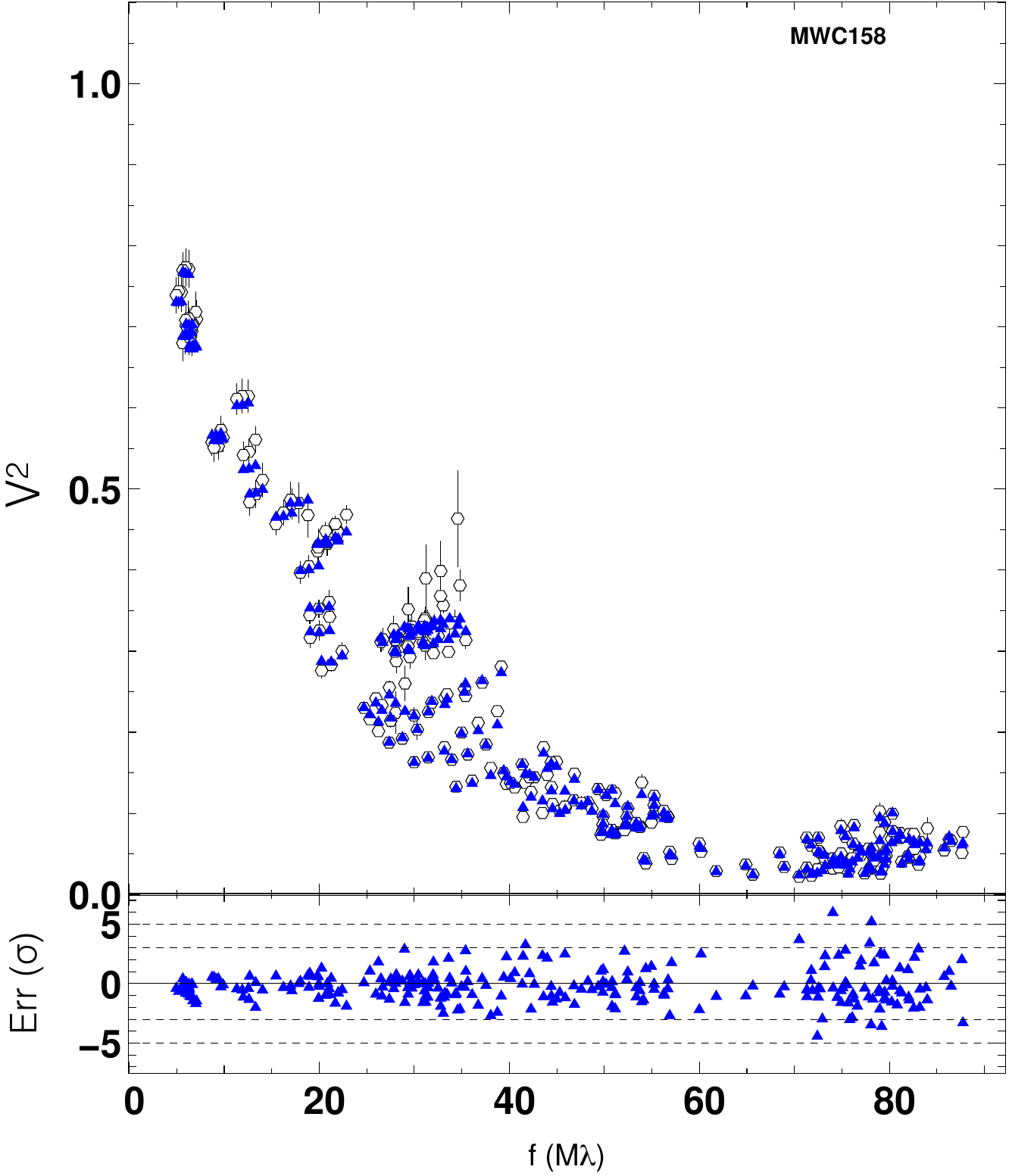} 
      \caption{Squared visibilities of the dataset and the images. For each target the top panel shows the squared visibilities for the data (black circles) and the image (blue triangles). The bottom panel shows the residuals normalized by the error bars on each data point.}
         \label{fig:data}
   \end{figure*}

\begin{figure*}
   \centering
\includegraphics[width=5cm]{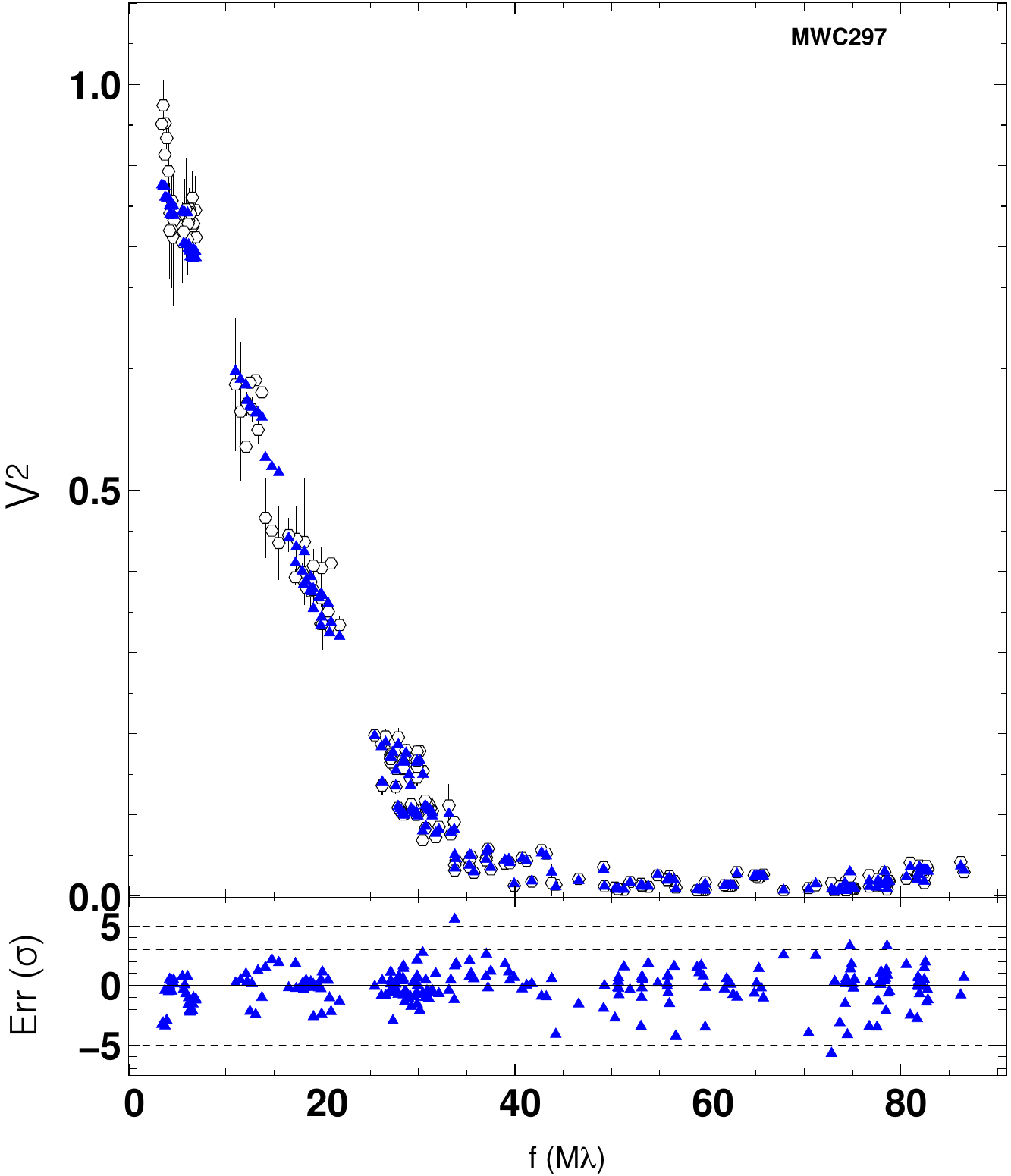} 
\includegraphics[width=5cm]{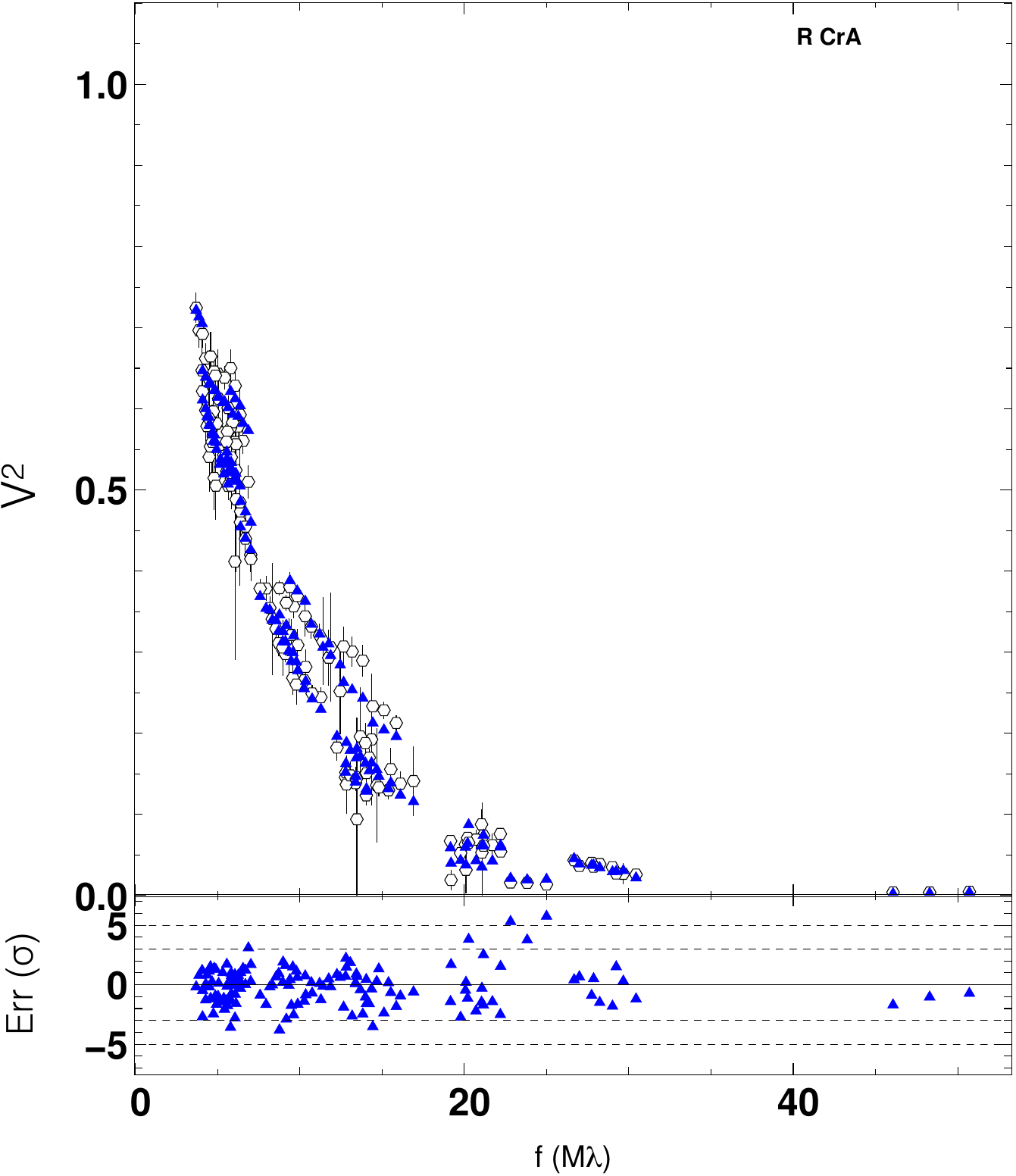} 
\includegraphics[width=5cm]{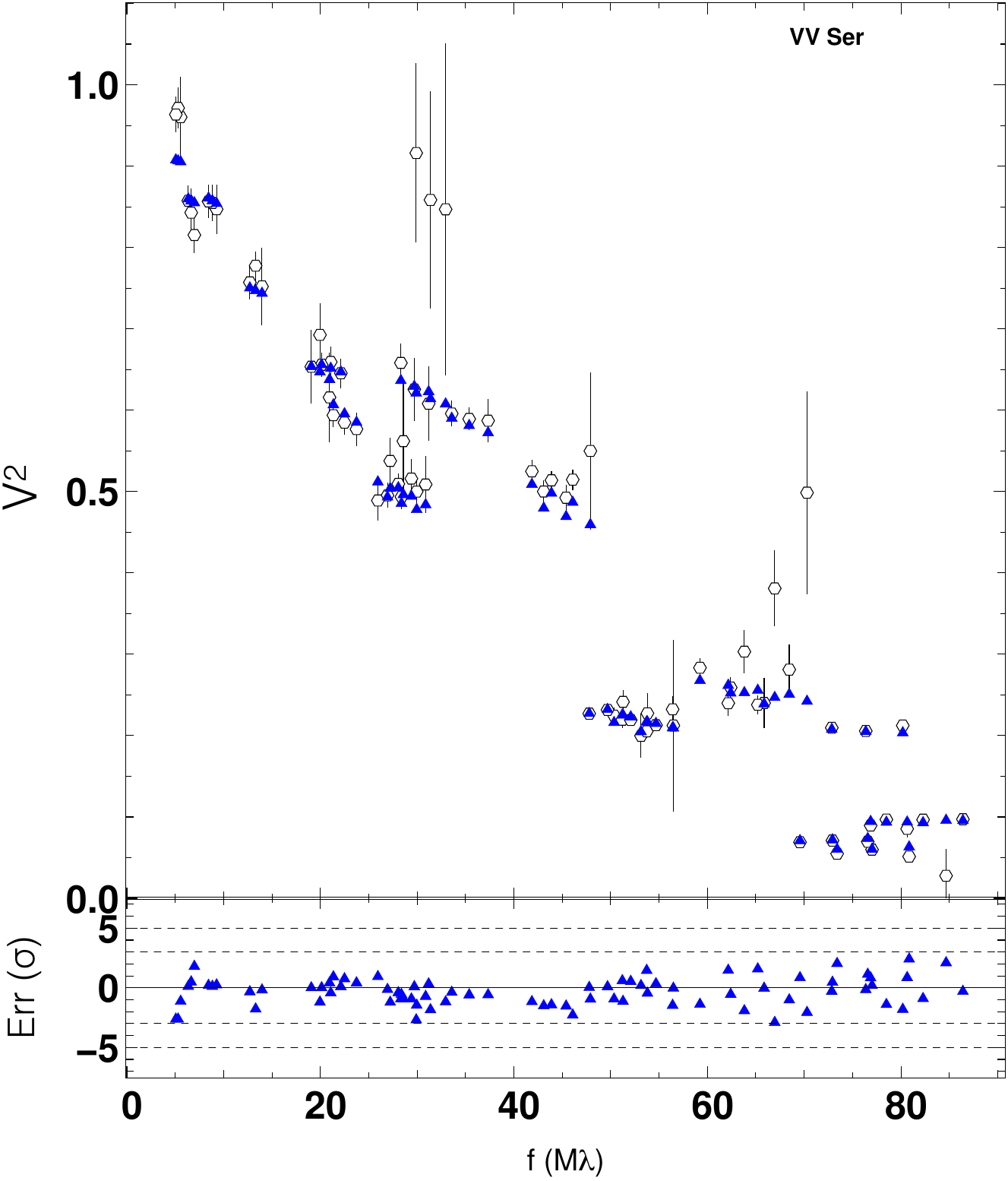}
\caption{Same as Fig.\,\ref{fig:data}}
         \label{fig:data2}
   \end{figure*}
   

    \begin{figure*}
   \centering
\includegraphics[width=4.5cm]{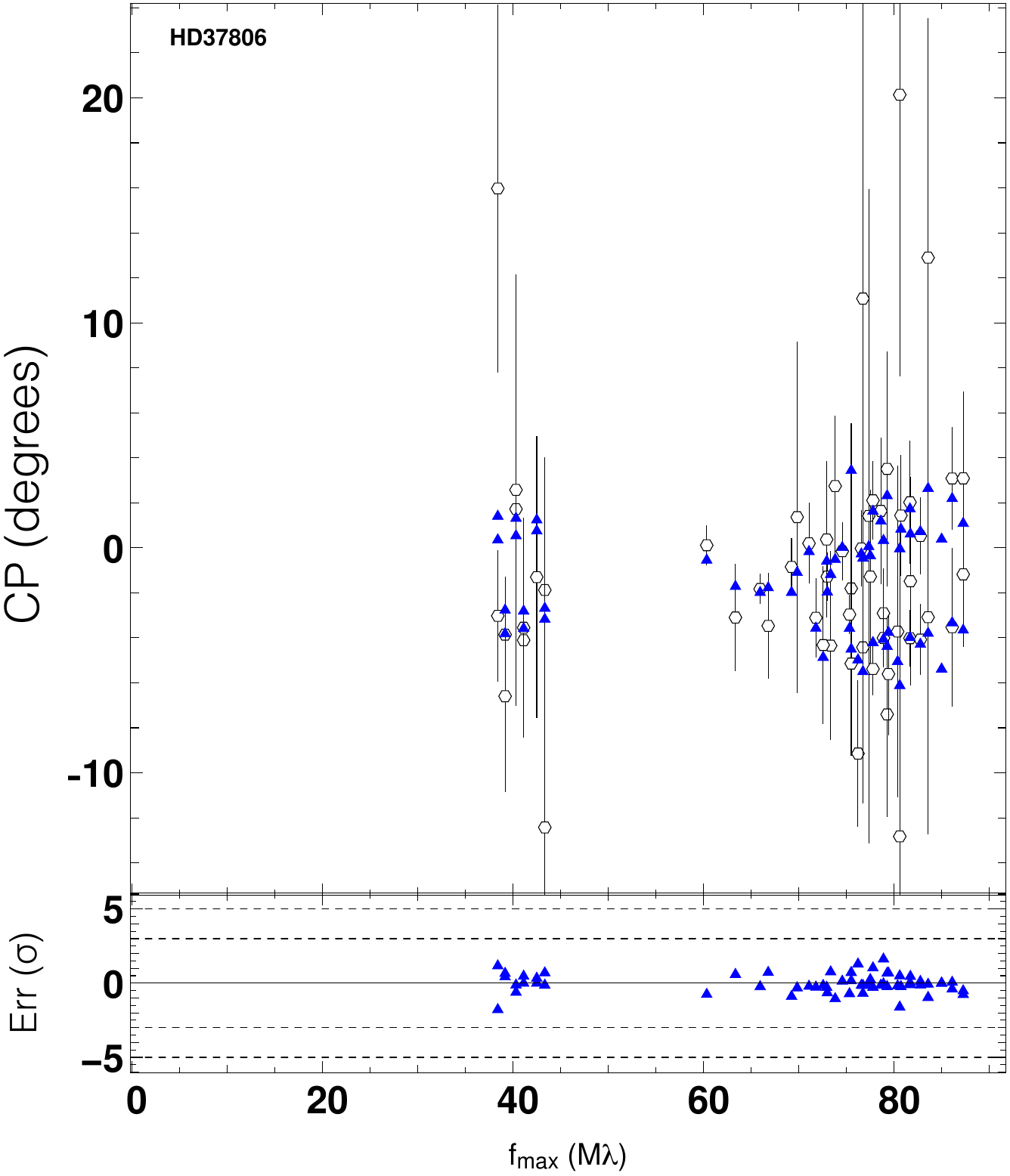} 
\includegraphics[width=4.5cm]{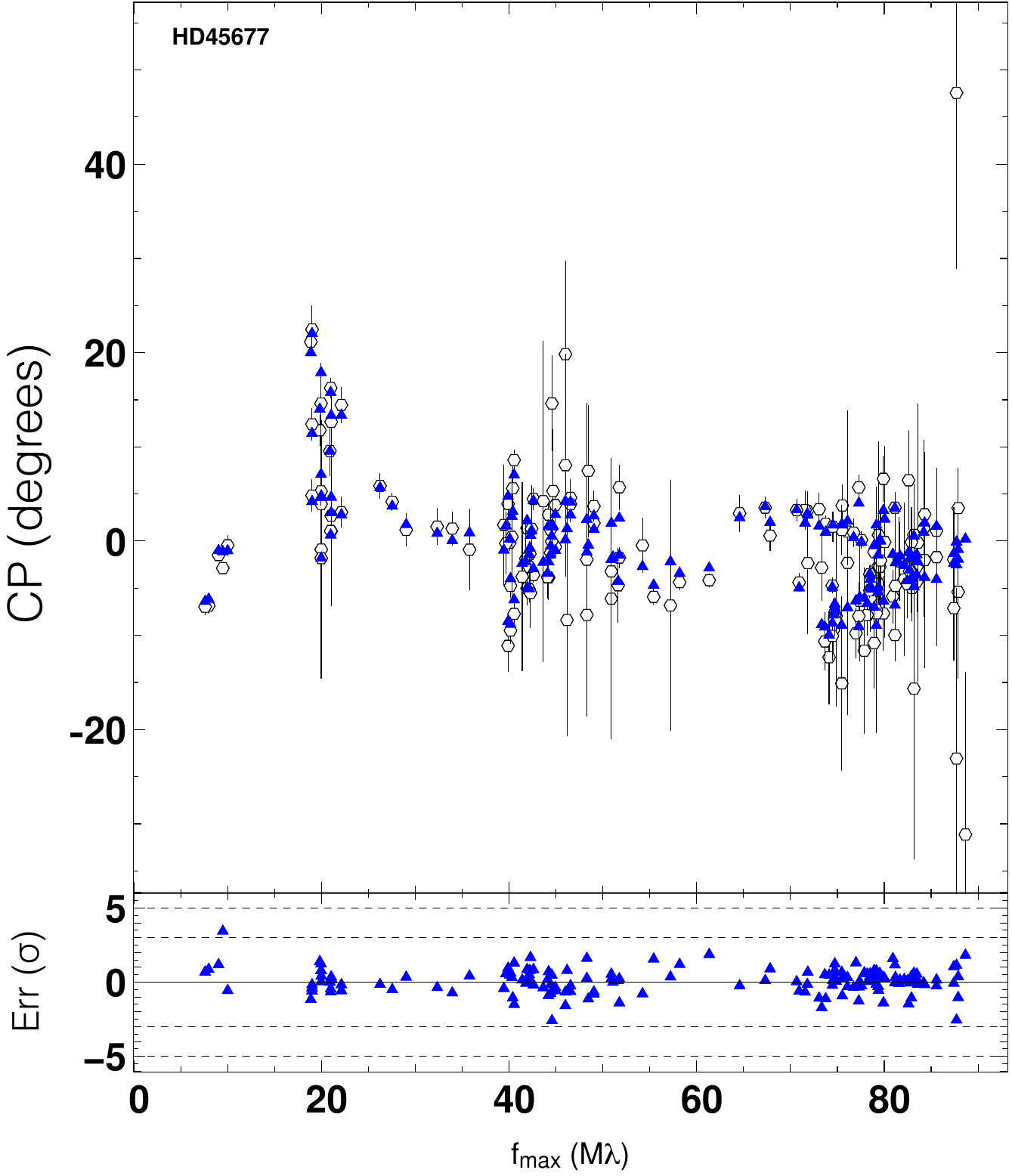} 
\includegraphics[width=4.5cm]{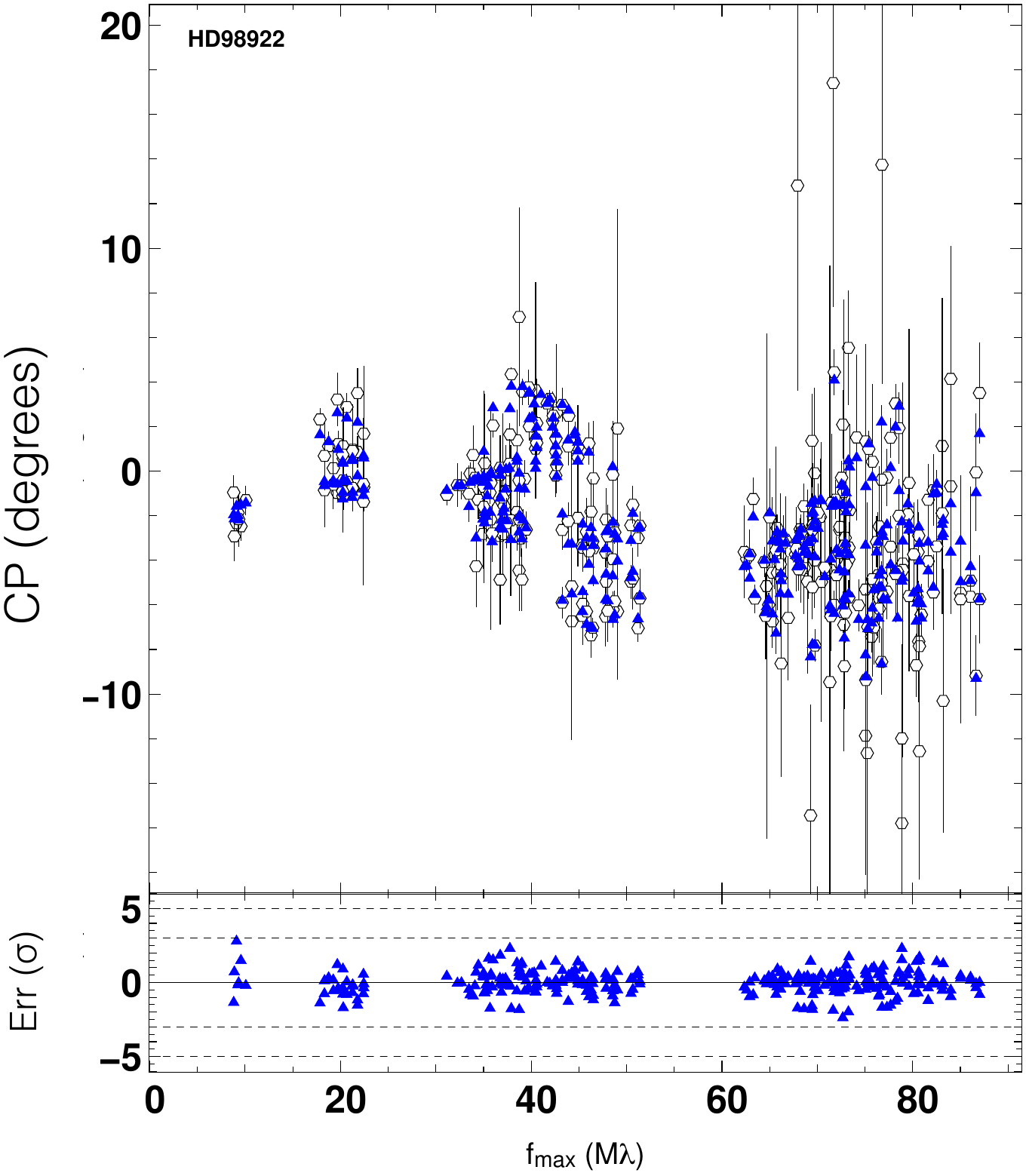} 
\includegraphics[width=4.5cm]{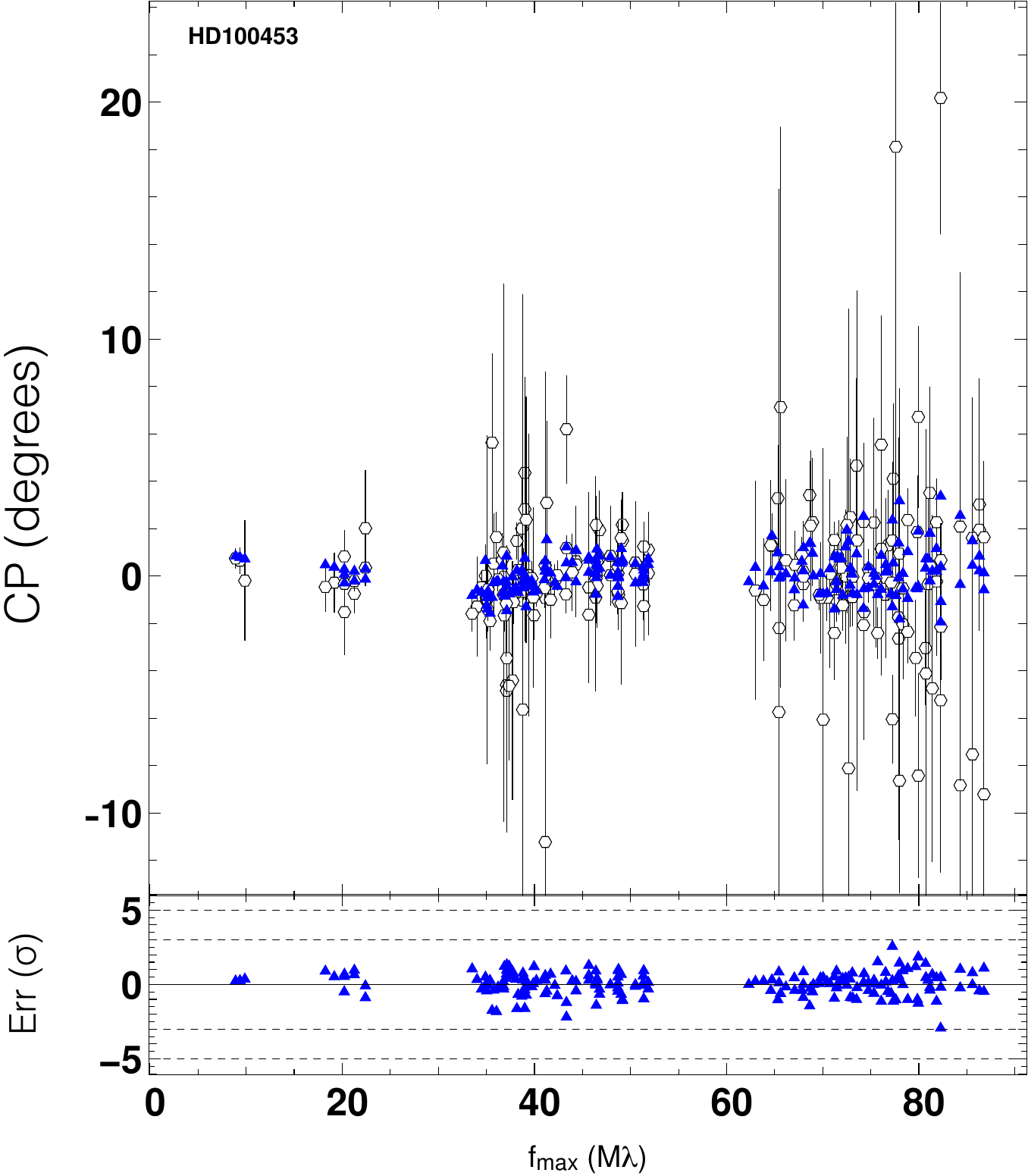} 
\includegraphics[width=4.5cm]{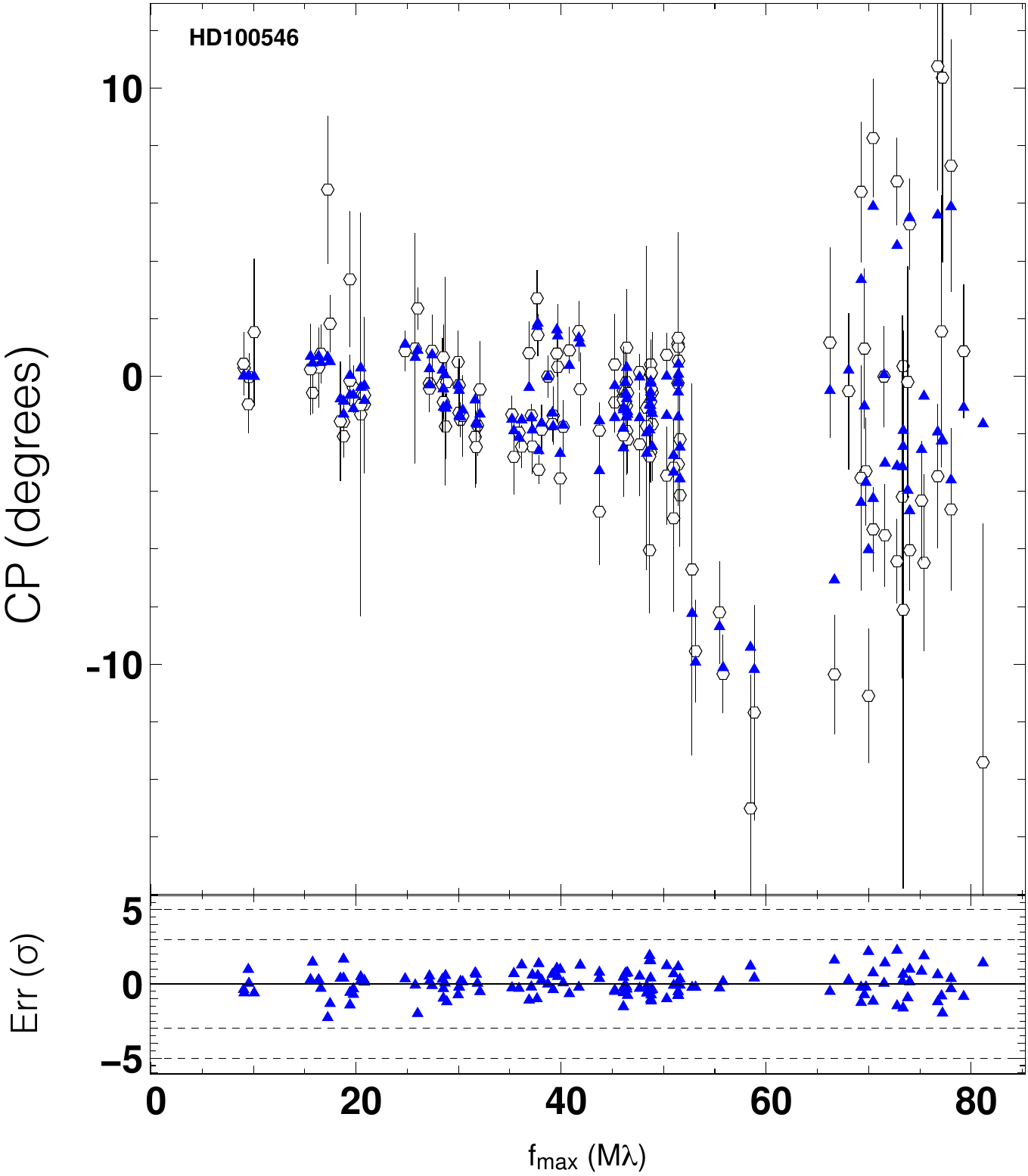} 
\includegraphics[width=4.5cm]{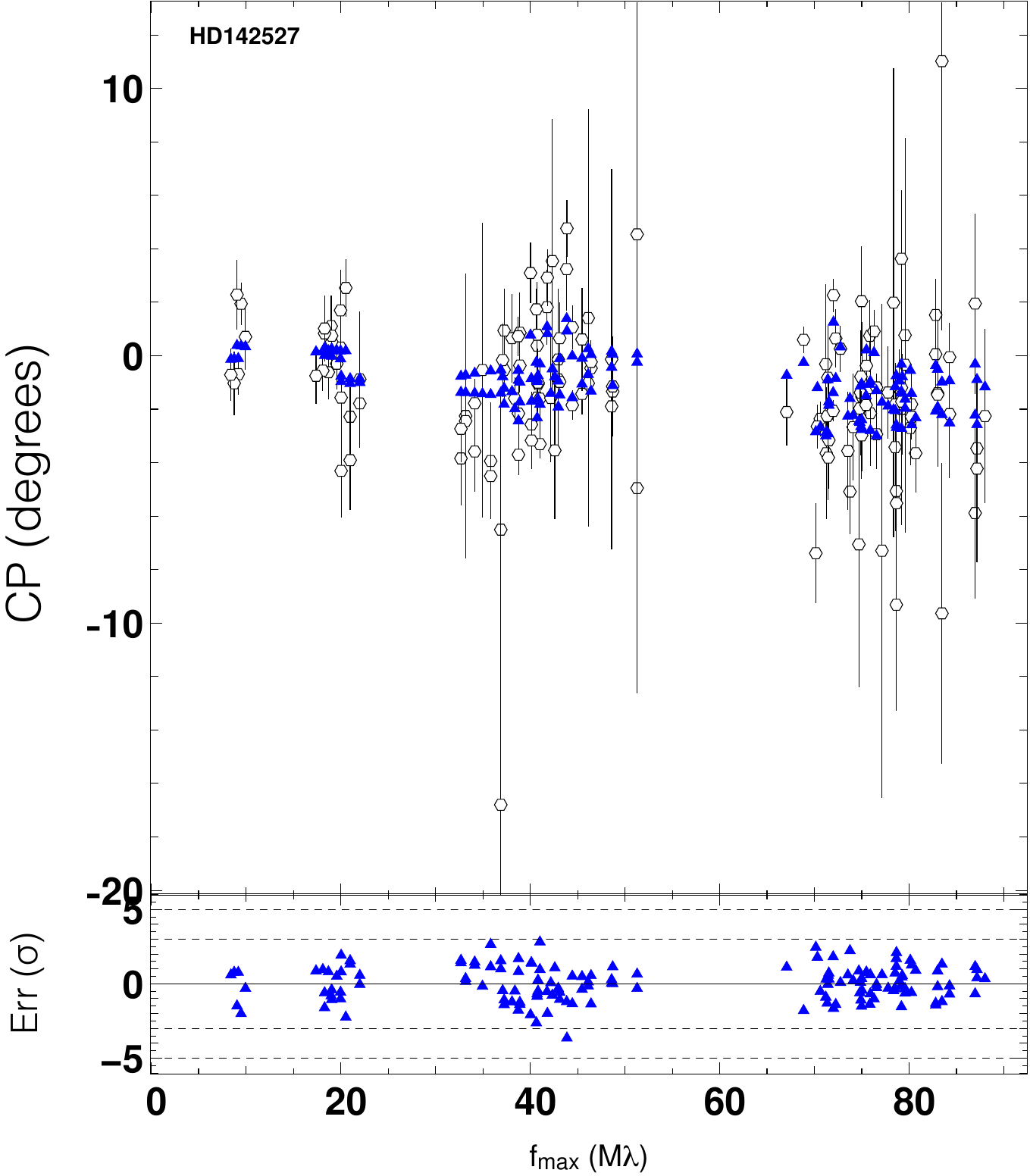} 
\includegraphics[width=4.5cm]{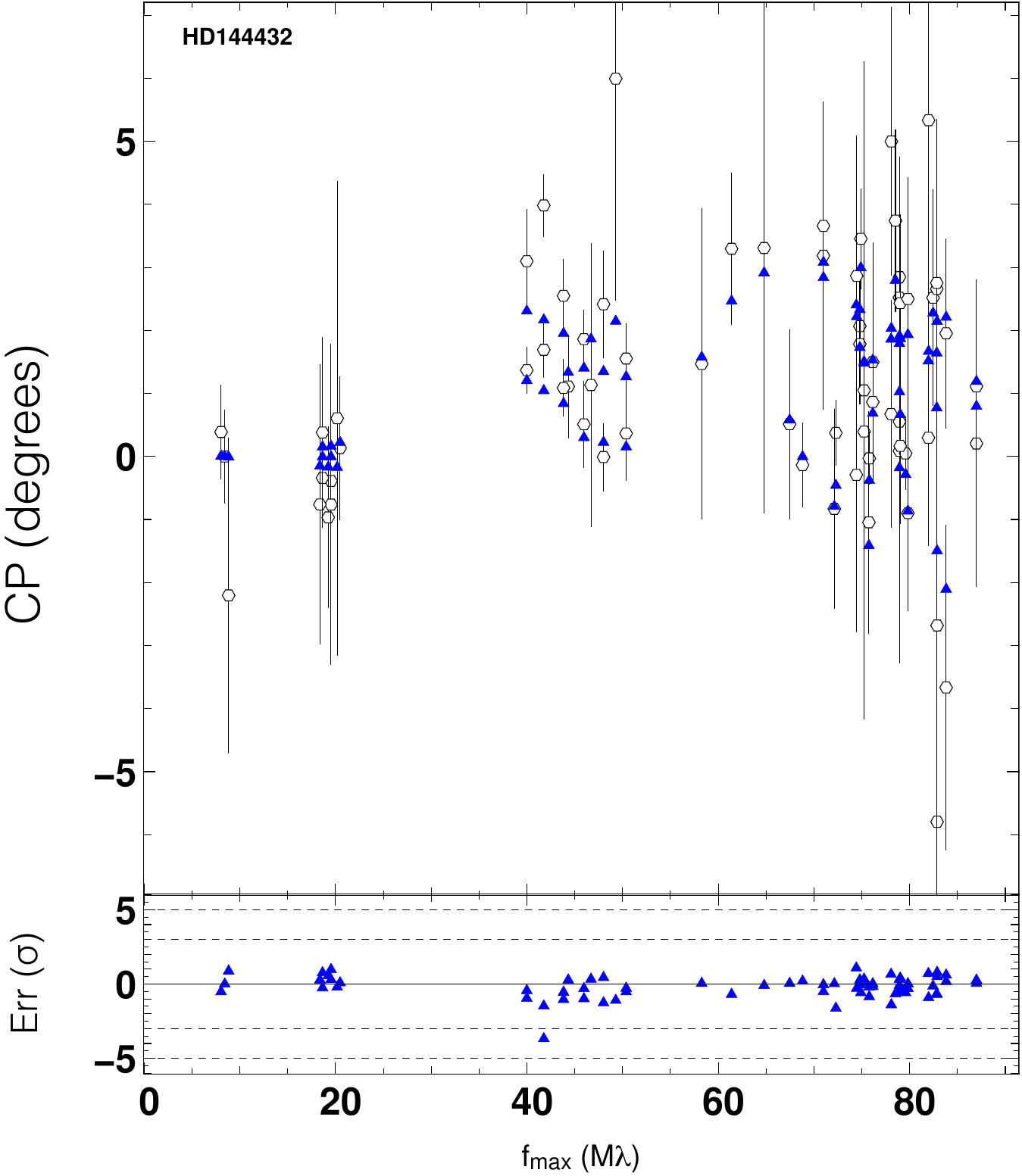} 
\includegraphics[width=4.5cm]{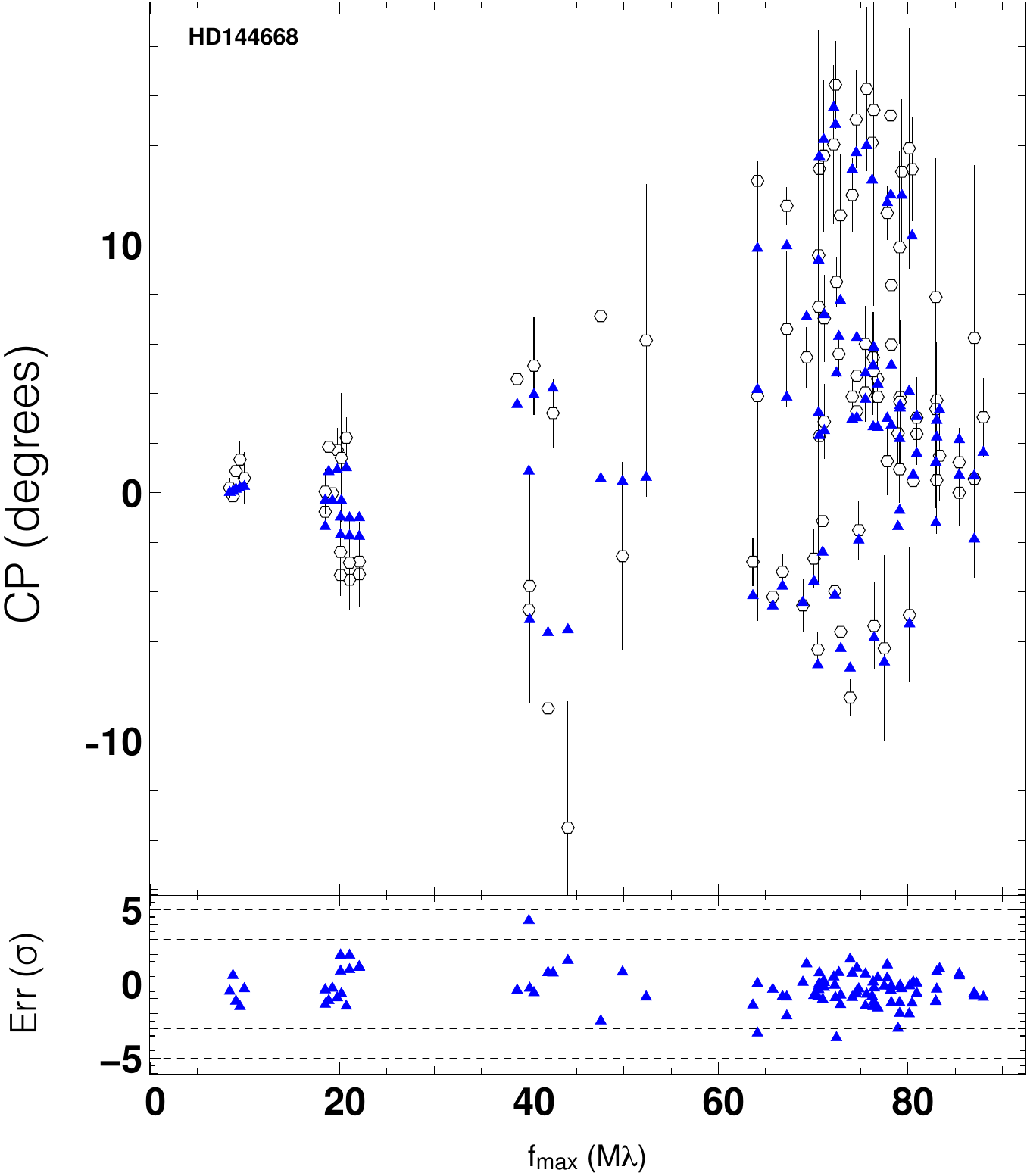} 
\includegraphics[width=4.5cm]{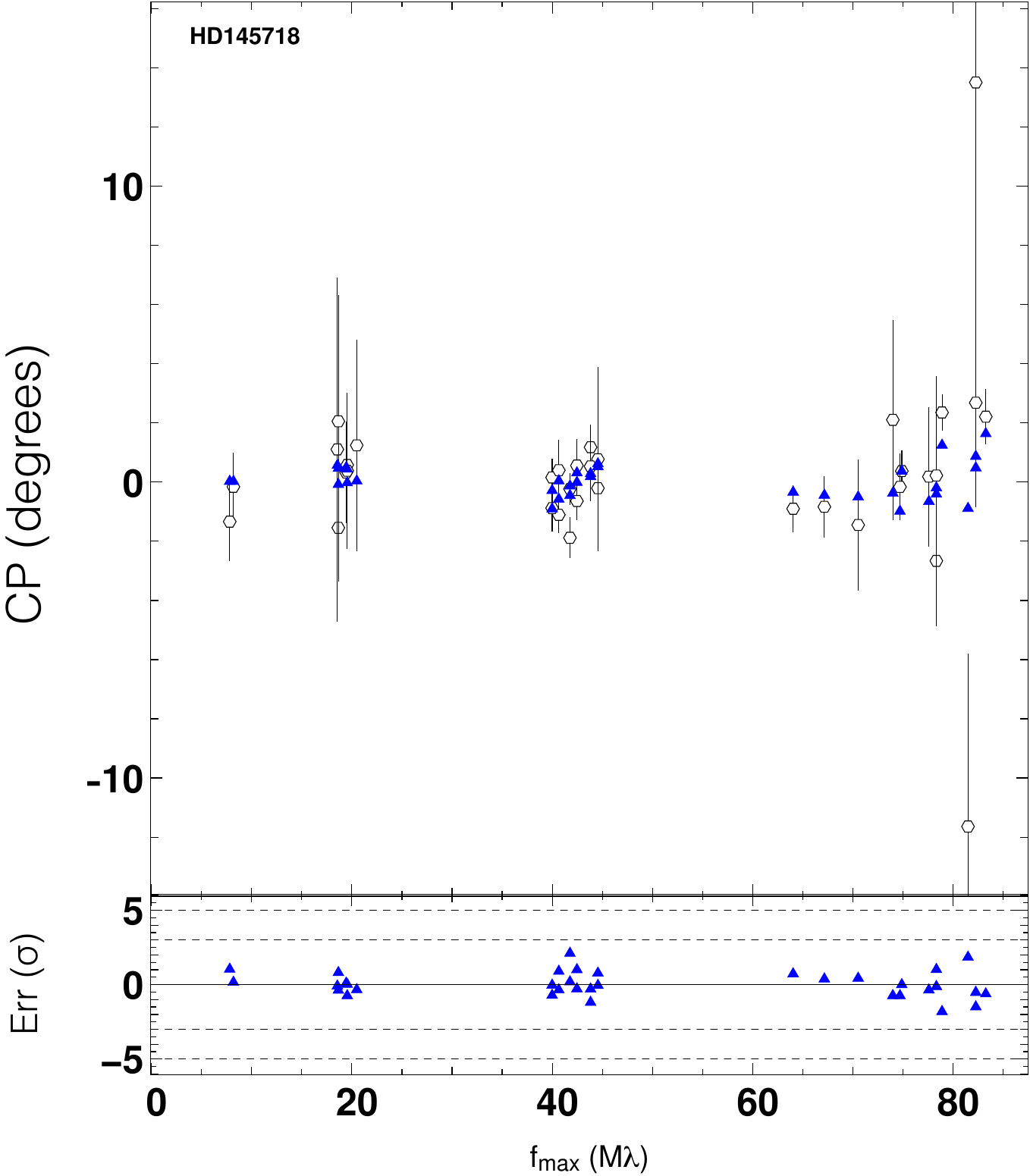} 
\includegraphics[width=4.5cm]{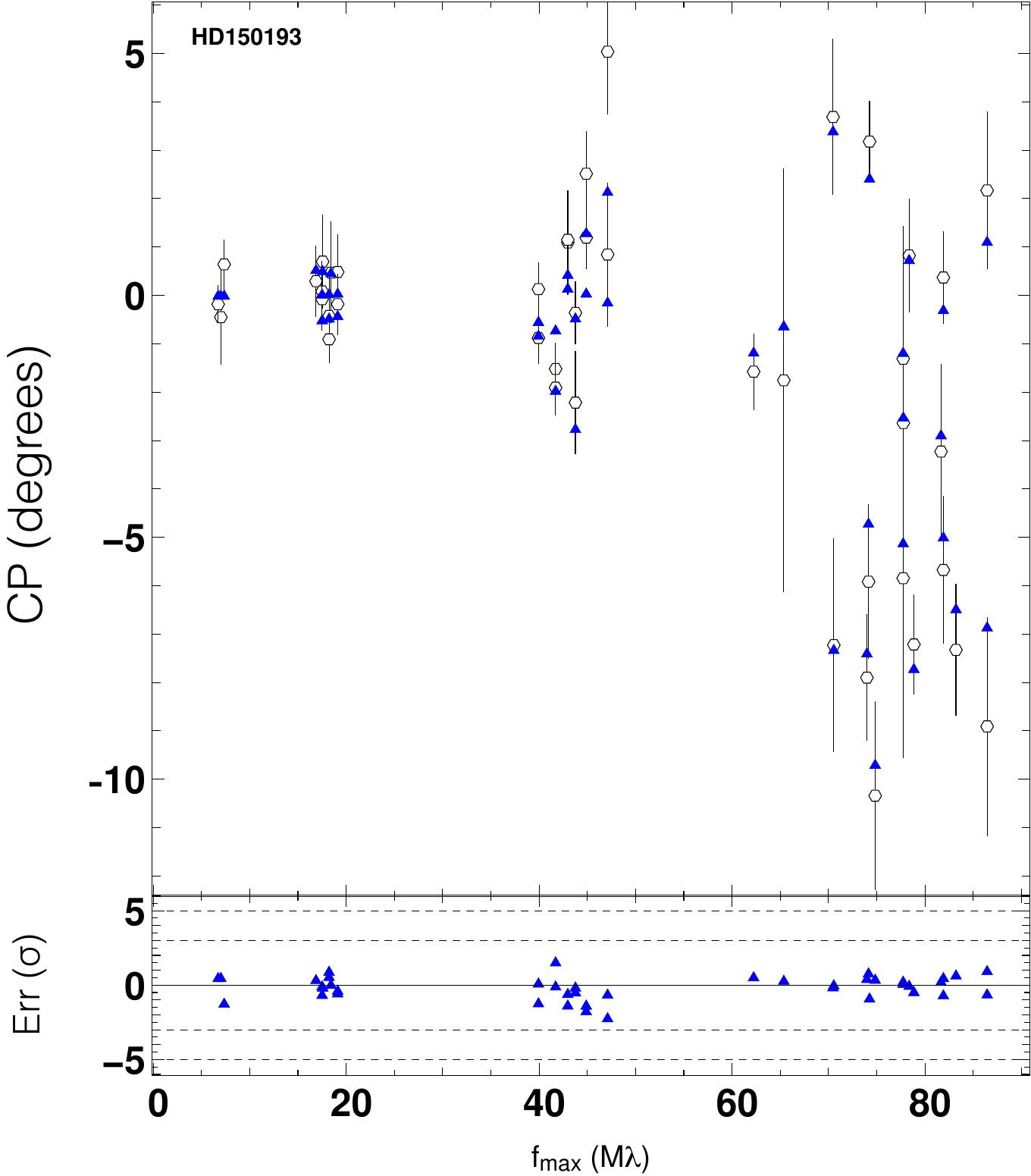} 
\includegraphics[width=4.5cm]{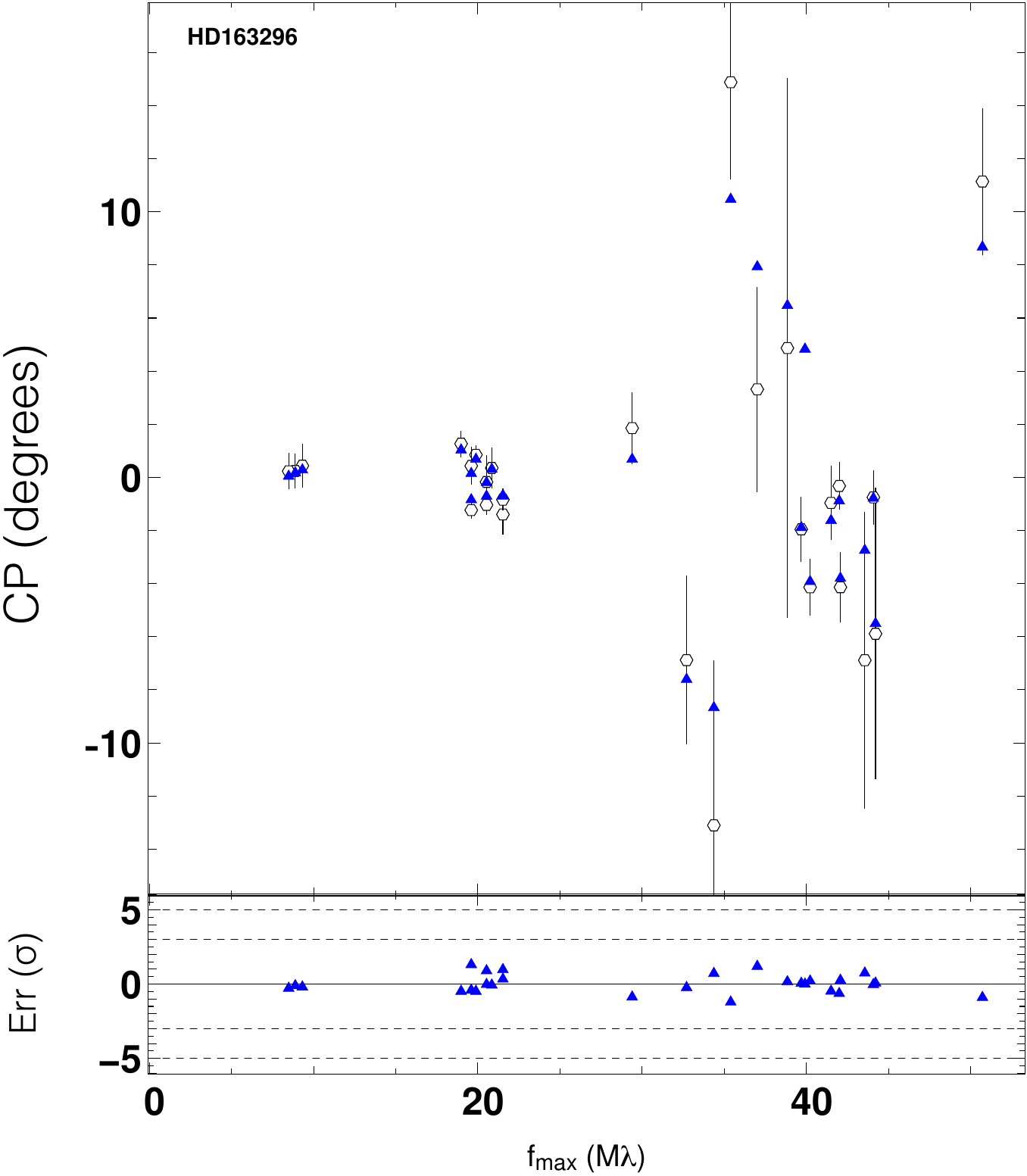} 
\includegraphics[width=4.5cm]{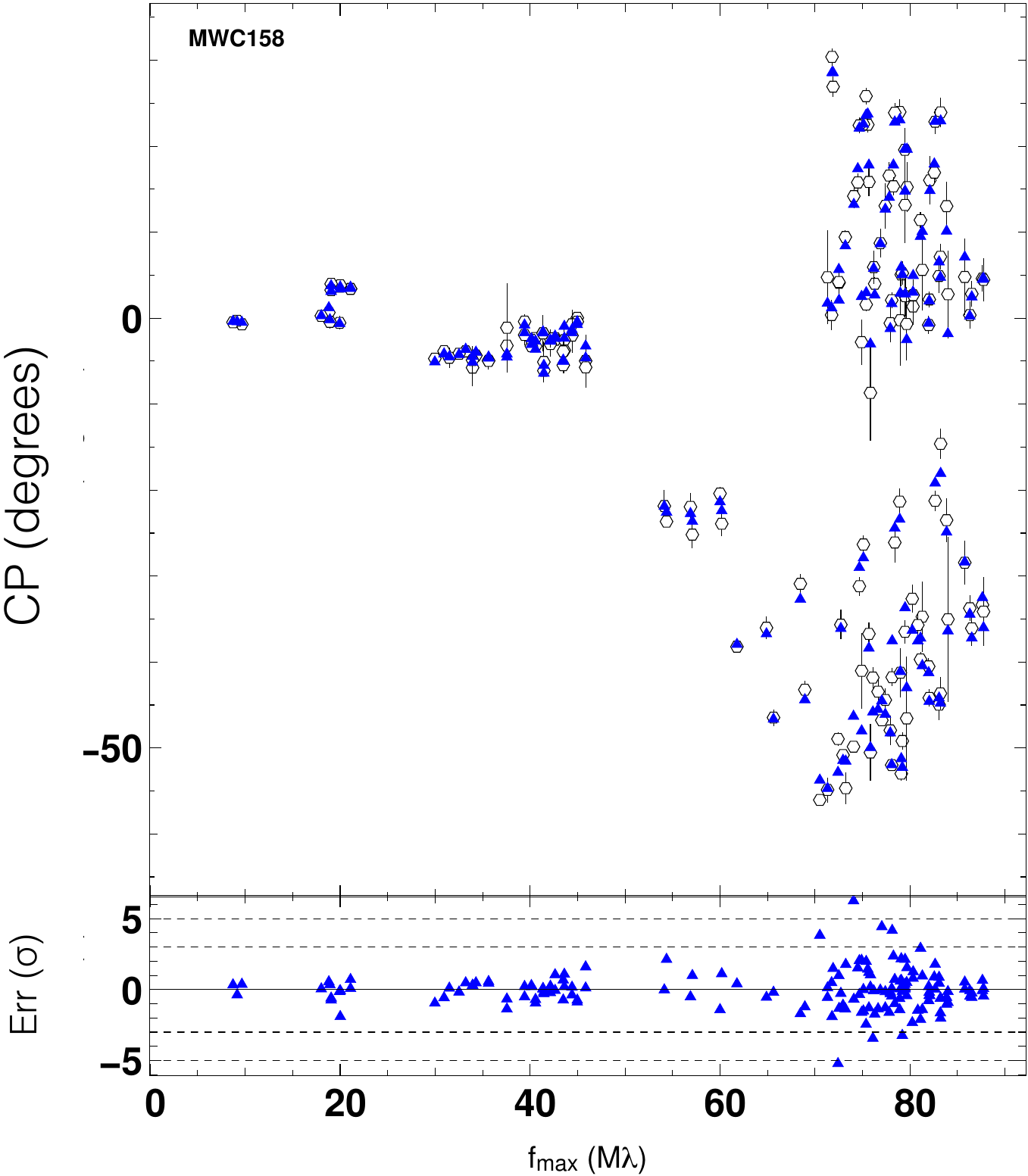} 
\caption{Closure phases of the dataset and the images. For each target, the top panel shows the closure phases for the data (black circles) and the image (blue triangles). The bottom panel shows the residuals normalized by the error bars on each data point.}
         \label{fig:data3}
   \end{figure*}

   \begin{figure*}
   \centering
\includegraphics[width=5cm]{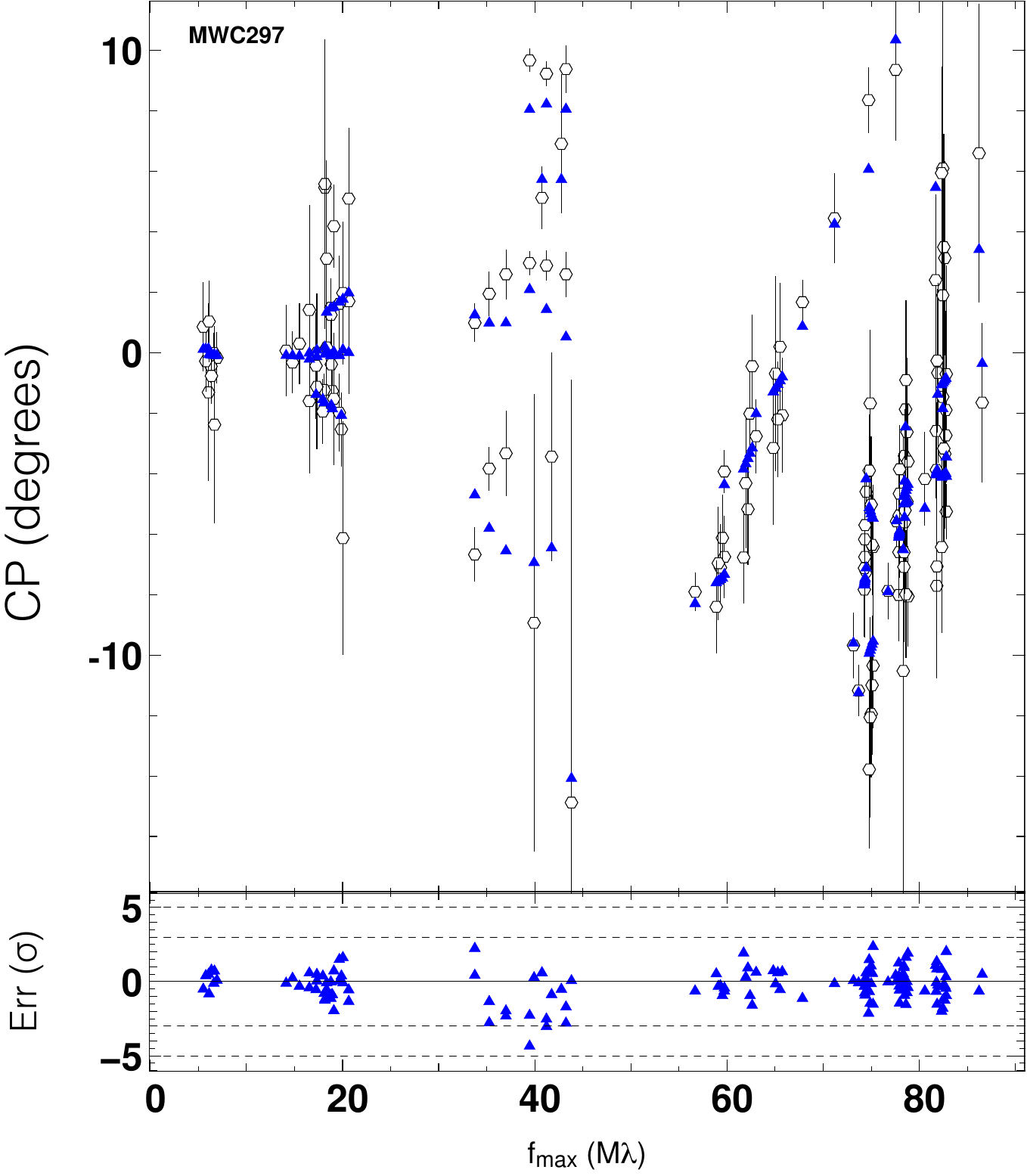} 
\includegraphics[width=5cm]{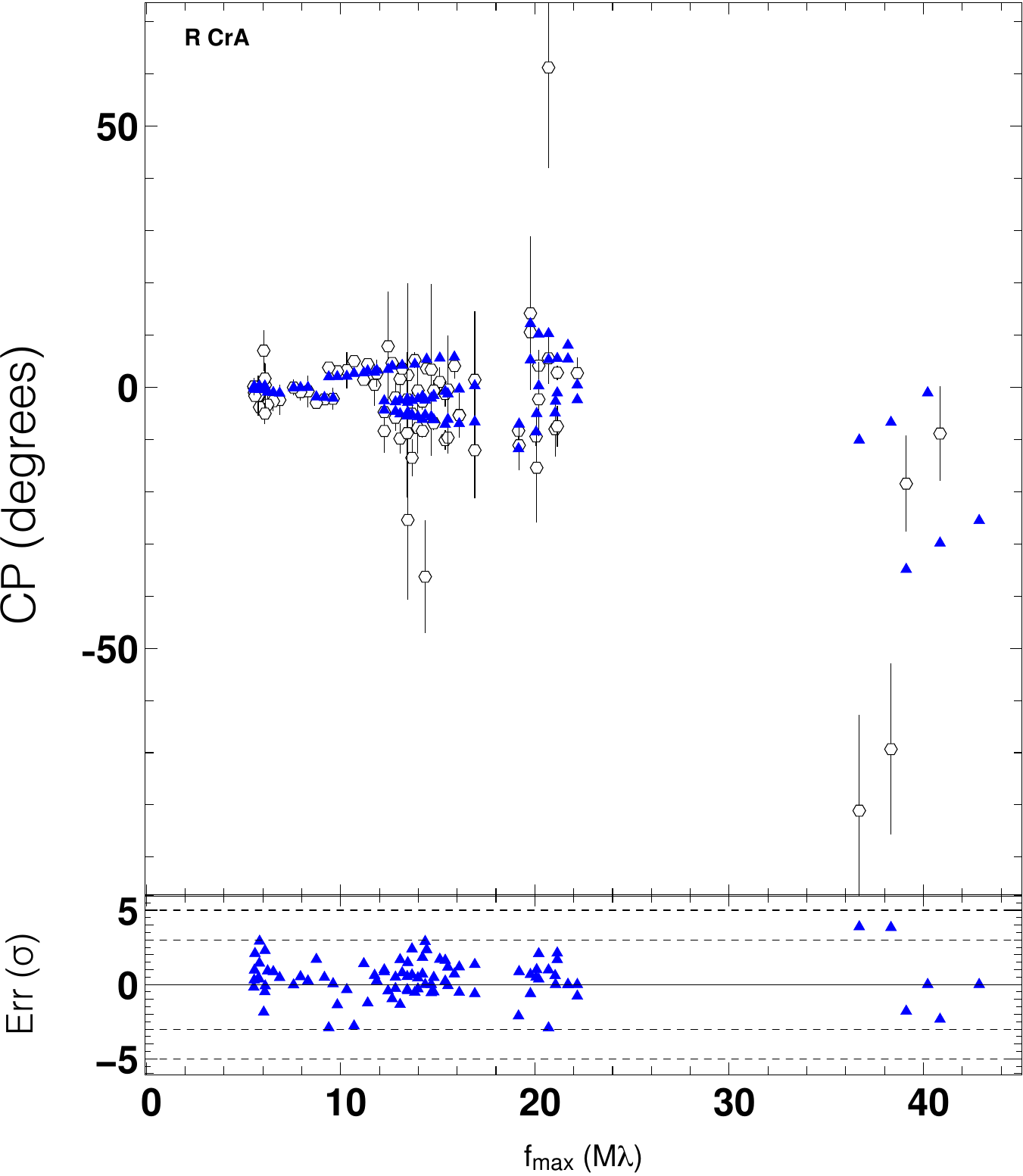} 
\includegraphics[width=5cm]{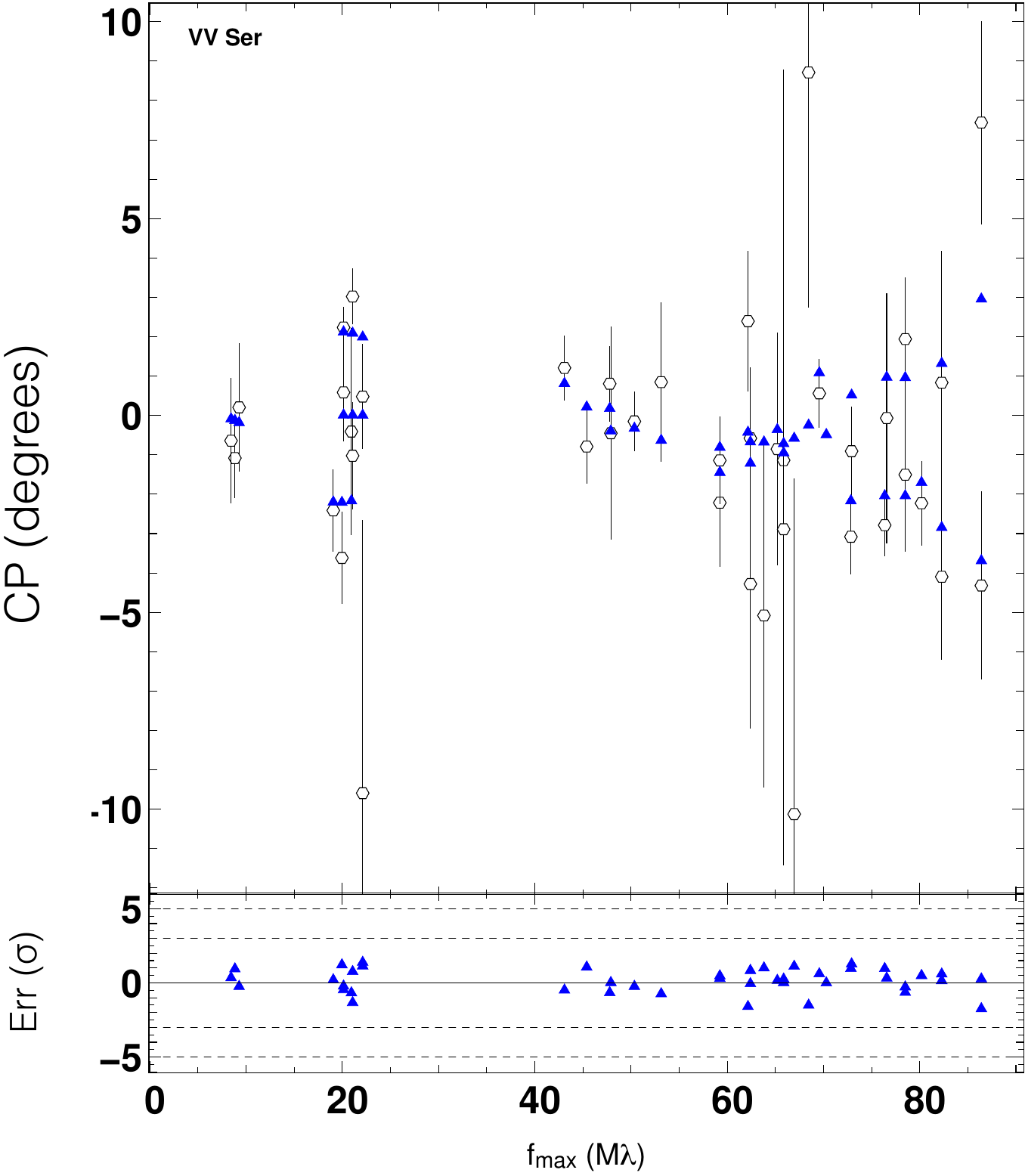} 
\caption{Same as Fig.\,\ref{fig:data3}}
         \label{fig:data4}
   \end{figure*}

\end{appendix}

\end{document}